\documentclass[pra,twocolumn]{revtex4}

\usepackage{graphicx}
\usepackage{epstopdf}
\usepackage{subfigure}
\usepackage{color}

\newcommand{\ket}[1]{\left | #1 \right\rangle}
\newcommand{\bra}[1]{\left \langle #1 \right |}
\newcommand{\proj}[1]{\ket{#1}\!\bra{#1}}

\newcommand{\eref}[1]{(\ref{#1})}

\begin{document}

\title{Quantum simulations and many-body physics with light}

\date{\today}

\author{Changsuk Noh$^{1,2}$ and Dimitris G. Angelakis$^{1,3}$}
\address{$^1$ Centre for Quantum Technologies, National University of Singapore, 3 Science Drive 2, Singapore 117543}
\address{$^2$ Korea Institute for Advanced Study, 85 Hoegiro, Seoul 130722}
\address{$^3$ School of Electronic and Computer Engineering, Technical University of Crete, Chania, Crete, Greece, 73100}

\begin{abstract}

In this review we discuss the works in the area of quantum simulation and many-body physics with light, from the early  proposals on equilibrium models to the more recent works in driven dissipative platforms. We start by describing the founding works on Jaynes-Cummings-Hubbard model and the corresponding photon-blockade induced Mott transitions and continue by discussing the proposals to simulate effective spin models and fractional quantum Hall states in coupled resonator arrays (CRAs). We also analyze the recent efforts to study out-of-equilibrium many-body effects using driven CRAs, including the predictions for photon fermionization and crystallization in driven rings of CRAs as well as other dynamical and transient phenomena. We  try to summarise some of the relatively recent results predicting exotic phases such as super-solidity and Majorana like modes and then shift our attention to developments involving one dimensional nonlinear slow light setups. There the simulation of strongly correlated phases characterising Tonks-Girardeau gases, Luttinger liquids, and interacting relativistic fermionic models is described. We review the major theory results and also briefly outline recent developments in ongoing experimental efforts involving different platforms in circuit QED, photonic crystals and nanophotonic fibers interfaced with cold atoms. 

\end{abstract}

\maketitle

\tableofcontents

\section{Introduction}
\label{sec:introduction}

Quantum simulators offer a promising alternative when analytical and numerical methods fail in analyzing models with strong correlations. Strong correlations characterise effects in many different fields of physical science. To date, a collection of such effects in different areas ranging from condensed matter physics to relativistic quantum theories and nanotechnology have been simulated \cite{CiracZoller2012a,HaukeLewenstein2012a}, using different platforms such as trapped ions \cite{BlattRoos2012} and cold atoms in optical lattices \cite{RevModPhys.80.885}.

In parallel with the progress in laser cooling that led to ultracold atoms and ion traps, the field of quantum nonlinear optics and cavity QED has seen great advances in the last two decades. This has motivated proposals to study many-body physics using strongly correlated photons (SCPs). Initially, coupled resonator arrays (CRAs) were considered where photons trapped in resonators were assumed to be interacting with real or artificial atoms. Pioneering works suggested the  simulation of Mott transitions in CRAs assuming either i) two-level dopants leading to what is now known as the Jaynes-Cummings-Hubbard (JCH) model \cite{AngelakisBose2007} or ii) four-level atoms with configurations used in electromagnetically induced transparency, giving rise to strongly interacting polaritons \cite{HartmannPlenio2006}, and calculated the phase diagram of the JCH model using mean field theory \cite{GreentreeHollenberg2006}.
These were followed soon after by a number of works investigating: beyond-mean-field effects using numerical methods \cite{RossiniFazio2007}, strongly correlated polaritons in photonic crystals \cite{NaYamamoto2007a}, quantum state transfer \cite{BoseBurgarth2007,HuSun2007a,ChakrabartiSreekumari2011}, propagation of photonic or atomic excitations \cite{MakinHollenberg2009}, quantum phase transition in the Tavis-Cummings lattice model \cite{Knapvon-der-Linden2010a}, simulation of spin models \cite{HartmannPlenio2007a,AngelakisKay2008,ChoBose2008}, and the fractional quantum Hall effect \cite{ChoBose2008b}. 
We would like to mention here that there also exists a line of works which bring the exotic physics of single particle relativistic effects into the laboratory. These have also been categorised as quantum or optical simulations and range from numerous theoretical works involving the Dirac equation and associated phenomena including  \textit{Zitterbewegung} and Klein-tunneling \cite{LamataSolano2007,VaishnavClark2008,JuzeliunasOhberg2008,OtterbachFleischhauer2009,Longhi2010,CasanovaSolano2010,NohAngelakis2012,AngelakisNoh2014} to experimental implementations of them in various platforms like the ones mentioned above \cite{GerritsmaRoos2010,GerritsmaRoos2011} and linear optics \cite{DreisowSzameit2010,DreisowSzameit2012}. In this review however, we will focus on simulations of strongly correlated phenomena with strongly interacting atom-photon systems.

These early proposals, and the numerous ones that followed studying the JCH system from different perspectives, marked the beginning of a novel direction in doing many-body physics and quantum simulations using light-matter systems. To mention a few of the works without trying to be exhaustive, we refer the reader to: proposals for entanglement generation \cite{HuoSun2008a,ChoBose2008c,IrishIrish2009a,GiampaoloIlluminati2010a}, polaritonic phase diagram  studies using numerical and approximate analytic techniques \cite{AichhornLittlewood2008,ZhaoUeda2008a,MeringSinger2009a,SchmidtBlatter2010a,Knapvon-der-Linden2010c}, works on different connecting geometries \cite{MakinGreentree2008a,ChenGuo2010}, two component models and the emergent solitonic behaviour \cite{HartmannPlenio2008a,PaternostroKim2009a}, the strong coupling theory for JCH \cite{SchmidtBlatter2009a}, and applications in quantum information processing \cite{KyosevaKwek2010a}. Implementations-wise, the coupled QED resonators found its ideal realization in  circuit QED platforms \cite{HouckKoch2012a}, enjoying widely-tunable coupling strengths and low decoherence rates. Other technologies such as photonic crystal structures and open cavities configurations have also been explored \cite{MajumdarVuckovic2012a,LepertHinds2011}.

The next major development in the field was proposals for simulations of 1D continuous interacting models with photons and polaritons in  quantum nonlinear optical setups. Here, light pulses trapped in hollow-core fibers interacting with cold atoms were shown to behave as a Tonks-Girardeau gas \cite{ChangDemler2008}. This was soon extended to Luttinger liquids exhibiting spin charge separation \cite{AngelakisKwek2011a}. More recently, a simulation of 1D interacting relativistic theories as described by the Thirring model was proposed, using polarised photons acting as effective fermions \cite{AngelakisKorepin2013}. 
 The ability to efficiently measure correlation functions of photonic states, the integrability of the proposed structures with other optical components on a chip, and the ability to operate at higher and even room temperatures, made quantum simulations with SCPs a quickly evolving field with many theoretical and experimental groups actively pursuing research.

We note here that the idea of using exciton polaritons (excitons coupled to photons in semiconductor materials) to realize quantum fluids has been developed in parallel, motivated by progress in controlling light-matter interactions in semiconductor structures \cite{CarusottoCiuti2013}. The interaction strength is typically weak in exciton-polariton systems but ways to enhance polariton-polariton interaction strength have been proposed (see for example \cite{CarusottoImamoglu2010}). Experimentally, the observation of Bose-Einstein condensation by Kasprzak et al.~in 2006 \cite{KasprzakDang2006} has stirred enormous interest in this system. The relationship between the paraxial propagation of a light beam in bulk-nonlinear media and the many-body quantum nonlinear Schr\"odinger equation has been pioneered by Lai and Haus \cite{LaiHaus1989a,LaiHaus1989b} and the possibility to use the latter type of system for quantum simulation has been recently addressed in \cite{LarreCarusotto2015}. Exciting developments in these fields have been extensively reviewed in \cite{CarusottoCiuti2013}, to which we refer the interested reader.

The aim of the current report is to review the evolution of SCPs as they have emerged from the early cavity QED approaches. The first part of this manuscript briefly reviews the early proposals on the simulations of Mott transition, the use of Mott state in quantum computing, quantum Heisenberg spin models, and the fractional quantum Hall effect in CRAs.  The second part presents more recent results in the direction of out-of-equilibrium properties of lossy driven CRAs. We discuss similarities and differences between the JCH and the Bose-Hubbard systems and review various phenomena arising in either or both of these systems. Fermionization and crystallization of photons are first reviewed, which is followed by the works on dynamical signatures of superfluid-Mott transition and the so-called localization-delocalization transition. We then review more exotic effects such as photonic supersolid-like phases and Majorana-like modes and finish off with a survey of studies on smaller arrays and possible experimental implementations. 
The final part of this review is devoted to reviewing the recent progress in simulating  strongly correlated one-dimensional systems using slow light in a hollow-core or tapered fiber coupled to an ensemble of cold atoms. In this part, we first review the proposal for realising a photonic Tonks-Girardeau gas, and then proceed to follow up works on simulating spin-charge separation and a photonic Luttinger liquid, the so-called pinning transition,  BCS-BEC crossover, and interacting relativistic theories with photons.  Although the focus in this part is on the main theoretical proposals, we also briefly refer to the most recent developments in ongoing experimental efforts in nonlinear fiber systems.

\section{Equilibrium many-body effects with coupled resonator arrays}


The early proposals on simulations of quantum many-body effects with coupled resonator arrays were partially motivated by: 1) advances in realising coupled resonator optical waveguides (CROWs) and 2) breakthroughs in achieving strong coupling between quantum emitters and single photons in various light-matter interfaces. CROWs were initially envisioned in photonic crystals structures \cite{Yariv:99} and were used mainly for classical microwave optics applications. These works were followed by studies in the optical frequencies for applications such as efficient guiding of light \cite{BayindirOzbay2000} and building Mach-Zehnder interferometers \cite{PalocziMookherjea2003,ShihMarshall2004}.  The latter were extended, among others, to 2D CROW structures in 2D photonic crystals  for tunable optical delay components and nonlinear optics \cite{AltugVuckovic2004}. Strong coupling between single emitters and single photons was also implemented soon after both in semiconductor \cite{BadolatoImamoglu2005} and circuit QED platforms \cite{WallraffSchoelkopf2004}. Motivated by these progress,  and the race for coming up with efficient QIP implementations, the use of cavity QED physics in CROWs doped with quantum emitters was proposed for implementing photonic phase gates \cite{AngelakisEkert2007}.

Next, the use of CRAs for quantum many-body simulations was proposed in an array of coupled QED resonators interacting strongly with {\em single} two level systems. Such a setup was envisioned as a photonic analogue of an optical lattice in atomic systems, that could allow both the trapping of photonic excitations and strong interactions between them. Taking advantage of the photon blockade effect--naturally arising when the atom is strongly coupled to the localized resonator photon \cite{ImamogluDeutsch1997,BirnbaumKimble2005}--it was shown that an insulator phase of the total (atomic$+$photonic or simply {\em polaritonic}) excitations was possible in an array of atom-resonator systems \cite{AngelakisBose2007}. In the same work, it was shown that, in parallel with optical lattices, one could mimic a quantum phase transition from the photonic insulator to the superfluid phase by varying the strength of the effective nonlinearity through the tuning of the atom-resonator detuning. Soon after, the complete phase diagram has been reproduced, first in 2D using mean field theory \cite{GreentreeHollenberg2006} and then in 1D using numerical methods based on DMRG techniques \cite{RossiniFazio2007}.  

The Jaynes-Cummings-Hubbard (JCH) Hamiltonian as it came to be known, involves a combination of bosonic (photons) and spin excitations (the atoms).  In general, the insulator-to-superfluid transition in this system is also accompanied by a transition in the nature of the excitations from polaritonic to photonic. Another important difference is that here quasiparticles are formed from {\em hybrid light-matter excitations}, rather than real physical particles such as atoms, and the former's internal structure makes the phase diagram extremely rich as we will discuss in detail later. In addition to the phase transition, the possibility to simulate the dynamics of XY spin chains with individual-spin manipulation was proposed in the same work \cite{AngelakisBose2007}, using a mechanism different from that of optical lattices \cite{DuanLukin2003}.  In parallel with the above, the possibility to use 4-level atoms and external fields to simulate strongly interacting dynamics was proposed by Hartmann et al.~\cite{HartmannPlenio2006}. In this case, a time-dependent approach was employed, where the hopping/interaction ratio is varied in time such that the system initially starts deep in the Mott regime and ends up in the superfluid regime. A good agreement between the actually system and the BH model is found upon comparing the dynamics of the number fluctuations.  This and subsequent works on simulating spin models using the four-level system approach have been reviewed in \cite{HartmannPlenio2008}.

\subsection{Photon-blockade induced Mott transitions and the Jaynes-Cummings-Hubbard Model}

In the following we describe in detail the initial proposal by Angelakis et al.~\cite{AngelakisBose2007} (first appeared in June 2006 and published in 2007), to implement the JCH model using two-level-system-doped CRAs in order to observe the Mott-superfluid transition of photons.

\begin{figure}
\begin{center}
\includegraphics[width=0.7\columnwidth]{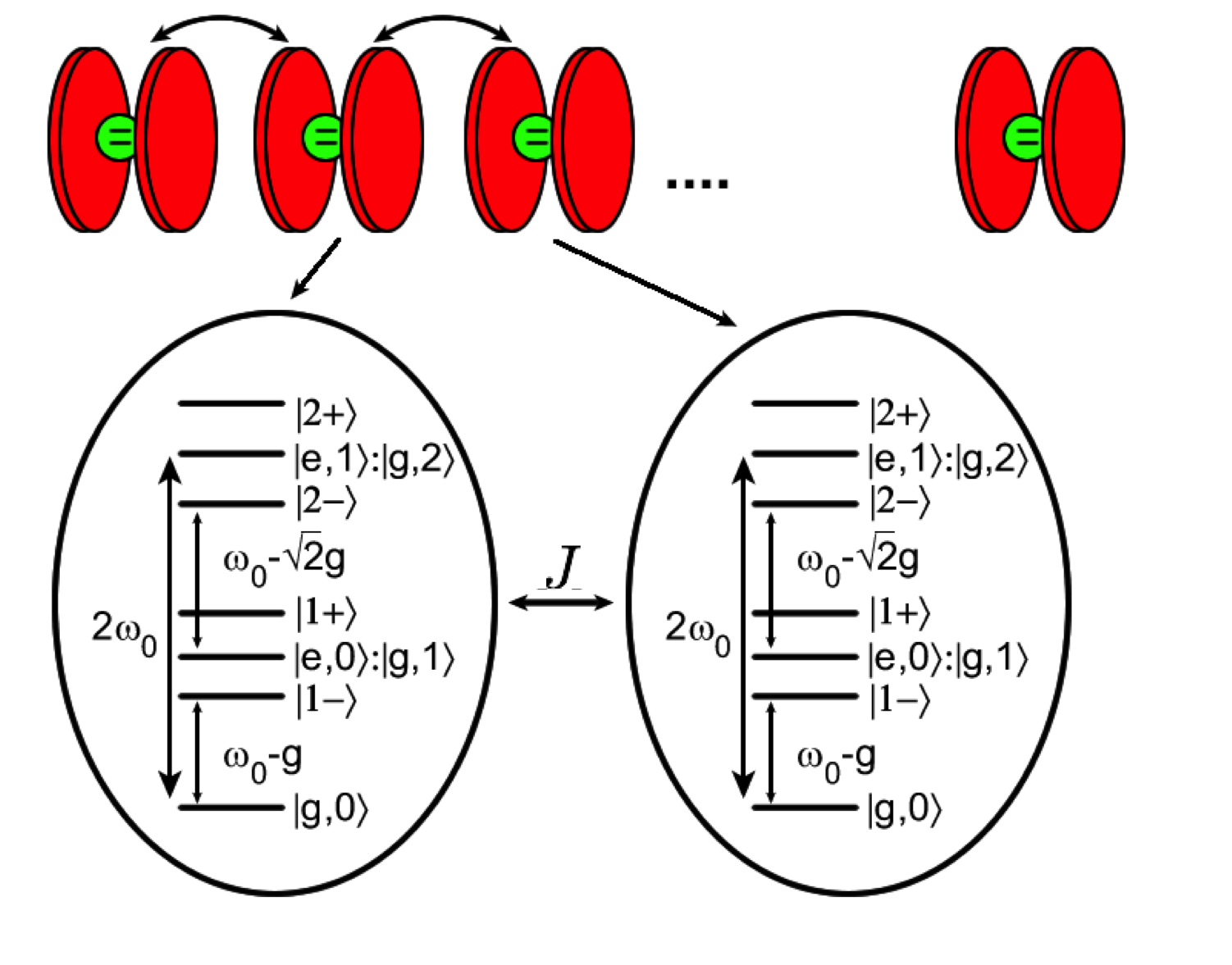}
\end{center}
\caption{ (Top) Schematic representation of a series of cavities coupled through photon hopping. (bottom) The local energy structure in each atom-resonator system where the system's bare states are denoted as  $\{ |e,n-1\rangle_k, |g,n\rangle_k \}$ and $|n\pm\rangle_k$ denotes the local eigenstates also known as dressed states. Reprinted from Angelakis et al.~\cite{AngelakisBose2007}.}
\label{coupled cavities}
\end{figure}

Consider a series of optical resonators coupled through direct photon hopping. The hopping could be generated through an overlap in the photonic modes as we assumed here or using mediating elements such as fibers \cite{LepertHinds2011}. The system is described by the Hamiltonian 
\begin{equation}
H=\sum_{k=1}^{N}\omega_da^{\dagger}_{k}a_{k}-\sum_{k=1}^{N}J(a^{\dagger}_{k}a_{k+1}+{\rm H.c.})
\end{equation}
which corresponds to a series of coupled quantum harmonic oscillators. The photon frequency and hopping rate is $\omega_{d}$ and $J$ respectively and no nonlinearity is present yet. In order to simulate strongly correlated effects, interaction is necessary and the most natural way to  induce it is to dope each resonator with a two level system. The latter could be an atom or a quantum dot or a superconducting qubit. Let us denote $|g\rangle_{k}$ and $|e\rangle_{k}$ as its ground and excited states at site $k$, respectively. The Hamiltonian describing the total system can be divided into three terms: $H_{free}$ the Hamiltonian for the free resonators and dopants, $H_{int}$ the Hamiltonian describing the local coupling of the photon and the dopant, and $H_{hop}$ for the photon hopping between cavities. Explicitly, these read
\begin{eqnarray}
H_{free}&&=\omega_{d}\sum_{k=1}^N a_k^\dagger a_k+\omega_{0}\sum_k|e\rangle_{k} \langle e|_{k},  \label{Hfree} \\
H_{int}&&=g \sum_{k=1}^N(a_k^\dagger|g\rangle_{k}\langle e|_{k}+{\rm H.c.}),\label{Hint} \\
H_{hop}&&= -J\sum_{k=1}^N(a_k^\dagger a_{k+1} + {\rm  H.c.}), \label{Hhop}
\end{eqnarray} 
where $g$ is the atom-photon coupling strength. 
The  local part of the Hamiltonian $H_{free}+H_{int}$ can be diagonalized in a basis of mixed photonic and atomic excitations. These excitations, known as dressed states or polaritons in quantum optics and condensed-matter communities respectively, involve a mixture of photonic and atomic excitations and are defined by annihilation operators
$P_{k}^{(\pm,n)}=|g,0\rangle_k \langle n\pm|_{k}$. The polaritonic states of the $k$th atom-resonator system are given by $|n+\rangle_k=(\sin\theta_{n}|g,n\rangle_k +\cos\theta_{n}|e,n-1\rangle_k)/\sqrt2$ and $|n-\rangle_k=(\cos\theta_{n}|g,n\rangle_k -\sin\theta_{n}|e,n-1\rangle_k)/\sqrt2$, where $|n\rangle_k$ denotes the $n$-photon Fock state in the $k$th resonator, $\tan2\theta_{n}= 2g\sqrt n/\Delta$, and $\Delta=\omega_0-\omega_d$ is the atom-light detuning. These polaritons have the energies $E^{\pm}_{n}=n\omega_{d} + \Delta/2\pm \sqrt{ng^2+(\Delta/2)^2}$ and are also eigenstates of the sum of the photonic and atomic excitation operators ${\cal N}_k=a_k^\dagger a_k+|e\rangle\langle e|_k$ with the eigenvalue $n$ (figure \ref{coupled cavities}).

Studying the energy eigenstates of the system consistent with a given number of net excitations per site (or filling factor), it was shown that the ground state of the system reaches a Mott state of the polaritons for integer values of the filling factor.  In contrast to the optical lattice systems where the lattice depth is tuned, here the detuning $\Delta=\omega_0-\omega_d$ is varied. The latter results in further shifts of the dressed states from the bare resonances, providing an effective control over effective nonlinearity. The hopping strength $J$ is usually fixed by the fabrication process and is harder to change experimentally. To see how the Mott state arises, it is convenient to rewrite the Hamiltonian in terms of the polaritonic operators (assuming $\Delta=0$):
\begin{eqnarray}
H &=&\sum_{k=1}^{N} \sum_{n=1}^{\infty} \bigg[ n(\omega_{d}-g)P_{k}^{(-,n)\dagger}P_{k}^{(-,n)} \nonumber \\ &+& n(\omega_{d}+g)P_{k}^{(+,n)\dagger}P_{k}^{(+,n)} +
g(n-\sqrt{n})P_{k}^{(-,n)\dagger}P_{k}^{(-,n)} \nonumber\\
&+&g(\sqrt{n}-n)P_{k}^{(+,n)\dagger}P_{k}^{(+,n)} \bigg ] \nonumber \\
&-& J\sum_{k=1}^{N}(a_k^\dagger a_{k+1} + {\rm H.c.} ). \label{nodet}.
\end{eqnarray}
Remember that the lower of the two energy eigenstates of the local system with a given number of excitations $\eta$ is the `-' dressed state $|\eta-\rangle$.
Therefore, assuming the regime $Jn<<g\sqrt{n}<<\omega_d$, one needs only consider the first, third and
last terms of the above Hamiltonian for determining the low-lying energy eigenstates of the whole system.
The first line corresponds to a linear spectrum, equivalent to that of a harmonic oscillator of frequency $\omega_{d}-g$. If only that part were present in the Hamiltonian, then it would not cost any extra energy to add an excitation (of frequency $\omega_{d}-g$) to a site already filled with one or more excitations, as opposed to an empty site.

The third term, on the contrary, provides an effective ``on-site" {\it photonic repulsion}, leading to the formation of the Mott state of polaritons as the ground state of the system for a commensurate filling. Reducing the strength of the effective nonlinearity (or the photonic repulsion) through detuning, one can then drive the system to the superfluid (SF) regime. This could be done, for example, by Stark shifting the atomic transitions from the resonator by an external field.  The shift being inversely proportional to the detuning, the new detuned polaritons are not as well separated as before, and in the case of large detuning, it costs little extra energy to add excitations (excite transitions to higher polaritons) in a single site. As a result, the system moves towards the SF regime with increasing detuning. Note here that
the mixed nature of the polaritons could in principle allow for excitations that are mostly photonic, resulting in a photonic Mott state \cite{HartmannPlenio2007b}.


\subsection{Mott and superfluid phases of light in the Jaynes-Cummings-Hubbard model}

Beyond the qualitative description reproduced above, an exact numerical simulation was performed in \cite{AngelakisBose2007}, for a different number of sites (from three to seven) using the Hamiltonian $H = H_{free}+H_{int}+H_{hop}$. Note that such finite-size calculations are often employed to study the physics of the transition between Mott and superfluid phases \cite{JakschZoller1998} because mean-field theories in low dimensions are not guaranteed to work.
We also note here that numerical simulations of the JCH Hamiltonian are more computationally costly than the ones for the BH Hamiltonian due to the existence of both bosonic (photon) and atomic operators in every site.  

\begin{figure}
\center
\includegraphics[width=1.0\columnwidth]{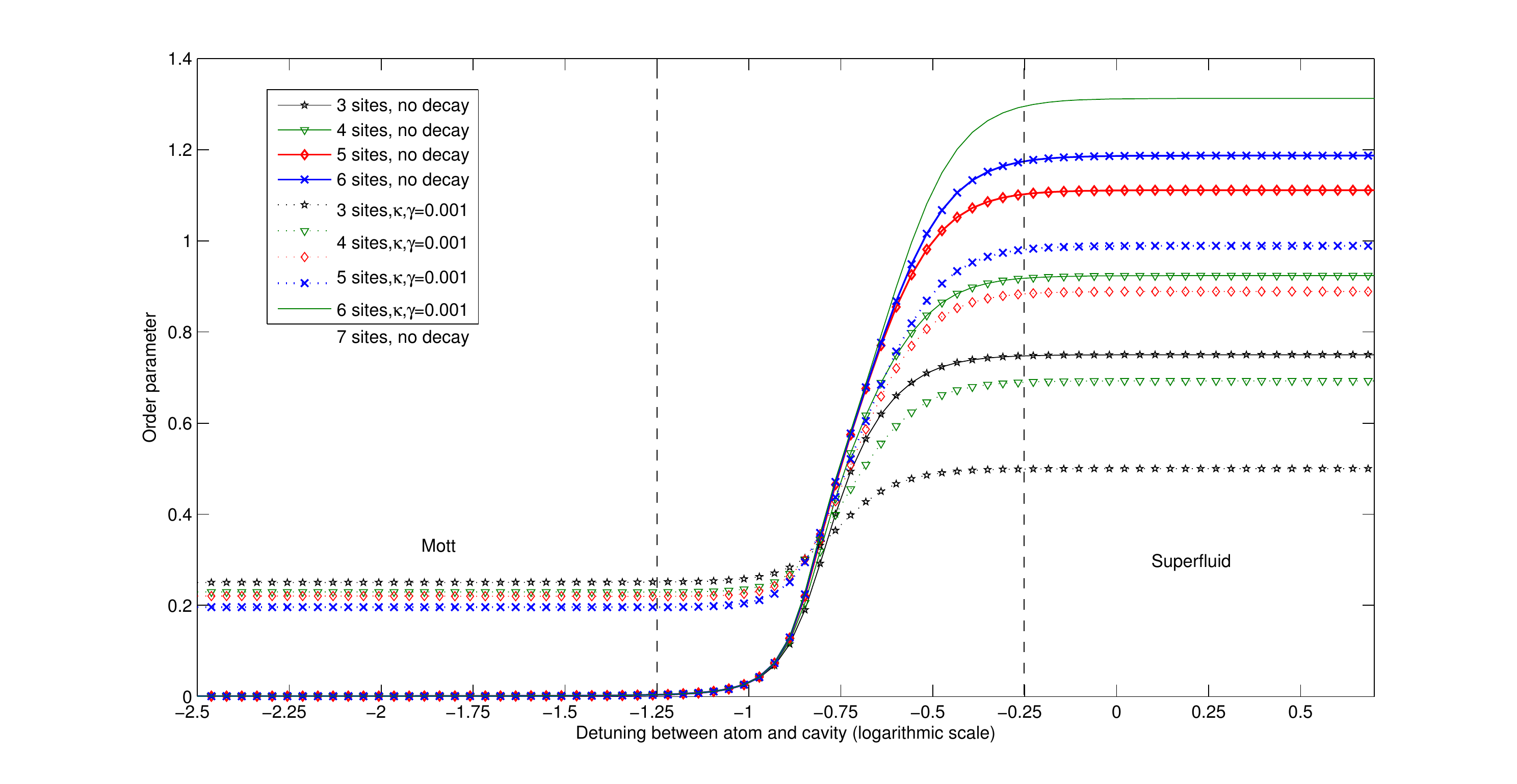}
\caption{The order parameter ($var({\cal N}_k)$) as a function of the detuning $\Delta$ for 3-7 sites, with and without atomic dissipation and resonator leakage. Close to the atom-photon resonance ($0 \le \Delta/g \le 10^{-1})$, where the effective on-site repulsion is maximum (and much larger than the hopping rate $A (=J)$), the system is forced into a polaritonic Fock state with the same number of excitations per site, i.e., the Mott insulator state. Increasing the detuning by applying external fields and inducing Stark shifts ($\Delta \ge g$) weakens the repulsion, leading to the appearance of different coherent superpositions of excitations per site, i.e., a photonic superfluid state. The increase in the system size leads to a sharper transition as expected. Reprinted from Angelakis et al.~\cite{AngelakisBose2007}.} \label{var}
\end{figure}

As discussed in \cite{AngelakisBose2007}, an appropriate order parameter in the JCH lattice is the variance of the polariton number per site, as the usual order parameter employed in other cases--like the expectation value of the annihilation operator in the BH model--is identically zero in a closed system. The variance $var({\cal N}_k)$ is plotted in figure \ref{var} as a function of $\log_{10}(\Delta/g)$ for a filling factor of one net excitation per site. For this plot, we have taken the parameter ratio $g/J=10^{2}$ ($g/J=10^{1}$ gives quantitatively similar results), with $\Delta$ varying from $\sim 10^{-3}g$ to $\sim g$ and $\omega_d,\omega_0 \sim 10^4g$. 
 The order parameter for dissipationless atoms and cavities is displayed as solid curves and  corresponding curves in the presence of a resonator dissipation and atomic spontaneous emission of $\sim 0.01g$ are depicted with dotted lines. 

\begin{figure}
\center
 \includegraphics[width=0.6\columnwidth]{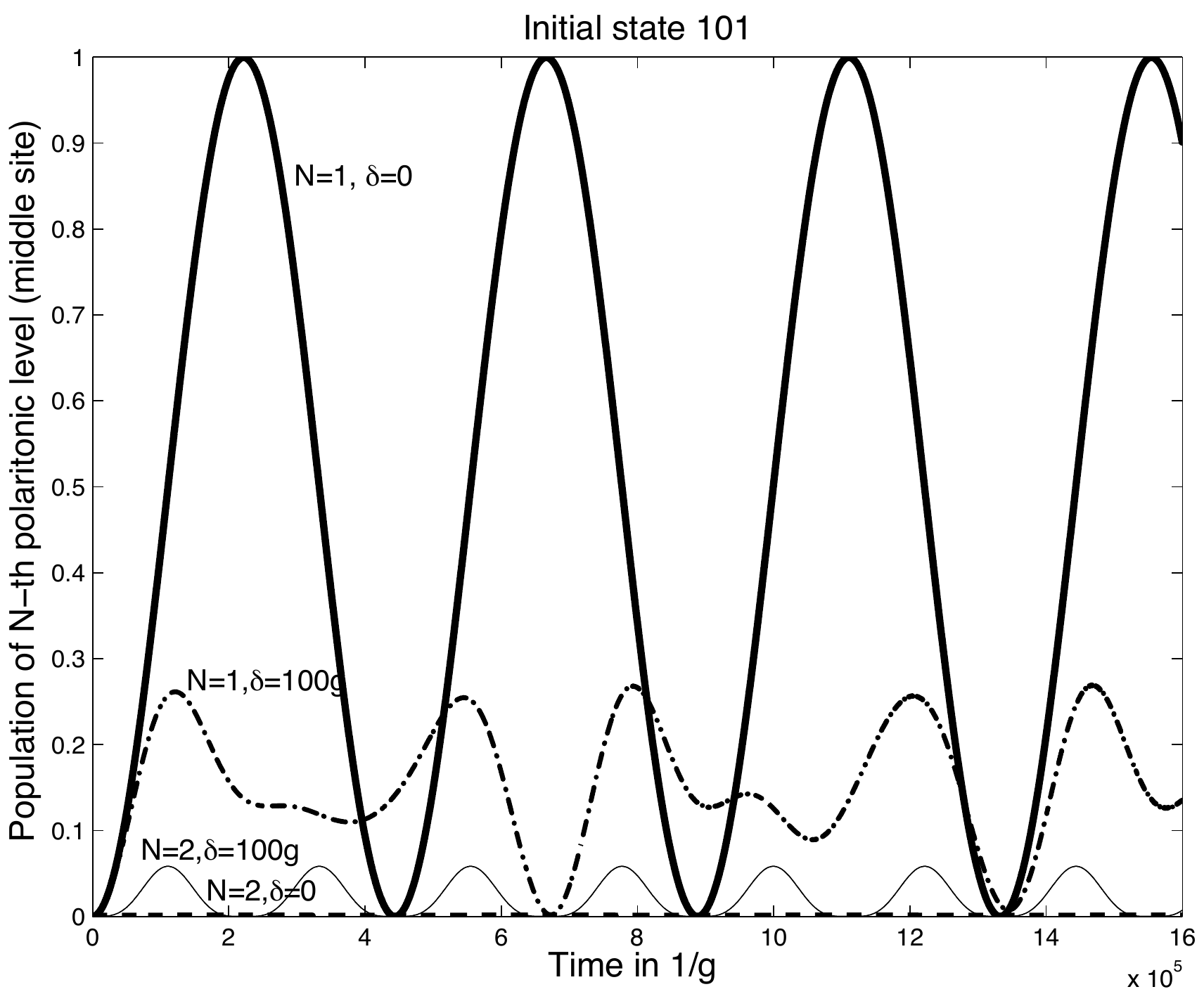}
 \caption{
The probability of exciting the polaritons corresponding to one (thick solid line) or two (thick dashed line)
excitations for the middle resonator which is taken initially to be empty (in the resonant regime). In this case, as we have started
with fewer polaritons than the number of cavities, oscillations occur for the single polariton excitation. The double excitation manifold, 
however, is never occupied due to the blockade effect (thick dashed line). For the detuned case, hopping of more than one polaritons is allowed (SF regime) which is evident as the higher polaritonic manifolds are being populated now (dot-dashed line and thin solid line). Reprinted from Angelakis et al.~\cite{AngelakisBose2007}.}
\label{Mott_dynamics}
\end{figure}

In figure \ref{Mott_dynamics} the quantum dynamics characterizing the excitation probability of the middle site in a chain of three cavities for different regimes are calculated assuming the middle resonator is initially empty. We observe the blockade of higher than one excitation in the limit of resonant interaction (where the Mott phase is expected) as discussed in figure \ref{var}.

\begin{figure}
\includegraphics[width=0.9\columnwidth]{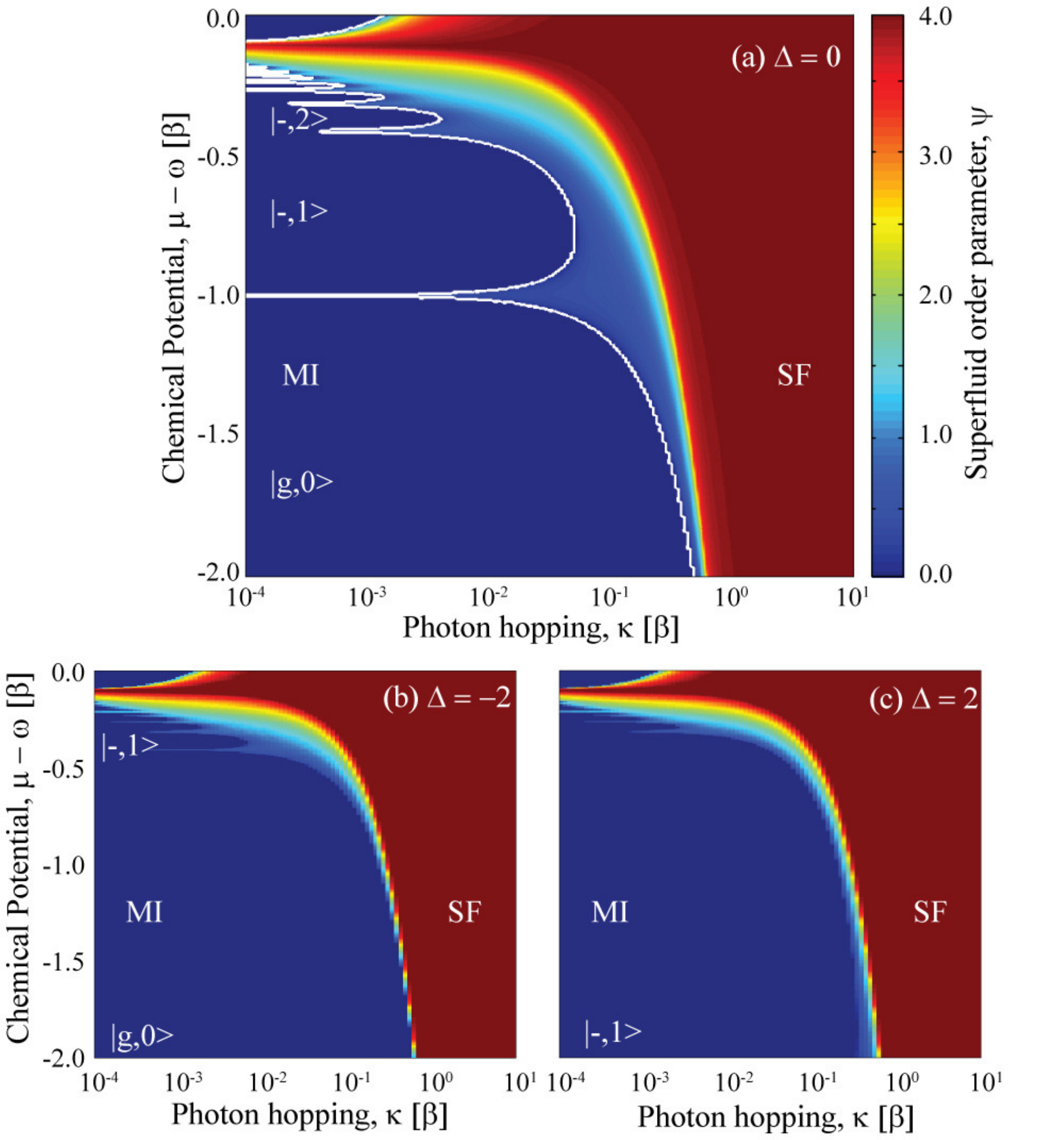} \vspace{0.5cm}

\includegraphics[width=1.0\columnwidth]{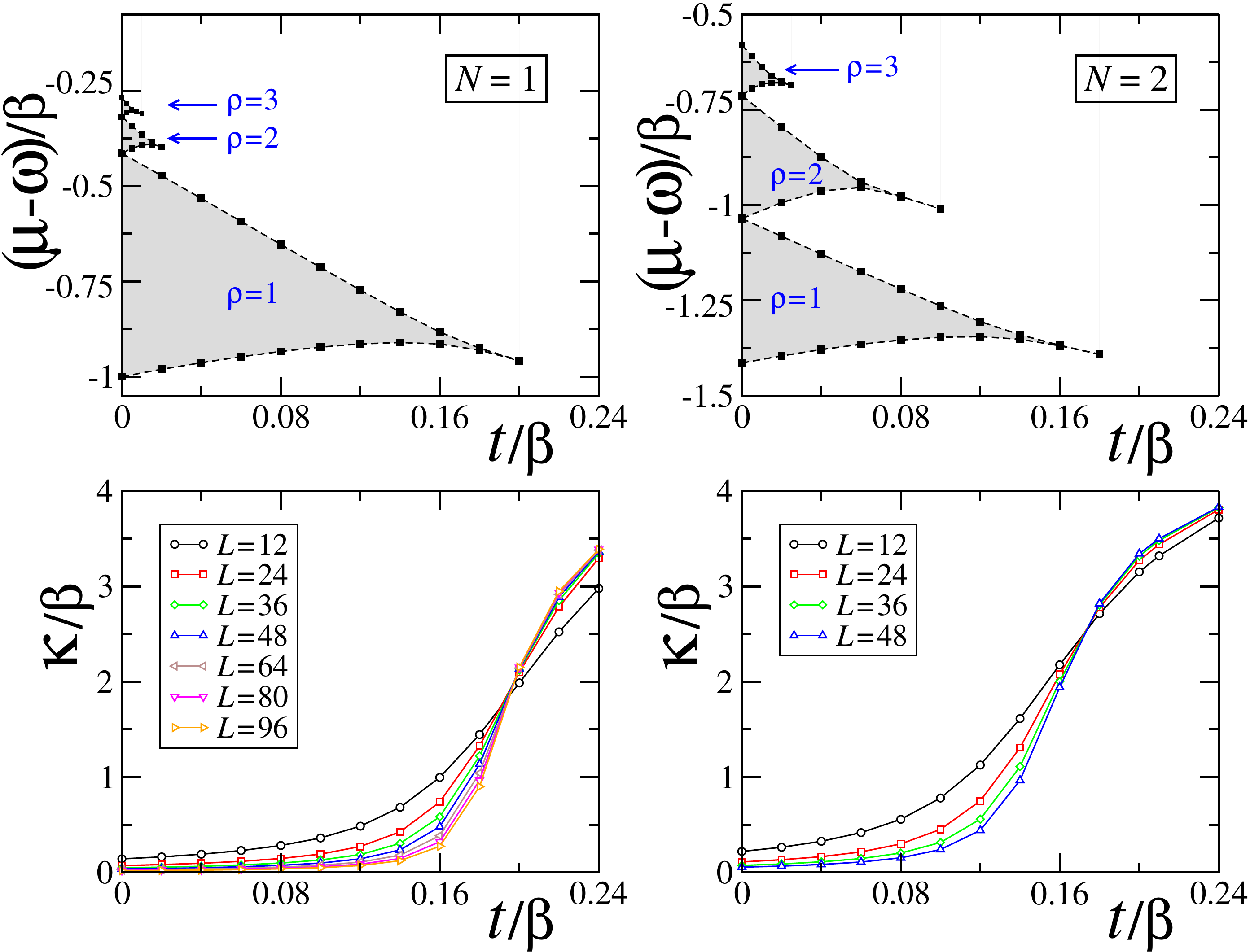}
\caption{(Top) The phases diagram of a 2D JCH system calculated using mean field theory for three different values of atom-photon detuning $\Delta = 0$ (a), $-2g$ (b), and $2g$ (c). $\mu$ is the chemical potential introduced separately to the Hamiltonian, $\kappa$ is the hopping strength ($=J$), $\beta$ is the atom-photon coupling strength ($=g$), and $\omega$ is the bare resonator frequency ($=\omega_d$) \cite{GreentreeHollenberg2006}. Reprinted with permission from Greentree et al.~\cite{GreentreeHollenberg2006}. (Bottom, top row) The phase diagram of a 1D JCH array, calculated using DMRG simulations. $\mu$ is the chemical potential, $\beta$ is the atom-photon coupling strength ($=g$), $t$ is the hopping strength ($=J$), and $\rho$ is the filling factor in the Mott phase. (Bottom, bottom row) The compressibility $\kappa$ as a function of hopping strength, for different system sizes $L$. $N=1$ (left column) and $N=2$ (right column) denotes the number of atoms inside each resonator \cite{RossiniFazio2007}. Reprinted with permission from Rossini and Fazio \cite{RossiniFazio2007}.}
\label{phase-diagrams-equil}
\end{figure}

The phase diagram of the JCH model has first been studied in 2D using mean field theory \cite{GreentreeHollenberg2006} and in 1D using a numerical method based on DMRG techniques \cite{RossiniFazio2007}. Figure \ref{phase-diagrams-equil}(top) shows slices of the superfluid order parameter $\psi = \langle a_i \rangle$ as a function of the photon hopping and the chemical potential for different values of atom-photon detuning. A triangular lattice was assumed and the usual decoupling approximation  $a_{i}^{\dagger}a_{j}= \langle a_{i}^{\dagger} \rangle a_{j}+a^{\dagger}_{i} \langle a_{j} \rangle+\langle a_{i}^{\dagger} \rangle\langle a_{j} \rangle$ was used \cite{GreentreeHollenberg2006}. The bottom panels of figure \ref{phase-diagrams-equil} show the numerically-calculated phase diagram (top row) and the compressibility ($\kappa = \partial \rho/ \partial\mu$, $\rho$ = filling factor) (bottom row) as functions of the hopping strength for a 1D chain. 

In both figures the corresponding Mott and SF phases are explored by changing the hopping strength and the effective photonic chemical potential (the free energy term in the JCH Hamiltonian), following the standard treatment of quantum phase transitions. We note that in CRA experiments the hopping strength is in general not tunable but fixed by the fabrication process. In these systems, it is easier to control the effective repulsion (the photon blockade), as discussed in the previous section. Nevertheless, the phase diagrams are useful in showing the existence of Mott lobes in this photonic system and how they depend on system parameters. These works have been supplemented using the variational cluster approach, where on top of the phase diagram, single-particle spectra and finite temperature effects have been investigated \cite{AichhornLittlewood2008}. Quantum Monte-Carlo method has also been employed to find various properties of the JCH-type systems \cite{ZhaoUeda2008a,PippanHohenadler2009a,HohenadlerPollet2011,HohenadlerSchmidt2012}.

\begin{figure}
\includegraphics[width=1.0\columnwidth]{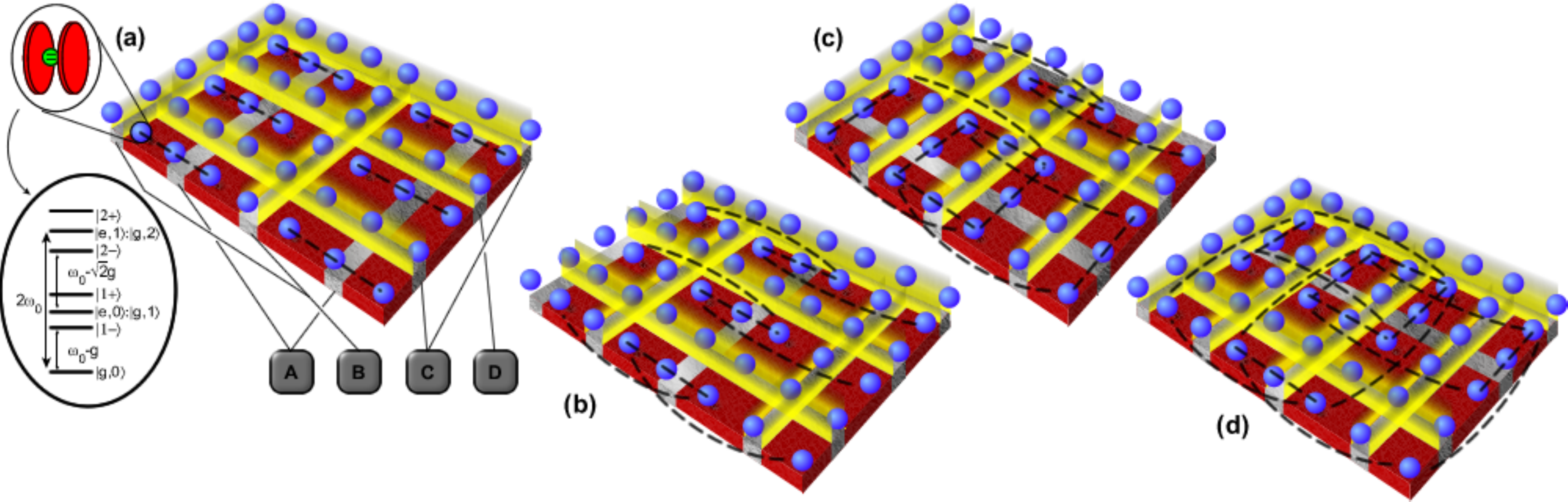}
\caption{
A 2D array of atom-resonator systems in the Mott state 
 can be used to implement cluster state quantum computing, utilizing the natural
 evolution, which in this case is of an XY spin type, and applying external global gates to
 switch on and off the dynamics.
by properly applying local external fields, utilizing the fact that
the cavities can be well separated. 
Reprinted from Angelakis and Kay \cite{AngelakisKay2008}.} 
\label{cluster}
\end{figure}

{\it An application of the Mott regime for implementing quantum information processing (QIP)}: 
The typical implementation of cluster state quantum computing requires initializing all qubits in
a 2D lattice in the $\ket{+}=(\ket{0}+\ket{1})/\sqrt{2}$ state and then performing controlled-phase gates ($CP$) between all nearest-neighbours \cite{Nielsen2006}. In the resonator array system the first proposal to create cluster states utilized the Mott state and the resulting XY dynamics \cite{AngelakisKay2008}. The ground state $|g,0\rangle$ and the first excited state $|1-\rangle$ of the combined atom-photon (polaritonic) system in each site was used as a qubit as depicted in figure \ref{cluster}. 

The need to disable the evolution was realized by combining the system's natural dynamics with a protocol where some of the available physical qubits are used as gate ``mediators"  and the rest as the logical qubits. The mediators can be turned on and off on demand by tuning the atoms in the corresponding sites on and off resonance from their cavities. This is done by Stark shifting them through the application of an external field (gates A, B, C D in figure \ref{cluster}), effectively shifting the site frequencies and inhibiting photon hopping, thereby isolating each logical qubit. 
Soon after there have also been other approaches focusing on the generation of the Ising interaction \cite{LiGuo2007,LiGuo2009a}  and also towards the simulation of a family of effective spin models in CRAs \cite{HartmannPlenio2007a,ChenGuo2010,KayAngelakis2008,BurgarthBurgarth2007,SarkarSarkar2012}. These will be analyzed in the next section.

\subsection{Effective spin models}

Quantum spin models have played a crucial role as basic models accounting for magnetic and thermodynamic properties of many-body systems. In implementations involving  arrays of Josephson junctions \cite{BruderSchon1993} or quantum dots \cite{LossDiVincenzo1998}, the spin-chain Hamiltonian naturally emerges from the spin-like coupling between qubits, albeit with limited control over the coupling constants. In optical lattice simulators, perturbative dynamics with respect to the Mott-insulator state allows for an effective spin-chain Hamiltonian \cite{DuanLukin2003,Garcia-RipollCirac2004}. This direction has the advantage that the spin-coupling constants can be optically controlled to a great extent.  In CRAs, a spin is represented by either polariton states or hyperfine  ground levels of the embedded emitter and the system was shown to generate an effective Heisenberg model with optically tunable interactions
\begin{equation}
H_{eff}= - \sum_{\langle j,k \rangle}\lambda_z \sigma_j^Z\sigma_k^Z+\lambda_x\sigma_j^X\sigma_k^X + \lambda_y\sigma_j^y\sigma_k^y + h\sum_j \sigma^Z_j,  \label{eqn:Heff}
\end{equation}
where the angled brackets denote the sum over nearest neighbours. We will use the same notation throughout this review.

\begin{figure}
  \centering
\includegraphics[width=0.8\columnwidth]{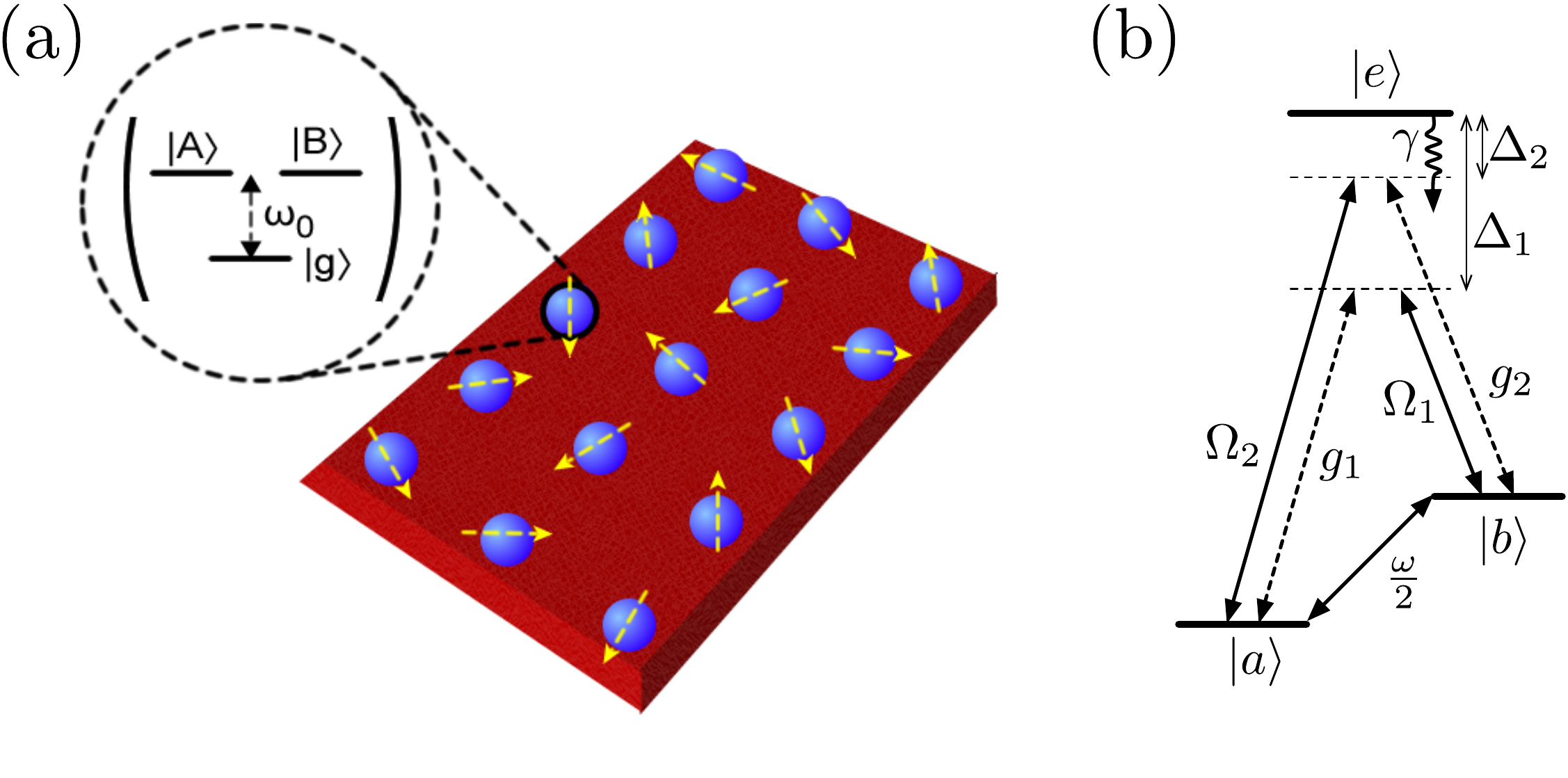}
\caption{
(a) Two-dimensional CRA for simulating Heisenberg spin models. Each resonator is doped with a V-type atom which is coupled to two orthogonal modes \cite{KayAngelakis2008}. Reprinted with permission from Kay and Angelakis \cite{KayAngelakis2008}. (b) The atomic level structure and optical transitions involved in order to simulate higher-spin Heisenberg models \cite{ChoBose2008a}. Reprinted from Cho et al.~\cite{ChoBose2008a}.}
\label{simple}
\end{figure}

In the work \cite{KayAngelakis2008}, an array of cavities was assumed on the vertices of a 2D lattice (note that any array is in principle possible) as depicted in figure \ref{simple}(a). Each resonator is doped with a single system (which we refer to as an atom), whose energy level structure comprises of a ground state $\ket{g}$ and two degenerate excited states $\ket{A}$ and $\ket{B}$. Within each lattice site, the Hamiltonian takes the form
\begin{eqnarray}
H_{0}=&&\omega_0\left( a_i^\dagger a_i+ b_i^\dagger b_i+\proj{A}_i+\proj{B}_i\right) \nonumber\\
&& +g\left(\ket{A}\bra{g}_i\otimes a_i+\ket{g}\bra{A}_i\otimes
a_i^\dagger \right) \nonumber\\
&&+g\left(\ket{B}\bra{g}_i\otimes
b_i+\ket{g}\bra{B}_i\otimes b_i^\dagger\right),   \nonumber
\end{eqnarray}
where $a_i^\dagger$ and $b_i^\dagger$ create photons of orthogonal polarizations at site $i$ and the atomic levels and the resonator modes are assumed to be on resonance. The strength of the coupling between the resonator and the atom is denoted by $g$. Coupling between nearest-neighbour resonators is described by the Hamiltonian
\begin{eqnarray}
H_{hop} =  \sum_{\langle i,j \rangle} J_a(a_i^\dagger a_j + \textrm{h.c.}) + J_b(b_i^\dagger b_j + \textrm{h.c.}).
\end{eqnarray}

Assuming strong atom-photon interaction regime, $g \gg J_{a,b}$, and restricting to the unit filling fraction, such that one excitation per site is expected, one obtains an effective Hamiltonian \eref{eqn:Heff} with
\begin{eqnarray}
 h&=&\frac{5}{8g}\left(J_a^2-J_b^2\right), \;\;\;\;  \lambda_z=\frac{9}{32g}\left(J_a^2+J_b^2\right), \nonumber \\ \lambda_x &=& \lambda_y =\frac{9J_aJ_b}{16g}.   \nonumber
\end{eqnarray}
Note that we have ignored the constant term $\kappa=\frac{31}{32g}\left(J_a^2+J_b^2\right)$.
The local magnetic fields can be manipulated by applying local Stark fields, thereby leaving an XXZ Hamiltonian where the coefficients $\lambda_z$ and $\lambda_x$ are independently tunable (at device-manufacture stage). 
This approach allows a stronger spin-spin coupling than an alternative one in \cite{HartmannPlenio2007a}, but lacks the optical control of the coupling. The latter exhibits greater optical controllability, but relies on rapid switching of optical pulses and the consequent Trotter expansion, which unavoidably introduces additional errors and makes the implementation more difficult. 

The question of simulating chains of higher spins, which may have a completely different phase diagram, was dealt with in \cite{ChoBose2008a}. In this proposal, a fixed number of atoms are confined in each resonator with collectively applied constant laser fields, and is in a regime where both atomic and resonator excitations are suppressed (figure \ref{simple}(b)). It was thus shown that as well as optically controlling the effective spin-chain Hamiltonian, it is also possible to engineer the magnitude of the spin.


\subsection{Fractional Quantum Hall states of photons}

Since its first discovery in a semiconductor heterostructure in the early 1980s~\cite{TsuiGossard1982}, systems exhibiting FQHE has now become routinely available in laboratories. In a different context, the achievement of trapping ultracold atomic gases in a strongly correlated regime has prompted an interest in mimicking various condensed matter systems with trapped atoms, thereby allowing one to tackle such complex systems in unprecedented ways~\cite{LewensteinSen2007}. The
great advantage of this approach is that such a tailored system is highly controllable at the microscopic level and is close to idealized quantum models used in condensed matter systems. In the context of FQHE, a 2D atomic gas is trapped in a harmonic potential~\cite{WilkinGunn2000,JuzeliunasOhberg2004} or in an optical lattice~\cite{JakschZoller2003,SorensenLukin2005}. Since the atoms in consideration have no real charge, the magnetic field is simulated artificially. For instance, it can be done by rapidly rotating the harmonic trap~\cite{WilkinGunn2000}, by exploiting electromagnetically induced transparency~\cite{JuzeliunasOhberg2004}, or by modulating the optical lattice potential~\cite{JakschZoller2003,SorensenLukin2005}.

\begin{figure}[ht]
\includegraphics[width=1.0\columnwidth]{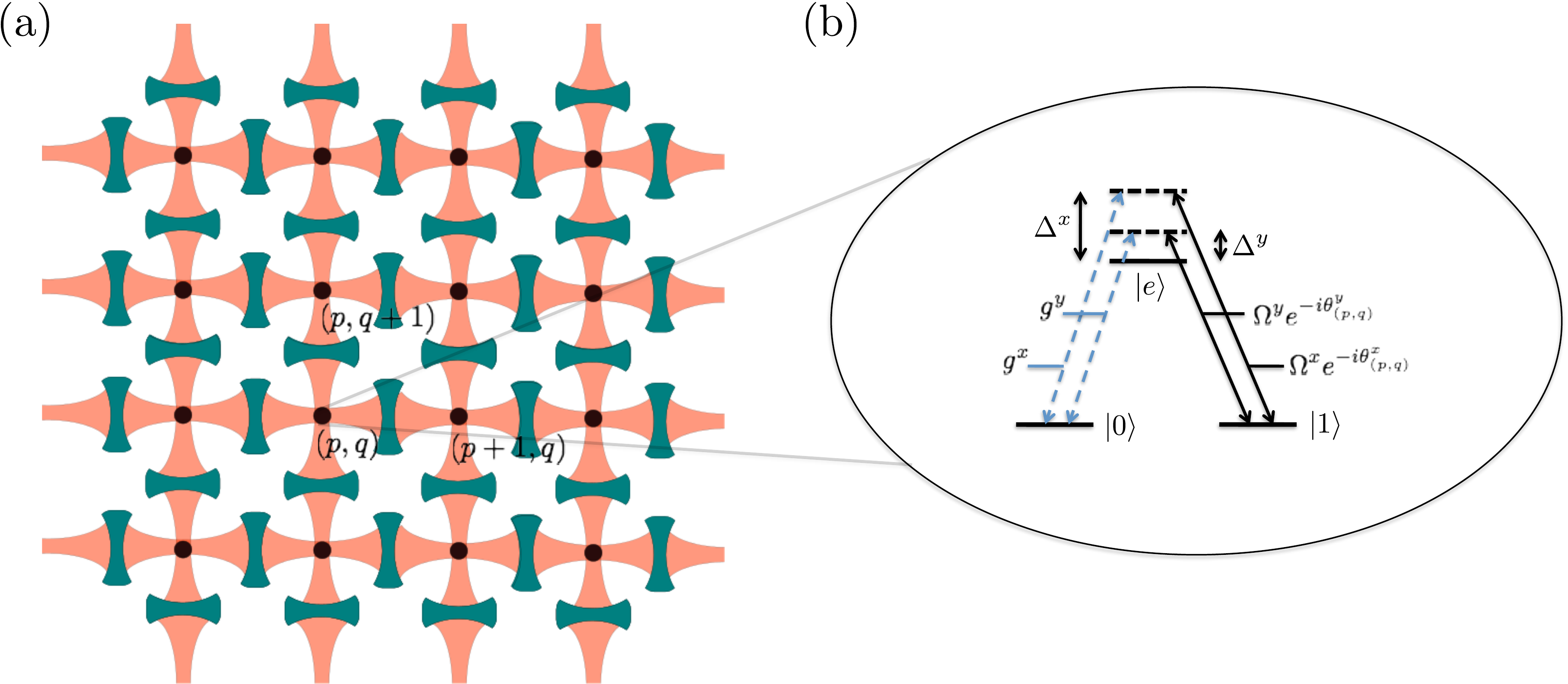}
\caption{ (a) Schematic representation of a 2D CRA which simulates a bosonic FQH states. Each site supports two orthogonal nondegenerate resonator modes that are coupled with a single atom.
(b) Involved atomic levels and transitions. The resonator mode $x$ ($y$) is coupled to the atomic transition $|e\rangle\langle 0|$ with coupling strengths $g^x$ ($g^y$) and detuning $\Delta^x$ ($\Delta^y$). The transition $|e\rangle \langle 1 |$ is driven by site-dependent laser fields with Rabi frequencies $\Omega^x e^{-i\theta^x_{(p,q)}}$ and $\Omega^y e^{-i\theta^y_{(p,q)}}$. The subscripts denote the position of the site. Adapted with permission from \cite{NohAngelakis2014}.}
\label{fig:FQH}
\end{figure}

The first proposal to realize the fractional quantum Hall system with strongly correlated photons was based on an optical manipulation of atomic internal states and inter-resonator hopping of virtually excited photons in a two-dimensional array of coupled cavities \cite{ChoBose2008b}. It was shown that the local addressability of coupled resonator systems allows one to simulate any system of hard-core bosons on a lattice in the presence of an arbitrary Abelian vector potential. The proposal is based on the realization of the Heisenberg spin model introduced earlier and is depicted in figure \ref{fig:FQH}. To facilitate a spatially varying gauge potential, an asymmetry in the resonator modes is assumed,
where two orthogonal resonator modes along the $x$ and $y$ directions have different resonant frequencies (this asymmetry is not an essential requirement as explained in \cite{ChoBose2008b}). Realizing this geometry would be viable in several promising platforms, such as photonic bandgap microcavities and superconducting microwave cavities. The frequency difference between the two modes is assumed to be much larger than the atom-resonator coupling rates. In this way, either direction of the spin exchange can be accessed individually. 

Similar works with different approaches and platforms within CRA have appeared since. Ways to introduce synthetic gauge fields--allowing one to break the time-reversal symmetry which is crucial in quantum Hall physics--have been devised in circuit QED \cite{KochGirvin2010,NunnenkampGirvin2011} and solid state cavities \cite{UmucalilarCarusotto2011}. A photonic version of the quantum spin Hall system, which does not break the time-reversal symmetry, has also been proposed \cite{HafeziTaylor2011} and experimentally realized \cite{HafeziTaylor2013a}. Interacting versions realizing fractional quantum Hall-type physics of photons were proposed in Kerr-nonlinear cavities \cite{UmucalilarCarusotto2012} and the JCH cavities \cite{HaywardGreentree2012a} and signatures of fractional quantum Hall physics in a non-equilibrium driven-dissipative scenario have been investigated \cite{HafeziTaylor2013}. 

Here we will take a closer look at the first proposal. But first, to set the stage, let us very briefly describe the model Hamiltonian and its ground state wave function that captures the physics of FQHE in its fermionic counterpart.

A system of bosonic particles confined in a 2D square lattice in the presence of an artificial perpendicular magnetic field $B$  is described by the Hamiltonian
\begin{equation}
H_{0}=-t\sum_{p,q}\left(c_{p+1,q}^{\dagger}c_{p,q}e^{-i2\pi\alpha q}+c_{p,q+1}^{\dagger}c_{p,q}+\textrm{h.c.}\right),\label{eq:FQH1}
\end{equation}
where the positions of lattice sites are represented by $a(p\hat{x}+q\hat{y})$ with $p$ and $q$ being integers and $\alpha$
is the spacing of the lattice. Here, $\alpha\equiv Ba^{2}/\Phi_{0}$, the number of magnetic flux quanta through a lattice cell, plays a crucial role in characterizing the energy spectrum, whose self-similar structure is known as the Hofstadter butterfly. In the hardcore (maximum one boson per site) and continuum limit ($\alpha\ll1$) the ground state is accurately described by the Laughlin state
\begin{equation}
\Psi_{m}(\{z_{j}\})=e^{-\frac{1}{4}\sum_{j}|{z_{j}}|^{2}}\prod_{j<k}(z_{j}-z_{k})^{m},\label{eq:laughlin}
\end{equation}
where $z_{j}=x_{j}+iy_{j}$ and the filling factor $\nu = 1/m$ corresponds to the ratio of the number of bosons to the number of
magnetic flux quanta.

Assume now a two-dimensional array of coupled cavities, each confining a single atom with two ground levels  as in figure \ref{fig:FQH}(b). Each resonator has two modes along the $x$ and $y$ directions, whose annihilation operators will be denoted by $a^{x}$ and $a^{y}$. The atom interacts with these resonator modes with coupling rates $g^{x}$
and $g^{y}$, and with detunings $\Delta^{x}$ and $\Delta^{y}$, respectively. Furthermore, to induce site-dependent hopping, two classical fields with (complex) Rabi frequencies $\Omega^{x}e^{-i\theta_{(p,q)}^{x}}$ and $\Omega^{y}e^{-i\theta_{(p,q)}^{y}}$ are applied. In the rotating frame, the Hamiltonian for the system reads
\begin{eqnarray}
H &=& \sum_{\mu=x,y\atop j=(p,q)}\left[g^{\mu}e^{-i\Delta^{\mu}t}a_{j}^{\mu}(\ket{e}\bra{0})_{j}\right. \nonumber \\
&& \left.+ \Omega^{\mu}e^{-i\theta_{j}^{\mu}}e^{-i\Delta^{\mu}t}(\ket{e}\bra{1})_{j}+\textrm{h.c.}\right] \nonumber \\
 &-&\sum_{p,q}\left(J^{x}a_{p+1,q}^{x\dagger}a_{p,q}^{x}+J^{y}a_{p,q+1}^{y\dagger}a_{p,q}^{y}+\textrm{h.c.}\right),
\end{eqnarray}
where $J^{x}$ ($J^{y}$) denotes the inter-resonator hopping rate of the photon along the $x$ ($y$) direction. We assume
$\Delta^{x}-\Delta^{y}\gg g^{x},g^{y}$, as mentioned above, and furthermore that $\Delta^{\mu}\gg g^{\mu}\gg\Omega^{\mu},J^{\mu}$.

In this strong coupling regime, the atomic excitation is suppressed and one can perform adiabatic
elimination to obtain an effective Hamiltonian
\begin{eqnarray}
H &=& \sum_{\mu=x,y\atop j=(p,q)}\left[\delta^{\mu}a_{j}^{\mu\dagger}a_{j}^{\mu}(\ket{0}\bra{0})_{j}  +\omega^{\mu}\left(e^{i\theta_{j}^{\mu}}a_{j}^{\mu}\sigma_{j}^{+}+\textrm{h.c.}\right)\right] \nonumber \\
&-&\sum_{p,q}\left(J^{x}a_{p+1,q}^{x\dagger}a_{p,q}^{x}+J^{y}a_{p,q+1}^{y\dagger}a_{p,q}^{y}+\textrm{h.c.}\right),
\label{eq:noexc}\end{eqnarray}
where $\delta^{\mu}=(g^{\mu})^{2}/\Delta^{\mu}$, $\omega^{\mu}=g^{\mu}\Omega^{\mu}/\Delta^{\mu}$,
and $\sigma^{+}=\ket{1}\bra{0}$. Here, we have ignored the ac Stark shift induced by the classical fields, which is negligible in the assumed regime (it may also be compensated by other lasers). 
We further assume $\delta^{\mu}\gg J^{\mu}\gg\omega^{\mu}$, which can be satisfied, along with the above conditions, when\begin{equation}
g^{\mu}/\Delta^{\mu}\gg J^{\mu}/g^{\mu}\gg\Omega^{\mu}/\Delta^{\mu}.
\label{eq:cond}
\end{equation}
Now the resonator photon is suppressed and an adiabatic elimination can be applied once more. Keeping the terms up to the third order and taking only the submanifold with no photons, the effective Hamiltonian further simplifies to
\begin{eqnarray}
H&=&-t\sum_{p,q}  \left[\sigma_{p+1,q}^{+}\sigma_{p,q}^{-}e^{i(\theta_{p+1,q}^{x}-\theta_{p,q}^{x})} \right. \nonumber \\
&+& \left. \sigma_{p,q+1}^{+}\sigma_{p,q}^{-}e^{i(\theta_{p,q+1}^{y}-\theta_{p,q}^{y})}+{\rm h.c.} \right],
\label{eq:final}
\end{eqnarray}
where $\delta^{\mu}$, $\epsilon^{\mu}$, $\omega^{\mu}$, and $J^{\mu}$ are chosen to yield
\begin{equation}
t=J^{x}\left(\frac{\omega^{x}}{\delta^{x}}\right)^{2}=J^{y}\left(\frac{\omega^{y}}{\delta^{y}}\right)^{2}.
\end{equation}
It is easy to see that this Hamiltonian reduces to the Hamiltonian $H_0$ in \eref{eq:FQH1} in the hardcore bosonic limit, if we adjust the phases of the classical fields such that $\theta_{p,q}^{x}=-pq\cdot2\pi\alpha$ and $\theta_{p,q}^y = 0$.
 The ground state of the Hamiltonian \eref{eq:final} is expected to be described faithfully by the Laughlin wavefunction \eref{eq:laughlin}, modulo changes due to the difference in gauge and boundary conditions, which can be calculated analytically \cite{HaldaneRezayi1985}. For a 4 by 4 lattice with $\alpha=1/4$, $m=2$, the fidelity between the numerically calculated ground state of the ideal Hamiltonian and this modified Laughlin state (with $m=2$) was shown to be excellent indeed ($\approx 0.989$). 
The numerical ground states of the effective Hamiltonian \eref{eq:noexc} for $\delta^\nu/10 = 10\omega^\mu = J^\mu$, corresponding to $\Delta^\mu/1000 = g^\mu/100 = \Omega^\mu = J^\mu$, has the fidelity of 0.976. The ground state can be prepared adiabatically by exciting two atoms at different sites initially without the laser fields ($t=0$) and then delocalizing these excitations via adiabatically tuning up the effective hopping rate. The quasi-excitations can also be generated via local tuning of the laser fields to mimic an infinitely thin solenoid, around which the excitations form. More details on these issues can be found in \cite{ChoBose2008b,NohAngelakis2014}.


\section{Out-of-equilibrium physics in CRAs}
\label{sect3}

All the works we have reviewed so far were on equilibrium physics of CRAs. However, as photonic systems are intrinsically dissipative, they are naturally operated in a non-equilibrium setting where some sort of external pumping is often introduced. This is evident for example in recent experiments on BECs of exciton-polaritons \cite{KasprzakDang2006}, Dicke quantum phase transition with a superfluid gas of atoms in an optical cavity \cite{BaumannEsslinger2010}, and BECs of photons in an optical microcavity \cite{KlaersWeitz2010}. In fact, one can go way back and regard lasing as a non-equilibrium BEC phase transition. The nature of Bose-Einstein condensation in such non-equilibrium settings have attracted much interest, because of its intrinsic differences to the equilibrium condensation phenomena \cite{ChiocchettaCarusotto2015}. Investigations of many-body effects in CRAs in such experimentally relevant driven-dissipative regime has only recently begun \cite{TomadinFazio2010, CarusottoCiuti2013, SchmidtKoch2013}. The open nature of CRAs makes it a particularly interesting platform to study non-equilibrium many-body physics. The external driving here can easily be tuned to be incoherent, coherent, or quantum, opening the road for exploration of novel many-body phases out of equilibrium. 

Before discussing the efforts in this area, we would like to mention that investigations on driven dissipative regimes in the context of semiconductor polaritons and cold atoms and ions systems have also been carried out, employing field-theoretical approaches \cite{SiebererDiehl2013}, or more conventional quantum optical approaches \cite{MullerZoller2012}. These have mostly focused on problems with specifically engineered baths \cite{DiehlZoller2010} or (quasi-) exactly solvable spin and fermionic models  or dynamical phenomena \cite{ProsenPizorn2008,HorstmannGiedke2013}.

In this section, we review the early and more recent works on non-equilibrium physics in dissipative JCH or BH models, with or without driving. While there has been a proposal to implement the attractive BH model directly in a circuit-QED system \cite{LeibHartmann2012}, the majority of experimental realizations of a CRA are of the JCH type where the two-level system could be a real or artificial atom. Therefore, we begin by looking at similarities and differences between the two models in a non-equilibrium setting, which was first investigated by Grujic and coworkers \cite{GrujicAngelakis2012}. 

After setting the stage, we look at fermionization and crystallization of photons in driven dissipative settings. We then move on to dynamical (transient) effects, first looking at the dynamical signatures of superfluid-insulator transition and then at the so-called localization-delocalization transition of photons due to self-trapping effect. The experimental verification of the latter in a circuit-QED platform is also reviewed. Then comes proposals to implement exotic out-of-equilibrium phases, viz photon supersolid and Majorana-ilke modes. The next section gives a brief survey of further interesting works that investigate photon correlations and entanglement in small-size lattices or using mean field as well as those that investigate phases in 2D lattices with or without artificial gauge fields. Finally, we close the section with a brief discussion on possible experimental platforms.

\subsection{Comparison between JCH and BH models}

Let us begin by describing how to model the driven-dissipative systems. Firstly, coherent driving by laser fields can be modelled by introducing a driving term in the Hamiltonian, $H_{\rm drive} =  \sum_i \left [\Omega_i(t) a_i^\dagger + \Omega_i^*(t) a_i \right]$, where $\Omega_i(t) = \Omega_i \exp (-i\omega_L t)$ is the time-dependent driving field and $a_i$ is the driven resonator mode. As we have noted earlier, we will encounter two types of model Hamiltonians. One is the Bose-Hubbard Hamiltonian which, including the driving term, can be written as
\begin{eqnarray}
H_{BH} &=& \sum_i \left[ \omega_c a_i^\dagger a_i + \frac{U}{2} a_i^\dagger a_i^\dagger a_ia_i \right] - \sum_{\langle i,j \rangle} J a_i^\dagger a_j \nonumber \\
&+& \sum_i \left [\Omega_i(t) a_i^\dagger + \Omega_i^*(t) a_i \right],
\end{eqnarray}
while the other is the JCH Hamiltonian 
\begin{eqnarray}
H_{JCH} &=& \sum_i \left[ \omega_c a_i^\dagger a_i + (\omega_c - \Delta )\sigma_i^+\sigma_i^-  \right. \nonumber \\ 
&+& \left. g\left( a_i^\dagger \sigma_i^-+a_i\sigma_i^+\right) \right] \nonumber \\
&-& \sum_{\langle i,j \rangle} J a_i^\dagger a_j + \sum_i \left [\Omega_i(t) a_i^\dagger + \Omega_i^*(t) a_i \right].
\end{eqnarray}
In both models, $\omega_c$ is the bare resonator frequency and $J$ is the photon hopping rate between the nearest neighbour resonators--denoted by the angled brackets. $\Delta$ is the detuning between the resonator and atom and $g$ is the atom-photon coupling strength. $\sigma_i^+$ and $\sigma_i^-$ are the atomic raising and lowering operators at site $i$. 

Secondly, photon losses are modelled by assuming that the system is (weakly) interacting with a bath of harmonic oscillators. Taking the Born-Markov approximation, one obtains a quantum master equation describing the dynamics of the reduced density matrix of the system $\rho$ \cite{Carmichael}. In this review, we will mostly deal with master equations of the form
\begin{eqnarray}
\label{mastereqn}
\dot{\rho} &=& -i[H_X,\rho] + \frac{\gamma}{2}\sum_{i=1}^N \left( 2a_i\rho a_i^\dagger - a_i^\dagger a_i \rho - \rho a_i^\dagger a_i\right) \nonumber \\ &&+\frac{\kappa}{2}\sum_{i=1}^N \left( 2\sigma_i^-\rho \sigma_i^\dagger - \sigma_i^\dagger \sigma_i \rho - \rho \sigma_i^\dagger \sigma_i\right),
\end{eqnarray}
where $X$ can be $JCH$ or $BH$; $\gamma$ and $\kappa$ are the dissipation rates of photons and atoms, respectively, assumed here to be identical for all sites.

\begin{figure}[ht]
\begin{center}
\includegraphics[width=1.0\columnwidth]{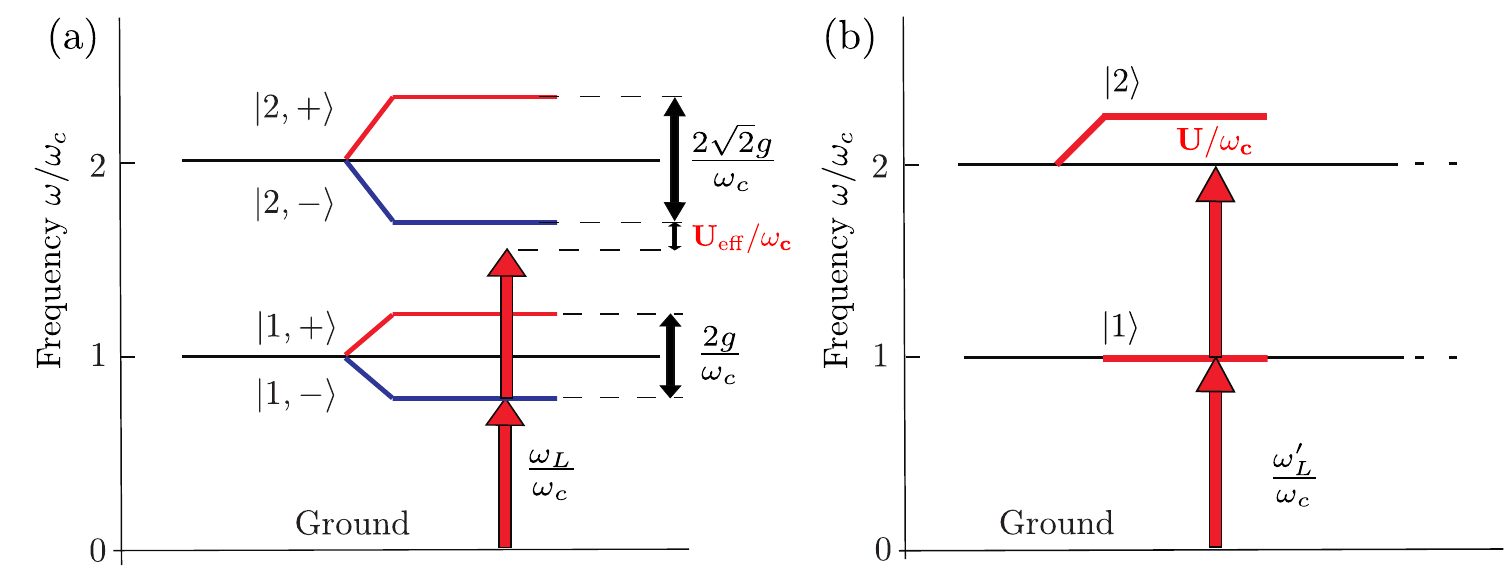}
\caption{Comparison between the lowest eigenmodes of (a) the Jaynes-Cummings model for $\Delta=0$ and (b) the Kerr-nonlinear model. Reprinted from Grujic et al.~\cite{GrujicAngelakis2012}.}
\label{Grujicfig1}
\end{center}
\end{figure}

Now we are ready to compare the two models. Let us start from the simplest case: a single-site system. The eigenstates of the Jaynes-Cummings Hamiltonian are the `dressed' states $|n,\pm\rangle$, where $n$ is the total excitation number and $\pm$ labels two polaritonic (hybrid atom-photon) modes with energies $\omega_n^\pm = n\omega_c - \Delta/2 \pm \sqrt{(\Delta/2)^2+ng^2}$. Because of the dependence on $n$, the energy of $|2,-\rangle$ is not twice that of $|1,-\rangle$, as illustrated in figure \ref{Grujicfig1}(a) for $\Delta = 0$. Comparing with the energy levels of the Kerr-nonlinear model shown in figure \ref{Grujicfig1}(b), one thus concludes that if `$-$' polaritons are driven and if the driving is sufficiently weak such that only the lowest states are occupied, an effective interaction strength $U_{eff}$ can be defined to formally connect the two models. That is, one can assign an effective BH parameter
\begin{equation}
\frac{U_{eff}}{g} = \frac{\omega_2^- - 2\omega_1^-}{g} = \frac{\Delta}{2g} + 2\sqrt{(\frac{\Delta}{2g})^2+1} - \sqrt{(\frac{\Delta}{2g})^2+2},
\label{ueff}
\end{equation}
to the JCH model. 

\begin{figure}[ht]
\begin{center}
\includegraphics[width=0.45\columnwidth]{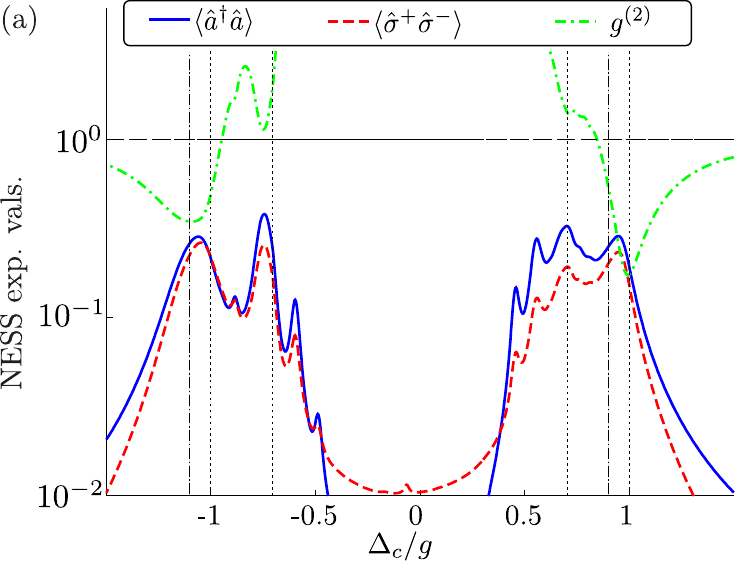}
\includegraphics[width=0.45\columnwidth]{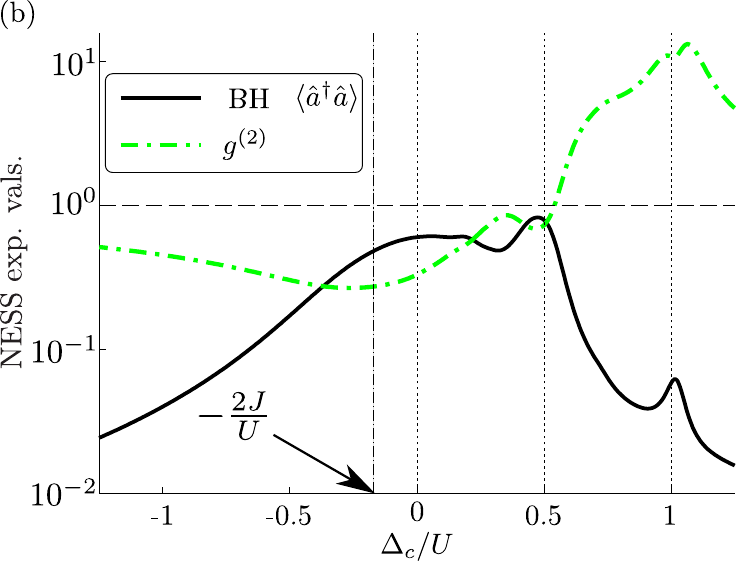}

\includegraphics[width=0.45\columnwidth]{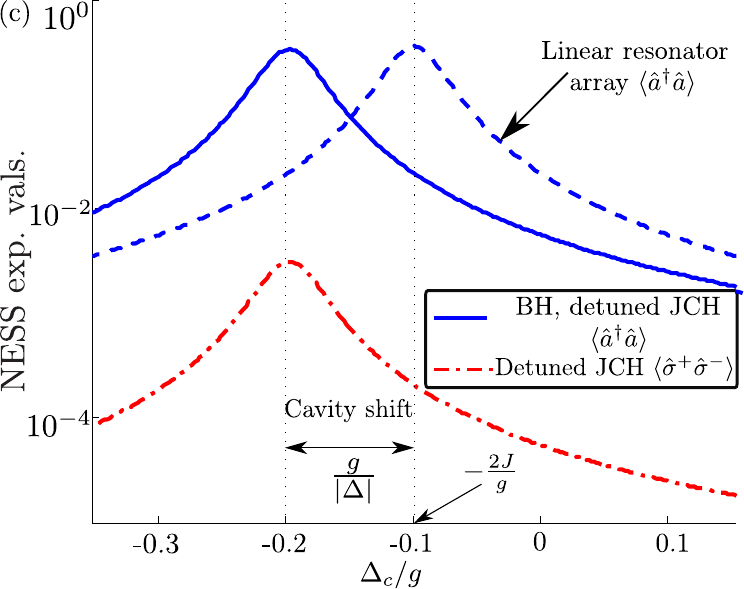}
\includegraphics[width=0.45\columnwidth]{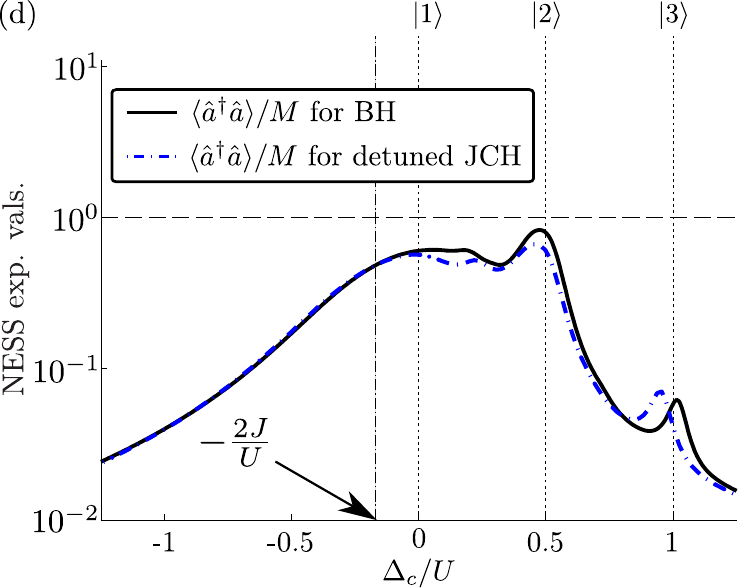}
\caption{Steady state expectation values for a homogeneously driven, cyclic three-site coupled resonator array. (a) A JCH-type system with $\Delta = 2J = 2$ and (b) a BH-type system with the effective nonlinearity set by $g = 20$.  (c) $\Omega = 0.3$, $g = 20$, $\Delta/g = -10$ (d) $g = 1.6\times 10^4$. $\Omega = 2$ in all figures except (c) and all parameters are in units of $\gamma$. Reprinted from Grujic et al.~\cite{GrujicAngelakis2012}.}
\label{Grujicfig2}
\end{center}
\end{figure}

Deep in the strongly interacting regime, we would thus expect the two models to yield similar qualitative behaviour: the polaritons will delocalize to minimize the kinetic energy, but the double occupation of any site is inhibited at the same time. On the other hand, we also know that the two models should converge in the opposite limit of linear CRAs. These two observations give a heuristic argument as to why we can see the Mott-superfluid transition in a JCH system. However, this does not mean that the two types of systems are similar in all aspects. After all, there are additional atomic degrees of freedom in the JCH systems. To illustrate similarities and differences in the two models, figure \ref{Grujicfig2} plots various expectation values of the non-equilibrium steady state (NESS) for a homogeneously-driven periodic three-site CRA.

Upon comparing the JCH (figure \ref{Grujicfig2}(a)) and the BH (figure \ref{Grujicfig2}(b)) results, we immediately see that the JCH has a more complex structure. Broadly, the spectrum is divided into two parts. The left (right) wing is mostly composed of the `$-$' (`$+$') polaritons for the small value of $J$ used here ($U=20J$). The vertical dash-dotted line indicates the driven single-excitation eigenstate while the dotted lines indicate the eigenstates of a single resonator ($J=0$ case). A BH system by comparison has a relatively simple spectrum with peaks near (but not exactly at) the single-site $n$-photon energies indicated by the dotted vertical lines. Note that the spectrum is quite broad as we have assumed $\gamma = J$. Despite these differences, there is a notable similarity when the systems are driven at the single-particle resonance (vertical dash-dotted lines)--enhanced steady-state photon number is accompanied by a dip in the same-time intensity-intensity correlation function $g^{(2)} = g^{(2)}(i_1,i_1)$. The latter is defined as
\begin{equation}
g^{(2)}(i_1,i_2) = \langle a_{i_1}^\dagger a_{i_2}^\dagger a_{i_2} a_{i_1}\rangle / \langle a_{i_1}^\dagger a_{i_1}\rangle\langle a_{i_2}^\dagger a_{i_2}\rangle,
\label{gtwo}
\end{equation}
where the expectation values are taken with respect to the steady state density operator. The dip in $g^{(2)}$ signifies that two excitations do not want to occupy the same site simultaneously. This quantifies the photon blockade effect mentioned earlier, arising due to the strong effective repulsion between photons.

At this point, the reader might be wondering whether it is possible at all to achieve a good quantitative agreement between the two models. It is indeed possible in the `photonic' regime, in which the atom-photon detuning is so large that the polaritons are mostly photonic. In this limit, however, the effect of atom-photon interaction is to merely shift the resonance (figure \ref{Grujicfig2}(c)). To counteract the reduced nonlinear effect, one needs a very large value of $g$. In figure \ref{Grujicfig2}(d), $\Delta/g = -10$ and $g = 6 \times 10^4 \gamma$ were used. Since these kind of numbers lie well outside the experimental realm, we conclude that differences between the two types of systems are there in realistic systems and one must be careful in choosing the correct model to describe a given CRA. In particular, one should check whether a phenomenon found in a BH system can also be observed in a JCH system in a reasonable parameter regime. We will come across such examples below.

\subsection{Fermionization and crystallization of photons and polaritons in BH and JCH models}

The first investigations on many-body physics in driven-dissipative CRAs employed 1-dimensional BH models. Signatures of strong on-site interaction leading to fermionization of the excitations (much as in Tonks-Girardeau gases) were shown to be visible in the NESS of driven-dissipative CRAs systems \cite{CarusottoImamoglu2009,Hartmann2010}. JCH versions of these phenomena have also been studied by Grujic et al.~\cite{GrujicAngelakis2012}--with large enough atom-photon coupling, the predicted  fermionization in the BH system was shown to occur in driven-dissipative JCH systems. 

\subsubsection{Fermionization of photons}
\begin{figure}[ht]
\begin{center}
\includegraphics[width=0.3\columnwidth]{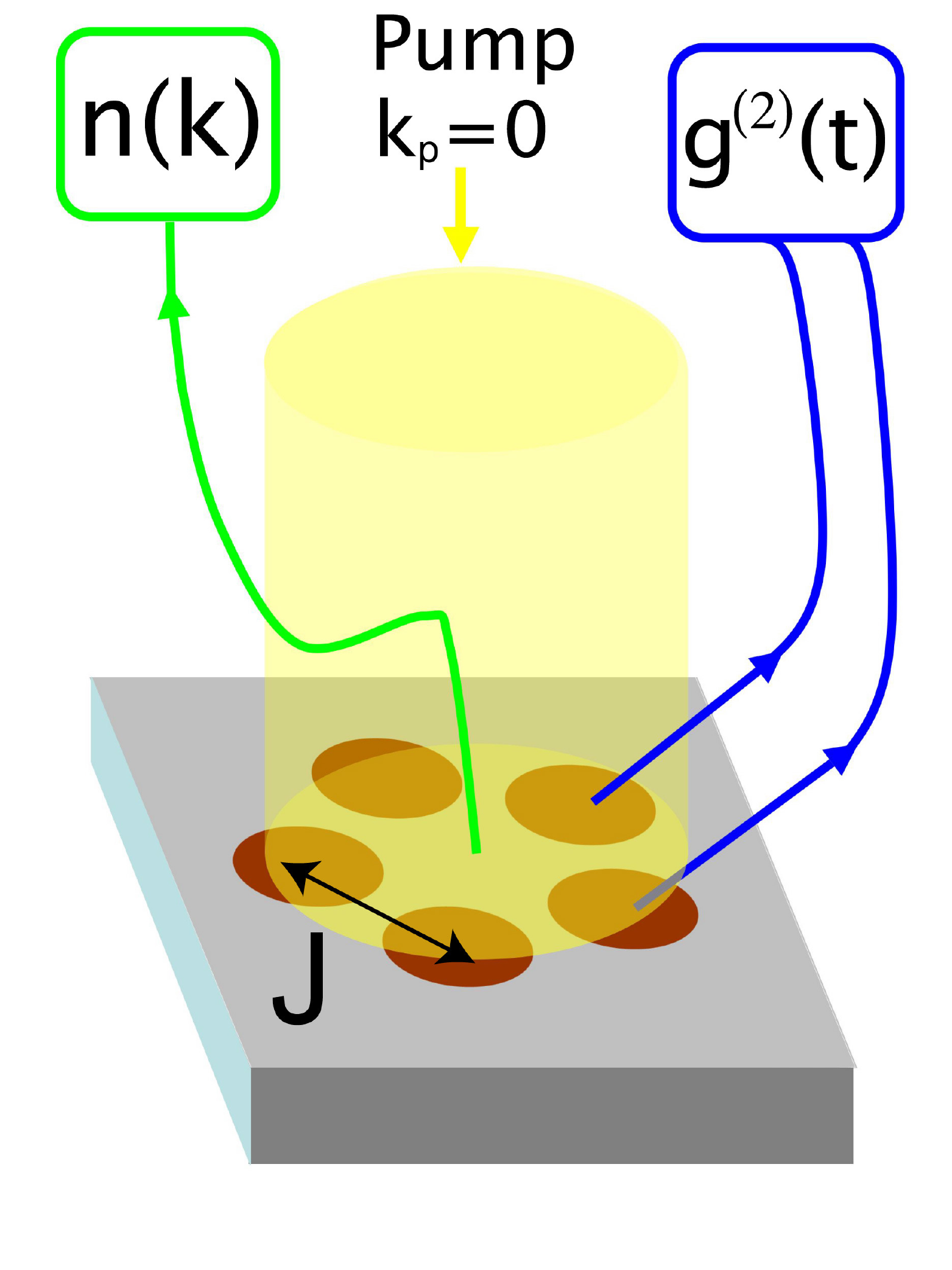}
\includegraphics[width=0.5\columnwidth]{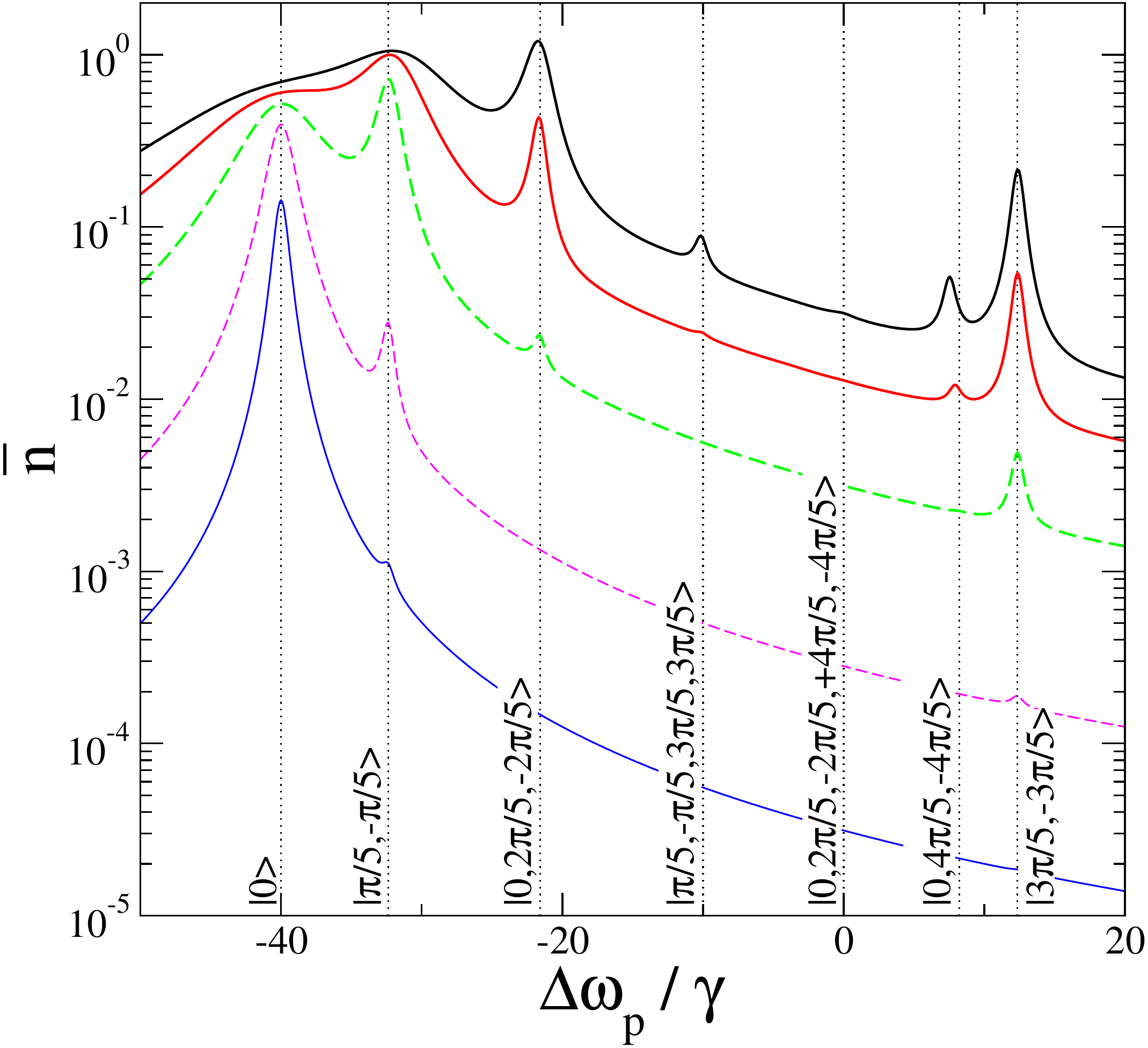}
\caption{Left: Schematic diagram of the system. Right: Total transmission spectra as a function of driving frequency for 5 cavities in the impenetrable boson limit, $U/J = \infty$ and $J/\gamma = 20$. Decreasing pump amplitude from top to bottom: $\Omega/\gamma = 3,2,1,0.3,0.1$. The vertical dotted lines indicate the predictions of the Bose-Fermi mapping where the states corresponding to the spectral peaks are written in the momentum basis. Reprint with permission from Carusotto et al.~\cite{CarusottoImamoglu2009}.}
\label{Carusottofig1}
\end{center}
\end{figure}

We begin with the work of Carusotto and coworkers~\cite{CarusottoImamoglu2009}. The system is composed of Kerr nonlinear cavities coupled in a cyclic fashion as illustrated in figure \ref{Carusottofig1}. The master equation describing the system is given by (\ref{mastereqn}) with $X = BH$, with homogeneous driving fields such that $\Omega_i = \Omega$. As shown on the right panel of figure \ref{Carusottofig1}, the `transmission' spectrum of the total `transmitted' intensity $n_T = \sum_i \textrm{Tr} [ \rho a_i^\dagger a_i] \equiv \langle a_i^\dagger a_i \rangle$ show the traces of strong interaction between photons. For the case of impenetrable bosons ($U \gg J$), the Bose-Fermi mapping can be employed \cite{Girardeau1960}, which translates an interacting Boson model into a free Fermion model to enable calculations of the energy eigenvalues. These eigenvalues with their eigenstates are shown in the figure, showing excellent agreement between the spectra of the driven-dissipative system and the eigenvalues of the closed system. 

\begin{figure}[ht]
\begin{center}
\includegraphics[width=0.5\columnwidth]{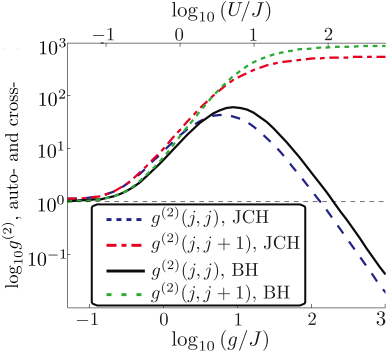}
\caption{Auto- (black, solid line) and cross- (red, dashed line) correlations as functions of the interaction strength. JCH systems with different couplings $g$ are compared with BH systems with $U$ calculated from (\ref{ueff}). $J/\gamma = 20$, $\Omega/\gamma = 0.5$, and $\Delta= 2J$. Reprinted from Grujic et al.~\cite{GrujicAngelakis2012}.}
\label{Carusottofig2}
\end{center}
\end{figure}

To further illustrate the fermionization of photons, figure \ref{Carusottofig2} plots the on-site ($g^{(2)}(i_1 = i_2)$) and nearest-neighbour ($g^{(2)}(i_1\neq i_2)$) correlations as functions of $U/J$, when the pump lasers are kept on resonance with the lowest two-photon transition. Clearly, both the JCH and BH systems yield qualitatively similar results when the effective nonlinearity is matched via equation (\ref{ueff}). In the weak interaction limit, all the levels are harmonic and the emitted light inherits the Poissonian nature of the pump. In the intermediate regime, $U/J \approx 1$, the (driven) two photon peak starts to separate from the single photon peak leading to bunching. Finally, in the large U/J limit, the two photons fermionize, resulting in strong antibunching with enhanced cross-correlations. 

A detailed investigation on the fluorescence spectrum and the spectrum of the second-order correlation function under coherent and incoherent driving for a two coupled cavities of JCH type was carried out by Knap et al.~\cite{KnapCole2011}. Similar to the work described above, the resonance peaks of the measured spectra of the dissipative system were matched to the eigenenergies of the closed system. 

\subsubsection{Crystallization of polaritons}

In an independent work, Hartmann has studied the effects of strong interaction between polaritons in coupled resonators (described by the BH model in which $a_j$ is the polariton annihilation operator) \cite{Hartmann2010}. In that setup, instead of driving the system homogeneously, a flow of polaritons is created via modulation of the phase of the driving laser such that $\Omega_j = \Omega e^{-i\phi_j}$, where $\phi_j = j\pi/2$. The interplay between this flow and the nonlinearity then results in polariton crystallization, where excitations are predominantly found within a specific distance from one another. 
To describe this phenomenon in detail, we will use the spatial correlation function $g^{(2)}(i,j)$ as well the momentum correlation function defined by
\begin{equation}
g_{\rm BM}^{(2)}(p,q) = \langle B_p^\dagger B_q^\dagger B_q B_p\rangle/\langle B_p^\dagger B_p\rangle\langle B_q^\dagger B_q\rangle, 
\end{equation}
where the momentum eigenmode operator is defined as $B_p = 1/\sqrt{N}\sum_j \exp(ipj)a_j$ with $N$ the total number of cavities. 

There are two interesting regimes determined by the relative strength between the nonlinearity $U$ and the driving strength $\Omega$: 1) Weakly nonlinear regime, $\Omega \gg U$, and 2) Strongly nonlinear regime, $U \gg \Omega$.
Let us start with the weakly nonlinear regime. In this strongly-driven regime, a semiclassical treatment is valid, i.e., $B_p = \sqrt{N}\beta\delta_{p,\pi/2} + b_p$, where $\beta$ is a mean amplitude for the driven mode at $p=\pi/2$ and $b_p$ is an annihilation operator for the quantum of fluctuations. To second order in fluctuations the Hamiltonian, neglecting the driving term for now, can be written as 
\begin{equation}
\tilde{H} = \sum_p \left[ (\omega_p + 2Un)b_p^\dagger b_p + \left( \frac{U\beta^2}{2}b_p^\dagger b_{\bar{p}}^\dagger + \textrm{H.c.}\right)\right],
\end{equation}
where $\bar{p} = \pi p/|p| - p$ and $n = |\beta|^2$ is the mean number of polaritons. This Hamiltonian is known to exhibit two-mode squeezing between the modes $p$ and $\bar{p}$, arising due to the scattering between $p$ and $\bar{p}$ modes induced by the nonlinear term. The latter also induces density-density correlations between the modes, which can be seen from the intensity correlation function between the Bloch modes evaluated to the leading order in $1/(Nn)$:
\begin{eqnarray}
g_{\rm BM}^{(2)}(p,q) &\approx& 1+ \delta_{p\neq \pi/2}(\delta_{p,q}  +\delta_{p+q,\pm \pi}\frac{|g|^2}{m^2})\nonumber \\
 &-& \delta_{p,\pi/2} \delta_{q,\pi/2}\frac{2m}{Nn},
\end{eqnarray}
where $m = 2U^2n^2/(12U^2n^2+\gamma^2)$ is the contribution to the mean photon number from quantum fluctuations and $g = \langle B_p B_{\bar{p}} \rangle = -\beta^2(4U^2n+iU\gamma)/(12U^2n^2 + \gamma^2)$. The term $\delta_{p+q,\pm \pi}\frac{|g|^2}{m^2}$ clearly shows the correlations between the modes $p$ and $\bar{p}$ for non-driven modes, while the driven mode $p=\pi/2$ exhibits slight antibunching $g_m^{(2)}(\pi/2,\pi/2) \approx 1 - 2m /Nn$.

\begin{figure}[ht]
\begin{center}
\includegraphics[width=0.85\columnwidth]{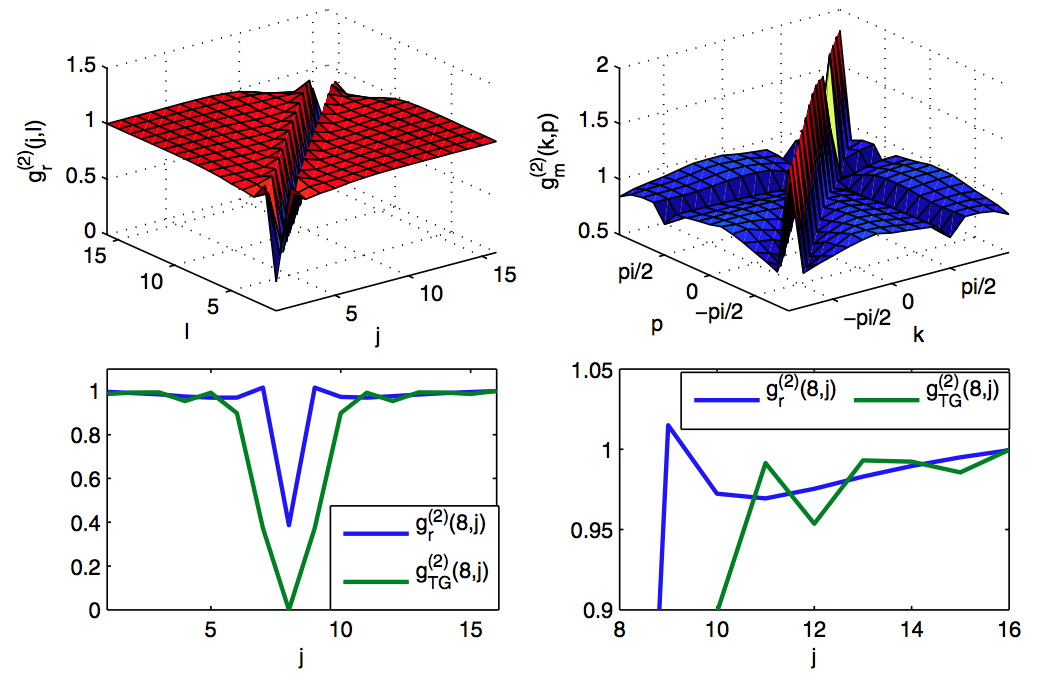}
\caption{ Intensity correlations of the steady state for 16 cavities in the strong interaction regime driven at resonance. $U/\gamma = 10$, $J/\gamma = 2$, $\Omega/\gamma = 2$ and $\Delta_L = 0$, where $\Delta_L$ is the polariton-laser detuning. Top row:  steady state intensity correlations between spatial modes (left) and Bloch modes (right). Bottom row: comparison with the correlation functions of a Tonks-Girardeau gas. Reprinted with permission from Hartmann \cite{Hartmann2010}.}
\label{Hartmannfig2}
\end{center}
\end{figure}

In the strongly nonlinear regime, the semiclassical description is no longer valid. Instead, the time evolving block decimation (TEBD) algorithm (see \cite{VerstraeteCirac2008,Schollwock2011} for reviews) was used to calculate the steady state for an array of 16 cavities (for other methods to deal with larger systems, see e.g., \cite{ValleHartmann2013,Degenfeld-SchonburgHartmann2014}). Contrary to the weakly nonlinear regime, there is now strong correlations between the same Bloch modes ($g^{(2)}_m(p,p) > 1$), while different modes are strongly anticorrelated ($g^{(2)}_m(p,q) < 1$, $p \neq q$), as shown in figure \ref{Hartmannfig2}. 

\begin{figure}[ht]
\begin{center}
\includegraphics[width=0.6\columnwidth]{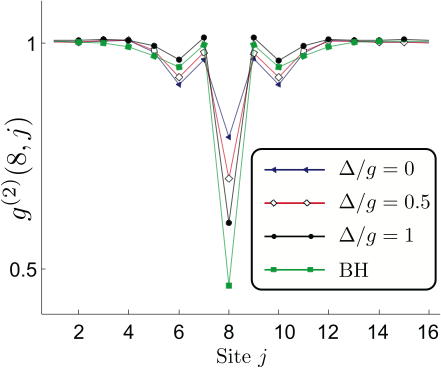}
\caption{Steady state photon density-density correlations for the JCH model for various values of $\Delta/g$ as compared with that from the BH model. $g/\gamma = 10$ ($U_{eff} \approx 6$), $J/\gamma = 2$, and $\Omega = 2$.  Reprinted from Grujic et al.~\cite{GrujicAngelakis2012}. }
\label{tomfig6}
\end{center}
\end{figure}

The same figure also shows an evidence of photon crystallization, by which we mean the following: if a polariton is found in one resonator, a second polariton is most likely to be found in an adjacent resonator and this happens at the reduced probability of finding a polariton at farther sites. This is clearly observed in the polariton correlation function between the cavities, where $g^{(2)}(j,j+1) >1$ and $g^{(2)}(j,j+k) <1$ for $k=1,2,...$. The correlations fall below 1 only when the driving phase is $\ne 0$, indicating that the observed phenomenon emerges due to the interplay between the nonlinearity and the flow of polaritons.
For comparison, figure \ref{Hartmannfig2} also plots (green lines) the density correlations of a Tonks-Girardeau (TG) gas $g^{(2)}_{\textrm{TG}}(i,j) = 1 - (\sin(\pi\tilde{n}(i-j))/\pi\tilde{n}(i-j))^2$, where $\tilde{n}$ is the number of polaritons per resonator. While the TG gas shows an oscillation in the density-density correlation, no such oscillation is observed in crystallized polaritons. Similar behaviour is observed for the JCH model when the polaritonic mode $|1,-\rangle_{p=\pi/2}$ is driven, as depicted in figure \ref{tomfig6}. It is interesting to note that the on-site repulsion is strengthened with increasing $\Delta/g$--this can be attributed to an increase in the atomic portion of the JC polariton with the detuning. 

\subsection{Dynamical phenomena}
In the previous subsection we have seen that the steady states of driven-dissipative CRAs are capable of exhibiting signatures of underlying many-body states in the equilibrium system and furthermore that interesting phases can be created due to driving. However, this is not the only non-equilibrium scenario that reveals interesting physics. One well-known example is quenching, in which the system is initially excited in a particular initial state and subsequent dynamics is observed.
Here we review two early works taking this approach. We begin with the non-equilibrium study on superfluid-insulator transition by Tomadin et al.~\cite{TomadinImamoglu2010} and briefly summarize a similar study on a hybrid system by Hummer et al.~\cite{HummerZueco2012}. Then we review the `localization-delocalization' transition of photons \cite{SchmidtTureci2010}, which has been verified experimentally in a superconducting circuit setup \cite{RafteryHouck2014}.

\subsubsection{Superfluid-insulator transition}

In \cite{TomadinImamoglu2010}, Tomadin and coworkers investigated a dissipative BH system in 2 dimensions without any driving field, initially prepared in a Mott-like state. Much as in the quantum quench scenario, subsequent transient evolutions depend sensitively on the underlying equilibrium phase (without dissipation). A self-consistent cluster mean-field approach was employed with up to two sites per cluster, where the evolution within a cluster is exact but the interaction with the rest of the system is treated at the mean-field level.

\begin{figure}[ht]
\begin{center}
\includegraphics[width=0.7\columnwidth]{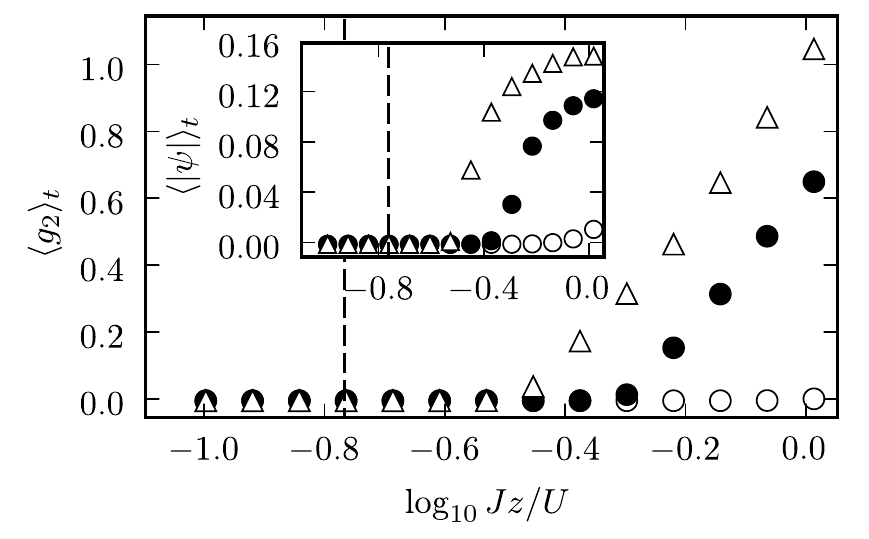}
\caption{Time averaged intensity correlations averaged in the interval $2/\gamma < t < 4/\gamma^{-1}$. The laser field is resonant to the bare cavity frequency while the dissipation rate takes the values $\gamma/U = 4\times10^{-2}$ (empty circles), $\gamma/U = 2\times10^{-2}$ (filled circles), and $\gamma/U = 1\times10^{-2}$ (empty triangles). The inset shows the mean-field order parameter. The vertical dashed line indicates the critical point where the Mott-superfluid phase transition occurs in the equilibrium case. Reprinted with permission from Tomadin et al.~\cite{TomadinImamoglu2010}. }
\label{Tomadinfig1}
\end{center}
\end{figure}

Let us first consider the dynamics of the initial Mott state. With negligible dissipation, the dynamics obviously depends sensitively on the value of $J/U$. For small $J/U$, the system is expected to stay more or less in the same state, whereas for large $J/U$, nontrivial dynamics are expected. As in the equilibrium case, a useful observable to distinguish between these distinct cases is the second-order correlation function, since strong antibunching is expected to persist for small $J/U$. In fact, in the transient scenario, a more useful observable was shown to be the time-averaged second-order correlation function, which we denote $\langle g \rangle_t$. Figure \ref{Tomadinfig1} illustrates the usefulness of this quantity, where the averaging is over $2/\gamma < t < 4/\gamma$. The inset shows the time-averaged mean-field order parameter $\langle |\psi | \rangle_t$ displaying similar behaviour as $\langle g \rangle_t$; $z$ is the coordination number denoting the number of nearest neighbours. For small $J/U$, the average intensity correlation is 0 indicating that the system remains in the initial Mott state. However, for large $J/U$, the initial state becomes unstable and a crossover towards Poisson statistics ($g^{(2)} = 1$) is observed, the latter occurring over a narrow region of $J/U$ near the critical point indicated by the vertical dashed line.

\begin{figure}[ht]
\begin{center}
\includegraphics[width=0.7\columnwidth]{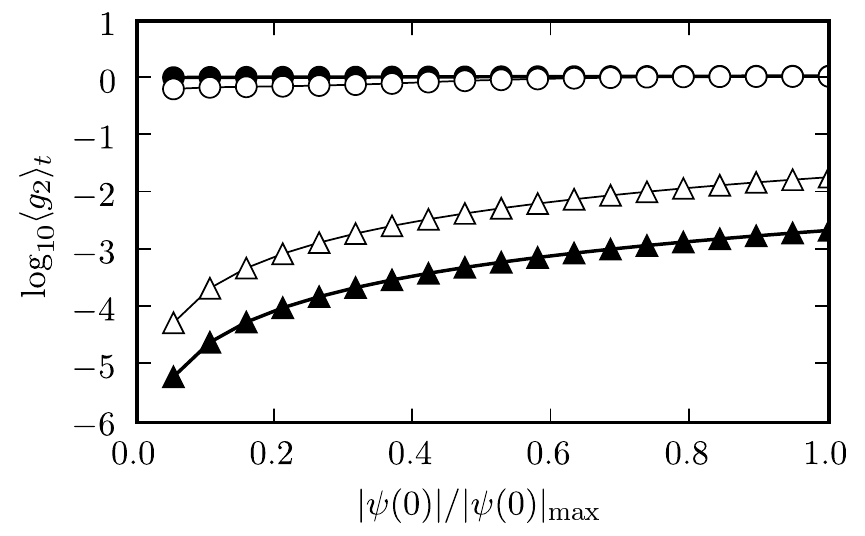}
\caption{Time-averaged intensity correlations, averaged over the interval $2/\gamma < t < 4/\gamma^{-1}$, for resonant driving and $\gamma = 2\times10^{-2}$. The vacuum population of the initial state is $\rho_0 = 0.02$ for the filled symbols and $\rho_0 = 0.2$ for the empty symbols. Results below the critical point ($zJ/U = 0.1$) are denoted by triangles while those above the critical point ($zJ/U = 1.0$,) are denoted by circles. Reprinted with permission from Tomadin et al.~\cite{TomadinImamoglu2010}. }
\label{Tomadinfig2}
\end{center}
\end{figure}

Experimentally, the initial state is most likely to be prepared by a series of laser pulses. In this case, there will inevitably be some imperfections in preparing the Mott state. Tomadin et al.~have studied the effect of such imperfections by considering non-ideal filling factors in the initial state.
Figure \ref{Tomadinfig2} shows the results. Clearly, a deviation from the ideal initial state does not destroy the stark difference between the superfluid (triangles) and Mott (circles) regimes as measured by $\langle g^{(2)}\rangle_t$. The same conclusion holds for a general initial condition, with up to 20\% depletion of the average filling, for the whole range of initial coherences allowed by the manifold spanned by 0 and 1 Fock states in each resonator.

A similar investigation on non-equilibrium signatures of the phase transition has been carried out by Hummer et al.~\cite{HummerZueco2012}, in a hybrid system of flux qubits coupled with NV centers. The model describing this system is different from the BH system in that i) each site comprises of qubits coupled to collective bosonic modes of NV centers surrounding it, and ii) the coupling is between the qubits, not bosons. Despite these differences, the time averaged probability to find two excitations in a single site is shown to be a very good indicator of the underlying phase transition in the equilibrium case.

\subsubsection{Localization-delocalization transition of photons}
\label{sect:FP}

The transient phenomenon to be reviewed here is not associated with the underlying phases of the equilibrium system, but with the dynamics of the system. The concept was initially introduced by Schmidt et al.~\cite{SchmidtTureci2010} essentially in a closed Jaynes-Cummings dimer, but the subsequent experimental demonstration \cite{RafteryHouck2014} showed that dissipation can play an important role in achieving the transition dynamically. As this experiment provides the state-of-the-art realization of a CRA, we discuss it in some detail. The essential physics of the localization-delocalization transition is similar to that of self-trapping: when two nonlinear systems are tunnel-coupled, the oscillation between the systems can be inhibited by the on-site nonlinearity. While previous studies in optical fibers \cite{Jensen1982}, molecules \cite{EilbeckScott1985}, cold atoms \cite{SmerziShenoy1997,AlbiezOberthaler2005,LevySteinhauer2007}, and polaritons \cite{SarchiSavona2008} have concentrated on the Kerr nonlinearity, Schmidt et al.~showed that the same phenomenon also occurs in a Jaynes Cummings dimer. The effect has been generalized to more than two sites by Schetakis and coworkers \cite{SchetakisAngelakis2013}. 

\begin{figure}[ht]
\begin{center}
\includegraphics[width=0.45\columnwidth]{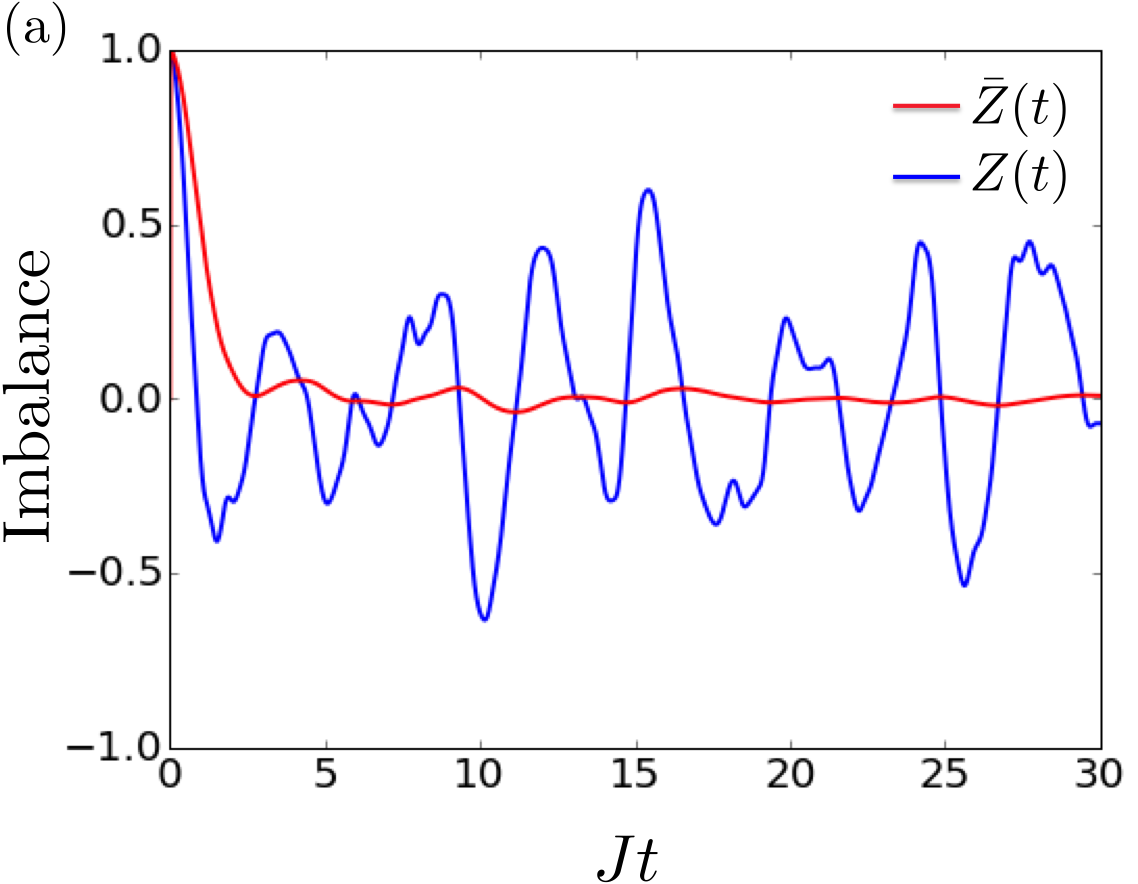}
\includegraphics[width=0.45\columnwidth]{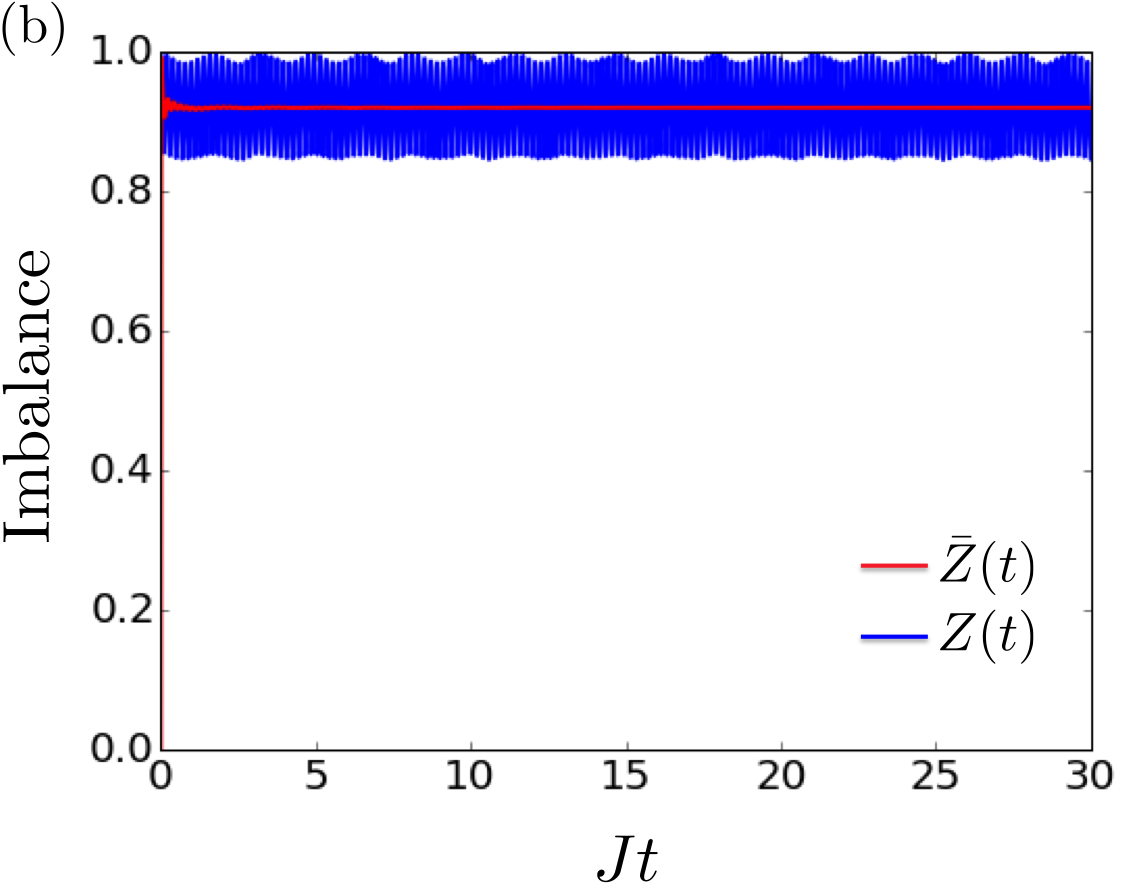}
\caption{Photon number imbalance (blue curves) and the time-averaged imbalance (red curves) for a lossless dimer containing 7 photons initially localized in the left resonator. (a) Below the critical coupling strength, $g=0.3g_c^{cl}$, exhibiting delocalized dynamics. (b) Above the critical coupling strength, $g=2.0g_c^{cl}$, exhibiting localized dynamics.}
\label{frozenfig1}
\end{center}
\end{figure}

As first explained in \cite{SchmidtTureci2010}, the physics of the localization-delocalization transition are nicely captured in the semiclassical picture without dissipation. One resonator is assumed to be initially in a coherent state with $N$ photons on average, while the other resonator is assumed to be empty. Then, the semiclassical analysis shows that the photon imbalance $Z(t) = (n_L(t) - n_R(t))/N$ undergoes an oscillation whose frequency decreases with increasing $g$ and furthermore goes to 0 at the critical coupling strength $g_c^{cl} \approx 2.8 \sqrt{N} J$. This behaviour survives in the quantum regime as shown in figure \ref{frozenfig1} for a lossless dimer with 7 photons initially placed in the left resonator. $\bar{Z}(t)$ is the time averaged imbalance up until time $t$. While the initial imbalance is washed out over time below the critical coupling strength, $g=0.3g_c^{cl}$ (figure \ref{frozenfig1}(a)), the signature of photon localization is evident above the critical coupling strength, $g=2.0g_c^{cl}$ (figure \ref{frozenfig1}(b)). The localization behaviour can be explained as follows. For a symmetric dimer, the two states with opposite imbalance 1 and $-$1 are approximate eigenstates of the system when $g \gg J$. Using the degenerate perturbation theory the period of oscillation between these states can be calculated as $1/(c_{N_0}J(J/g)^{N_0-1})$ with $c_{N_0}$ a constant that depends on the total photon number $N_0$. Clearly, this number diverges rapidly with increasing $N_0$ when $J/g \ll 1$.

A detailed analysis shows that there are features not captured by the semiclassical treatment such as a beating in the coherent oscillation that is observable in the weakly-nonlinear case, smoothening of the transition, and a shift in the critical coupling strength where $g_c^{qu} < g_c^{cl}$ \cite{SchmidtTureci2010}. However, rather than delving into these features further, let us move on to the experimental realization of the dissipative localization-delocalization transition, where dissipation plays a key role in driving the transition dynamically. The latter occurs due to the photon number dependence of the critical coupling strength. For an experimental setup with fixed $g$, the initial photon number $N_0$ decides whether the system is in the localized regime ($g>g_c^{qu}(N_0)$) or in the delocalized regime ($g<g_c^{qu}(N_0)$). In a closed system, the system has no choice but to remain in a given regime. However, in a dissipative system, the decrease in the total number of photons can take the system from a delocalized regime to a localized regime (because $g_c^{qu}$ decreases with $N$). 

\begin{figure}[ht]
\begin{center}
\includegraphics[width=0.7\columnwidth]{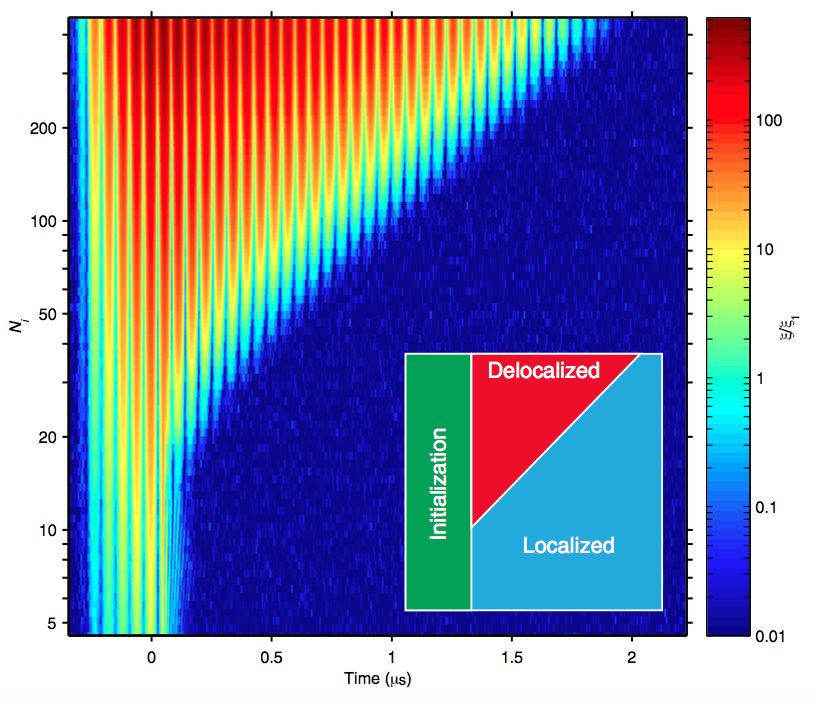}
\caption{Phase diagram of the dissipation-induced delocalization-localization transition for an open Jaynes-Cummings dimer. Homodyne signals from the initially unoccupied resonator as a function of initial photon number $N_i$ and time. For $N_i \gtrsim 20$ the system starts in the delocalized regime because $g < g_c^{qu}(N)$. However, as time progresses the system enters the localized regime as photon losses drive the system into the localized regime characterized by  $g\ge g_c^{qu}(N)$. Reprinted from Raftery et al.~\cite{RafteryHouck2014}.}
\label{frozenfig2}
\end{center}
\end{figure}

This effect has been experimentally demonstrated by Raftery and coworkers in a superconducting circuit platform \cite{RafteryHouck2014}. In the experiment, the initial state was prepared in a coherent state instead of a Fock state, but the described physics remain qualitatively unchanged. Initially, the qubits and the resonators are far-detuned so they are practically uncoupled. Then the system is driven by a coherent pulse and left to evolve a specific period of time, which prepares the system with the imbalance 1. The nonlinearity is then rapidly increased by bringing the qubits into resonance with the resonator mode. The results for various initial photon numbers ($N_i$) are plotted in figure \ref{frozenfig2}. 
It shows the homodyne signal in the initially unoccupied resonator as a function of the initial photon number and time. For a small enough $N_i$, the system starts off and remains in the localized regime, whereas for a larger $N_i$ the system starts off in the delocalized regime and crosses over to the localized regime after enough photons are lost from the system so that the critical coupling strength is smaller than the experimental coupling strength. 

\subsection{Exotic phases out of equilibrium}
Coupled resonator arrays, especially in the superconducting circuit platform, allow one to engineer different types of interactions that are difficult to achieve in other systems. Here, we look at two proposals utilizing this advantage to realize exotic phases in non-equilibrium scenarios. We begin by reviewing a proposal on a photonic superfluid phase in a driven-dissipative setup and then a proposal to achieve Majorana-like modes in a chain of driven cavities.
  
\subsubsection{Photon supersolid phase}

Let us start with the proposal by Jin and coworkers, which investigated the consequences of adding a cross-Kerr term between nearest neighbours \cite{JinHartmann2013,JinFazio2014}. Such a term naturally arises in electronic systems due to the Coulomb interaction between electrons, but are necessarily much smaller than the on-site interaction term. However, in CRAs the magnitude of the new term can be made arbitrarily large as compared to the on-site interaction term, giving rise to interesting new phases. The open {\it extended} Hubbard system studied by the authors is described by the master equation (\ref{mastereqn}) with
\begin{equation}
H_X = H_{BH} + V \sum_{\langle i,j \rangle} n_i n_j,
\end{equation}
which, as they show, can be realized using nonlinear coupling elements between the cavities. 

\begin{figure}[ht]
\begin{center}
\includegraphics[width=0.7\columnwidth]{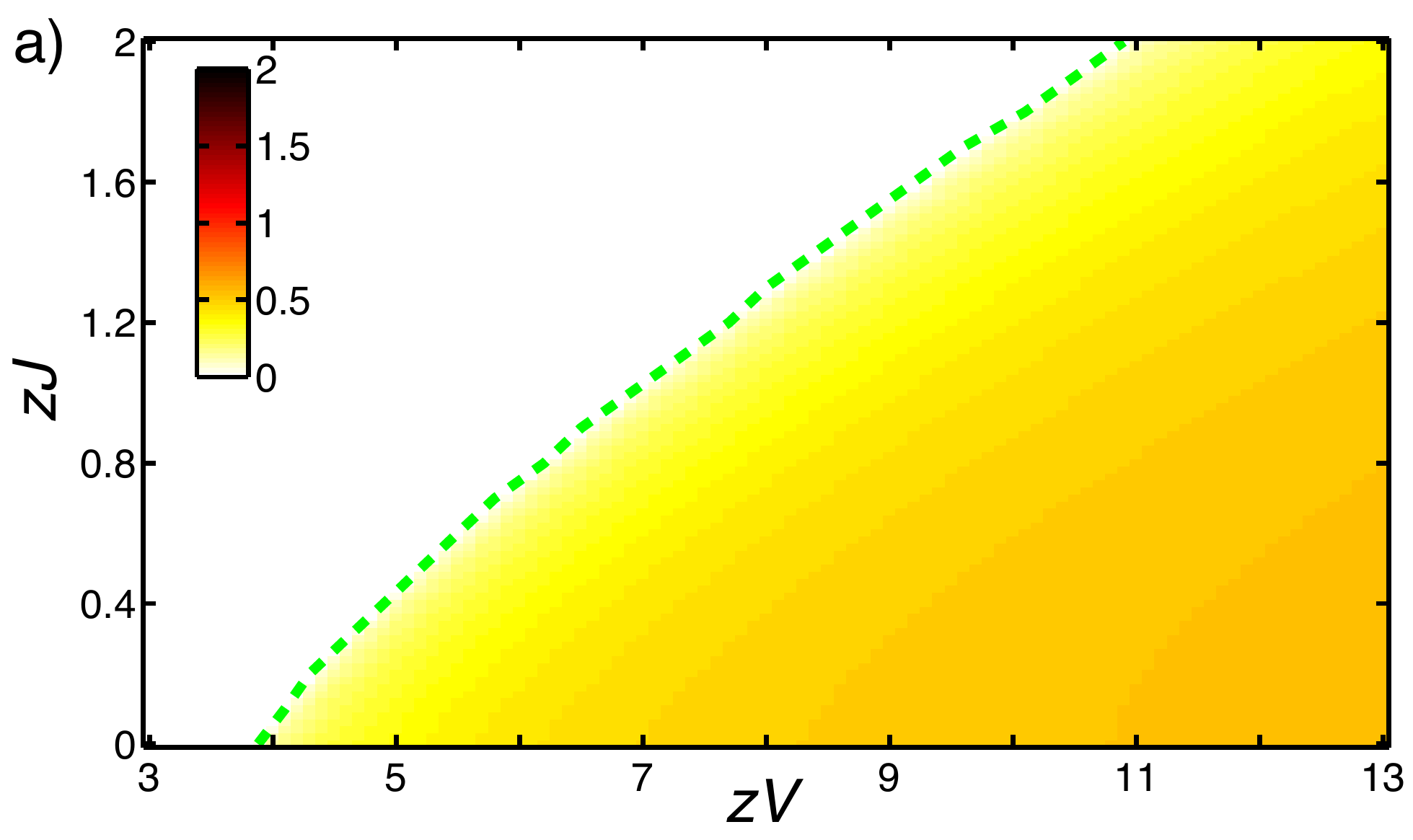}
\includegraphics[width=0.7\columnwidth]{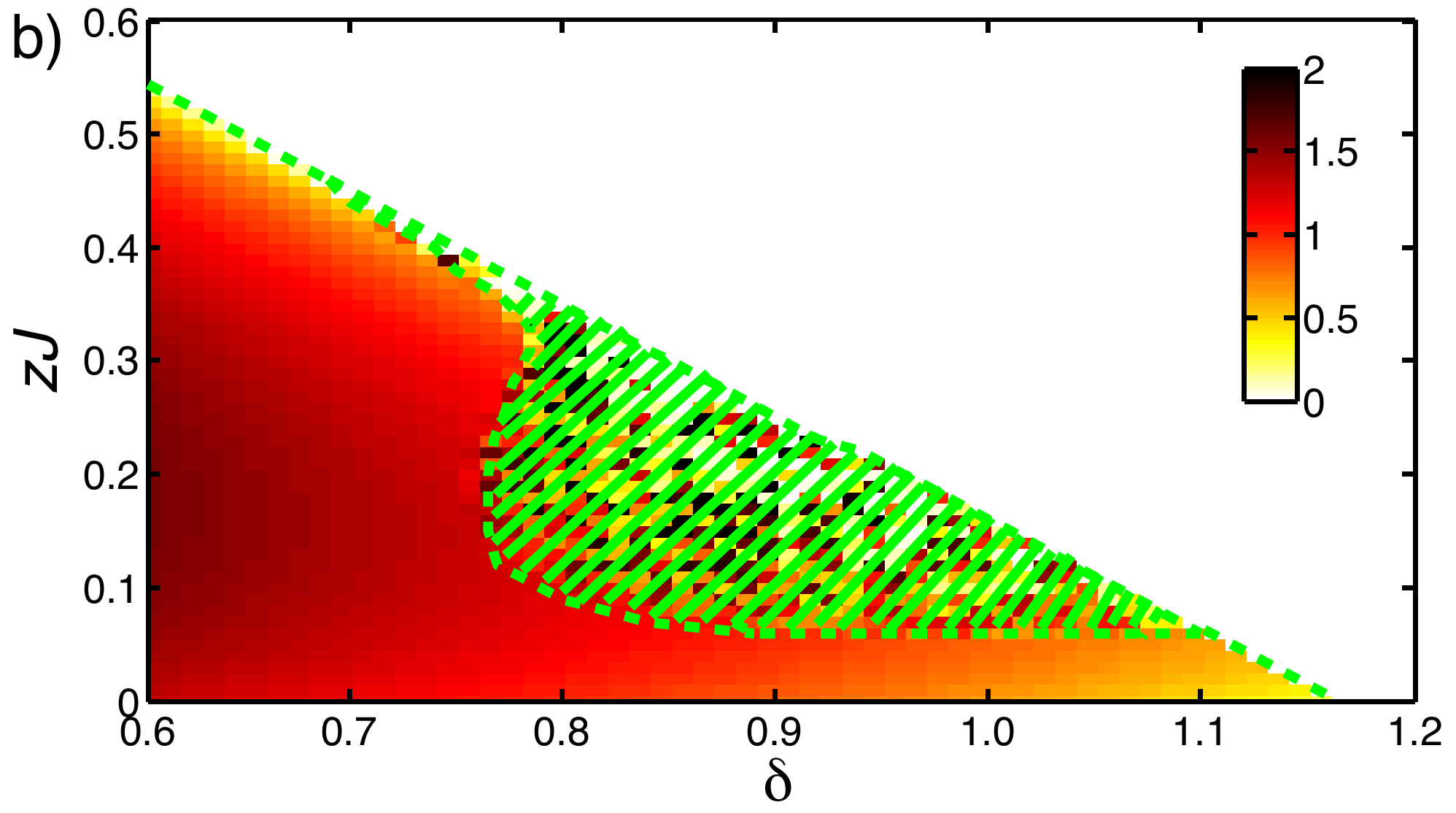}
\caption{ a) Order parameter as a function of $J$ and $V$, for $\delta = 0$ and $U=1$. b) Order parameter as a function of $J$ and $\delta$, for $zV = 0.6$ and $U=0$. $\Omega = 0.75$ for both diagrams. For finite values of the detuning (of the driving field), an intermediate regime between staggered and uniform phase appears, which can be seen as a nonequilibrium analogue of supersolid. Green dashed lines are there to provide a guide to the eye. Reprinted with permission from Jin et al.~\cite{JinHartmann2013}.}
\label{Jinfig1}
\end{center}
\end{figure}

Consider a 2D bipartite lattice--lattice that divides into two sublattices--with homogeneous coherent pumping ($\Omega_i = \Omega$). Due to the presence of the cross-site interaction, a non-zero photon number difference between the sublattices may develop. This broken symmetry phase is characterized by the order parameter $\Delta n = |\langle n_A\rangle-\langle n_B\rangle|$, where $A$ and $B$ are sublattice indices. Figure \ref{Jinfig1}(a) shows the order parameter as a function of the hopping strength and cross-Kerr nonlinearity for $U=1$. The coloured region indicates the broken-symmetry (or crystalline) phase, where the photon number distribution exhibits a checkerboard pattern. We see that the interaction term favours the broken symmetry phase, whereas the hopping tends to restore the symmetry. Remarkably though, having $V \ne 0$ is not a prerequisite to obtain the checkerboard phase, unlike in the equilibrium counterpart \cite{JinHartmann2013}: it is possible, although much harder, for the non-equilibrium steady-state to have non-zero $\Delta n \ne 0$ even with $V=0$. The requirements are a non-zero initial imbalance and a finite hopping strength.

In figure \ref{Jinfig1}(b), the effects of driving field detuning, $\delta$, is shown for $U=0$ and $zV = 0.6$. Here the situation gets more interesting. A new phase appears (shaded green area) where the steady-state never becomes truly stationary in that there is a residual time dependence of $\langle a\rangle$ (on top of a trivial time dependence due to driving). A closer inspection on the reduced density matrix of the sublattice shows that there is a global dynamical phase coherence on top of the checkerboard pattern, which led Jin and coworkers to call this phase a \textit{photonic supersolid phase}. Similar phases were also found in the equilibriums setting for an extended JCH model with the nearest neighbour interaction between the qubits \cite{BujnowskiMartin2014}. 

\begin{figure}[ht]
\begin{center}
\includegraphics[width=0.7\columnwidth]{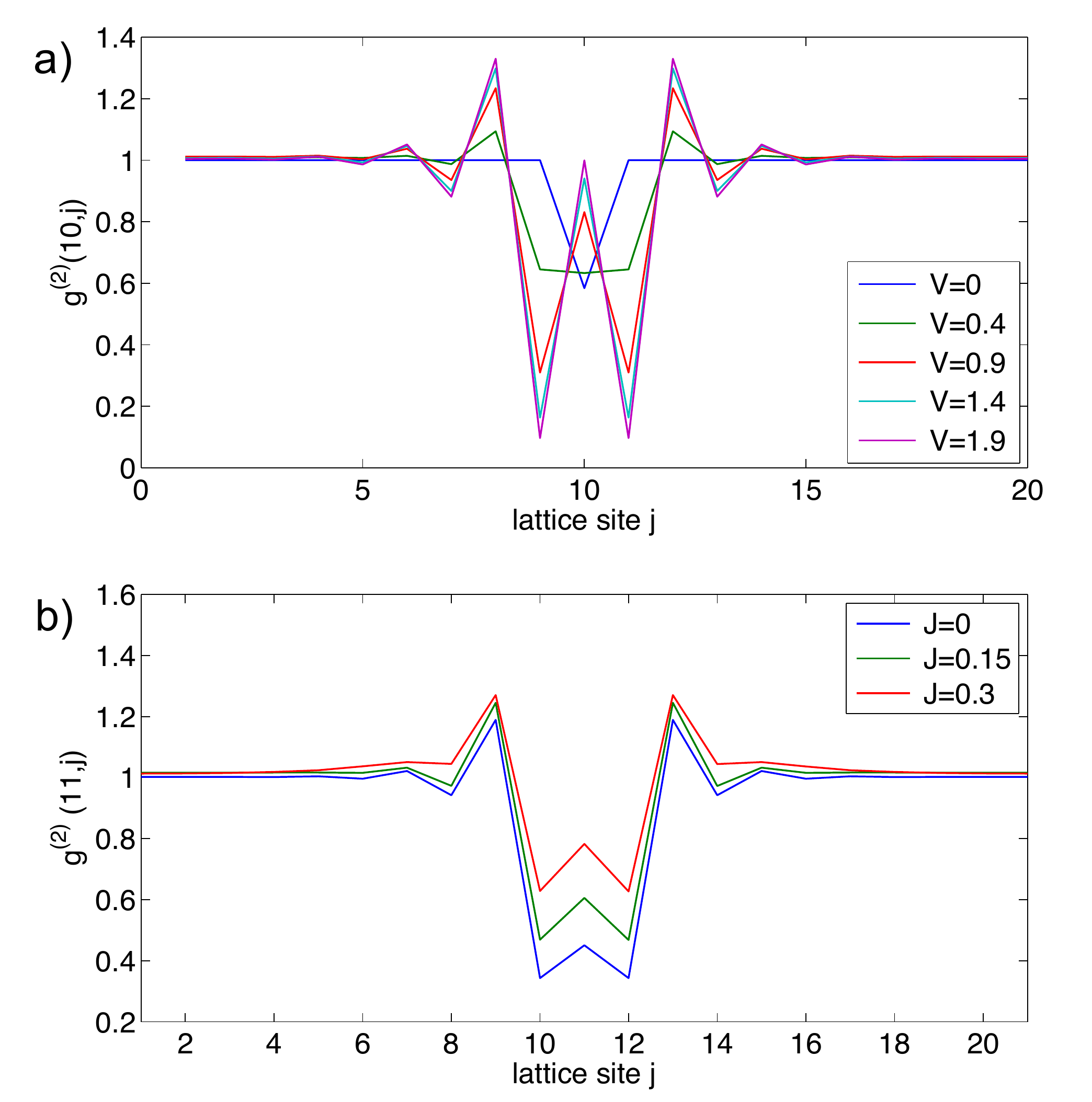}
\caption{Matrix product operator results for 1D CRA. (a) $g^{(2)}(10,j)$ for $\delta=0$, $J=0$, $U=0.5$, and $\Omega = 0.4$. (b) $g^{(2)}(11,j)$ for $\delta=0$, $V=1$, $U=1$, and $\Omega = 0.4$. Reprinted with permission from Jin et al.~\cite{JinHartmann2013}.}
\label{Jinfig2}
\end{center}
\end{figure}

The existence of the crystalline phase was corroborated by performing numerical simulations using the matrix product operator technique in 1D \cite{Hartmann2010}. This approach takes into account quantum fluctuations ignored in the mean-field analysis and is therefore much more reliable. The results are shown in figure \ref{Jinfig2}. The nature of the crystalline phase is displayed in the second-order intensity correlation function $g^{(2)}(i,j)$. Figure \ref{Jinfig2}(a) plots this function for 20 cavities in the absence of hopping ($J=0$). In this regime, the staggered distribution of photons is manifest, especially for larger values of $V$. Figure \ref{Jinfig2}(b) plots $g^{(2)}$ for 21 cavities for various values of hopping strengths, displaying the suppression of crystalline phase with increasing hopping strength. 

\subsubsection{Majorana-like modes}

As the last detailed review of this section we look at a proposal by Bardyn and Imamoglu on a photonic realization of Majorana-like modes (MLMs) in a 1D CRA \cite{BardynImamoglu2012}.  The non-equilibrium steady state of the driven-dissipative version of the model has been studied by Joshi and coworkers in a different context \cite{JoshiKeeling2013}. There the model was shown to be equivalent to a transverse field anisotropic XY model, whose NESS was calculated using a matrix product operator representation. Quantum entanglement and correlations near the critical point (of the equilibrium system) were found to reveal interesting similarities and differences to the equilibrium physics. For instance, singular behaviour was found in the isotropic limit where the range of entanglement diverges (with vanishing magnitude), which is very distinct from the ground state behaviour. We refer the interested reader to the original article. 

\begin{figure}[ht]
\begin{center}
\includegraphics[width=0.7\columnwidth]{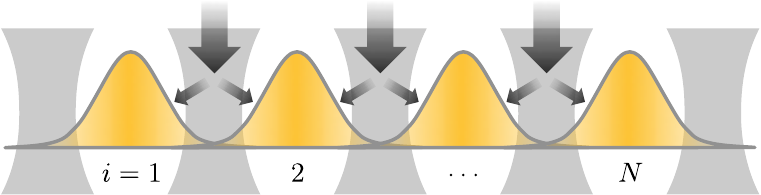}
\caption{Schematic diagram of a cavity array setup for realizing Kitaev's toy model. Cavity modes have strong Kerr-nonlinearity to inhibit occupation of multiple photons, while the parametric pumping of the inter-cavity fields produce the `p-wave' term. Reprinted with permission from Bardyn and Imamoglu \cite{BardynImamoglu2012}.}
\label{Bardynfig1}
\end{center}
\end{figure}

The Majorana modes are exotic `half-a-fermion' zero energy modes that can form, among others, in a 1D topological p-wave superconductor described by Kitaev's toy model \cite{Kitaev2001}. The latter involves the so-called p-wave paring term, $c_ic_{i+1} + \textrm{h.c.}$, where $c_i$ is the fermion annihilation operator at site $i$, and is crucial for the formation of Majorana modes. To realize Kitaev's model with photons, the authors employ the hard-core limit ($U>>J$) of the BH model and a parametric driving of the inter-cavity field as shown in figure \ref{Bardynfig1}. In the hard-core limit ($U>>|\Delta|$), the effect of the parametric driving (with an amplitude $\Delta = |\Delta|e^{i\phi}$) is to introduce the `squeezing' term 
\begin{equation}
H_{{\rm drive}} = |\Delta| \sum_i^{N-1}(e^{i(2\omega_pt + \phi)}\tilde{a}_i \tilde{a}_{i+1} + \textrm{H.c.}),
\end{equation}
where $\tilde{a}_i$s are the hard-core photon operators. Using the Jordan-Wigner mapping $c_i  = \prod_{j=1}^{i-1}(-\sigma_j^z)\sigma_i^-$ with $\sigma_i^- = \tilde{a}_i$, the total Hamiltonian is written as
 \begin{eqnarray}
 H &=& -J \sum_{i=1}^{N-1}(c_i^\dagger c_{i+1} + \textrm{H.c.}) + |\Delta|  \sum_{i=1}^{N-1}(e^{i\phi}c_ic_{i+1} + \textrm{H.c.})  \nonumber \\
 &&- \mu \sum_{i=1}^{N-1}c_i^\dagger c_i,
 \end{eqnarray}
 where  $c_i$'s are fermionic operators and  $\mu = \omega_p - \omega_c$ is the pump-resonator detuning.

The above Hamiltonian is identical to Kitaev's model of 1D p-wave superconductor of spinless fermions and hosts localized zero-energy Majorana modes \cite{Kitaev2001}. The latter can most readily be seen in the limit $J=\Delta>0$ and $\mu=0$, in which the Hamiltonian can be written as
\begin{equation}
H = iJ \sum_{i=1}^{N-1}d_{2i}^\dagger d_{2i+1} = -J \sum_{i=1}^{N-1}\sigma_{i}^x \sigma_{i+1}^x, 
\end{equation}
where the Majorana operators are defined as $d_{2i-1} = c_i + c_i^\dagger$ and $d_{2i} = -i(c_i-c_i^\dagger)$. The Majorana operators $d_1$ and $d_{2N}$ are absent from the Hamiltonian and therefore the energy does not depend on whether or not the non-local fermion $f = (d_1+d_{2N})/2$ is present. This means that the ground state is two-fold degenerate, characterized by the Majorana qubit operator
\begin{equation}
\sigma_M^z = id_1d_{2N} = \prod_{j=1}^N (-\sigma_j^z)\sigma_1^x\sigma_N^x.
\end{equation}
The stringlike operator $P \equiv \prod_{j=1}^N (-\sigma_j^z)$ corresponds to the parity operator of the total number of (hardcore) photons. It commutes with the Hamiltonian and is therefore conserved, meaning that the Majorana qubit is associated with the end qubits only, thus (topologically) protected from local perturbations that preserves parity. 

\begin{figure}[ht]
\begin{center}
\includegraphics[width=0.7\columnwidth]{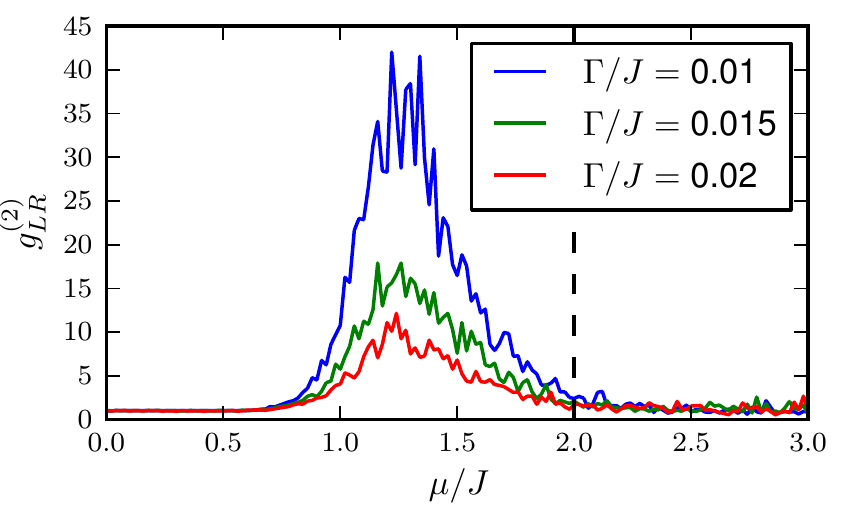}
\caption{Second-order cross-correlation function between the left and right probe cavities as a function of the detuning $\mu$ for 10 cavities (excluding probe cavities), $\Delta/J = 1$, and $J_{L,R}/J = 0.02$. $\Gamma/J = \Gamma_{L,R}/J$ denote cavity loss rates. Majorana-like modes are expected in topologically nontrivial regime (left of the vertical dashed line). Strong bunching is observed inside this region for large $\mu/J$ and for small dissipation rates. Reprinted with permission from Bardyn and Imamoglu \cite{BardynImamoglu2012}.}
\label{Bardynfig2}
\end{center}
\end{figure}

In the proposed photonic system, local dissipation breaks the parity and therefore also the topological protection of the Majorana qubit. However, the existence of localized Majorana modes and their characteristic non-Abelian braiding do not depend on the parity protection, and can therefore be simulated. The Majorana modes can be detected by attaching two probe cavities at the two ends of the array and measuring their second order cross-correlations. Due to the so-called Majorana mediated Cooper pair splitting \cite{NilssonBeenakker2008, LawNg2009}, the nonlocal Majorana qubit mediates a nonlocal coherent exchange of photons between the probe cavities, and the cross-correlation function reveals strong bunching in the topologically nontrivial regime as shown in figure \ref{Bardynfig2}. 

The braiding operation can also be simulated by using the `tunnel-braid' operation \cite{Flensberg2011}, where the Majorana edge mode in one end of the array is tunnel-coupled, via an intermediate cavity, to the Majorana edge mode in another array. A sequence of operations such as turning on and off the tunneling amplitude to the intermediate cavity can be performed to simulate the braiding operation.

\subsection{Further interesting works}

There are many more works that address interesting out-of-equilibrium physics arising in CRAs, which we did not have the space to cover in detail. Here, we provide a brief survey of these works for the interested reader.

Photon statistics in driven-dissipative BH or JCH dimers have been investigated by many authors. One of the main themes is how the photon statistics, as measured by the local intensity-intensity correlation function $g^{(2)}$, change with the nonlinearity-hopping ratio. Gerace and coworkers showed that photon correlations can be used to reveal strongly correlated physics in a photonic Josephson junction--a three-site array with two end cavities coherently driven and the middle site containing single-photon nonlinearity \cite{GeraceFazio2009a}. Ferretti et al.~have studied a homogeneously driven case for a BH dimer and found a crossover between coherent and anti-bunched regimes \cite{FerrettiGerace2010}. An independent study on the JCH dimer was carried out by Leib and Hartmann, by mapping the system to an attractive BH model \cite{LeibHartmann2010}. Driving one of the cavities and measuring $g^{(2)}$ on the other, they observed the same crossover. A more detailed study on photon `transport' was carried out by Biella et al.~for up to 60 Kerr-nonlinear resonators, in which both the transmission spectrum and second-order correlation functions were evaluated \cite{BiellaFazio2015}. Going beyond coherent driving, effects of quantized driving fields were investigated by sending two-photon input states into a BH dimer \cite{LeeAngelakis2015}. Effects of incoherent driving was also investigated by Ruiz-Rivas and coworkers, in which a spontaneous build-up of collective coherence was found \cite{Ruiz-RivasHartmann2014}.

The crossover from the weak to strong hopping for a dimer as well as the infinite lattice was investigated by Nissen and coworkers \cite{NissenKeeling2012}. There, in the dimer case, the antibunching was shown to persist up to a large value of hopping ($J/g \approx$ 10), for driving strengths comparable to or larger than the dissipation rate. However, a semiclassical analysis in the limit of large coordination number shows that the antibunching disappears with increasing system size. 
On the other hand, photon super-bunching was also shown occur in driven dissipative coupled resonator arrays, if the laser driving is tuned to the lowest-lying two-photon (polaritonic) mode in the BH (and JCH) system as shown by Grujic et al.~\cite{GrujicAngelakis2013}. In fact, we have already observed the same behaviour while discussing fermionization of photons. TEBD simulations of 2-7 coupled JC cavities with homogeneous driving shows that when the hopping strength $J$ is larger than a critical value, the system exhibits photon (super) bunching, the magnitude of which decreases with increasing system size. Lastly, we want to discuss an interesting effect called unconventional photon blockade. Normally, a large value of nonlinearity ($U,g >> \gamma$) is necessary to induce photon antibunching. However, Liew and Savona discovered that antibunching can occur in a weakly nonlinear ($U<\gamma$) dimer upon making a judicious choice of parameters \cite{LiewSavona2010}. Soon after, the origin of this effect was identified as a destructive quantum interference between excitation pathways \cite{BambaCiuti2011} and a similar effect has been shown to arise in resonators with a $\chi^2$ nonlinearity \cite{GeraceSavona2014}.

Quantum entanglement in driven-dissipative CRAs has also been investigated. Angelakis et al.~have studied entanglement between polaritons formed at two cavities, where the latter are coupled via an empty resonator mode \cite{AngelakisMancini2009}. Furthermore, a generalization to three sites that are interconnected via three coherently-driven waveguides modes allows the entanglement to be controlled \cite{AngelakisKwek2010,DaiMancini2011}. In an alternative scheme, Liew and Savona considered three-site BH system with the middle site driven and found steady-state entanglement between the spatially separated modes \cite{LiewSavona2012}. An interesting aspect of this work is that, just like in the authors' antibunching work, only a small amount of nonlinearity is required. Furthermore, by repeating the three site setup, a multimode entanglement can be generated \cite{LiewSavona2013}.

Finally, we would like to mention studies in 2D systems. We have already come across the latter when we reviewed non-equilibrium signatures of superfluid-insulator transition \cite{TomadinImamoglu2010} and the photonic supersolid phase \cite{JinHartmann2013,JinFazio2014}. The phase diagram of the driven-dissipative BH array (without the extra cross-Kerr interaction term) was studied by Le Boit\'e and coworkers \cite{Le-BoiteCiuti2013,Le-BoiteCiuti2014}. While the Mott-lobe-like structures were found, the phase diagram in the non-equilibrium setting exhibits clear differences to its equilibrium counterpart. This points to rich many-body physics beyond the equilibrium physics that arises due to the interplay between driving and dissipation. Verification of these mean-field results is an important future direction as predicted effects such as bistability may not survive the quantum fluctuations  (see, for example \cite{Mendoza-ArenasJaksch2016} where tensor networks simulations for 1D arrays up to 60 sites have been performed).  While one dimensional arrays can be numerically simulated efficiently, an efficient method for 2D arrays is still missing. Towards this direction, new numerical methods such as the `corner-space renormalization method' \cite{FinazziCiuti2015} and `self-consistent projection operator theory' \cite{Degenfeld-SchonburgHartmann2014} are currently being developed. 

An interesting possibility in 2D that is absent in 1D is to introduce artificial magnetic fields by inducing phase-dependent hopping terms. This allows one to implement quantum Hall-like models for photons as we have looked at earlier. Driven-dissipative signatures in such systems were studied by various authors. Umucal{\i}lar and Carusotto proposed a method to experimentally verify the fractional quantum Hall state hidden in the NESS of a weakly driven array \cite{UmucalilarCarusotto2012}. In a similar setting, Hafezi and coworkers showed how $g^{(2)}$ measurements can be used to infer the underlying fractional quantum Hall states \cite{HafeziTaylor2013}. Efforts to devise methods to experimentally measure topological invariants in integer quantum Hall systems have also been made. Ozawa and Carusotto showed that the Chern number and the Berry curvature can be measured by following the displacement of a localized excitation by a coherent driving field \cite{OzawaCarusotto2014}, while Berceanu et al.~showed that by adding a weak harmonic potential to the model, the `momentum-space Landau levels' can be observed \cite{BerceanuCarusotto2015}. Alternatively, Hafezi \cite{Hafezi2014} and Bardyn et al.~\cite{BardynZilberberg2014} showed that twisted boundary conditions can be implemented to experimentally measure the topological invariants. Methods to stabilize strongly-correlated states of light using auxiliary two-level atoms have also been suggested \cite{KapitSimon2014,LebreuillyWouters2015}. 

\subsection{Experimental implementations}

Before we finish this section, perhaps a few words on experimental realizations are in order.
There are a number of promising experimental platforms to realize the JCH (or the BH) model: superconducting circuits, photonic crystal devices with quantum dots, microcavities connected by optical waveguides, polaritons in microstructures, and trapped ions. The trapped ion implementation uses phonons, the quantized vibrational motions of trapped ions, instead of photons and have been experimentally demonstrated for the dimer case \cite{ToyodaUrabe2013}. 2D patterns of micropillars etched out of a semiconductor microcavity that can host exciton polaritons is a viable experimental platform especially if the polaritonic nonlinearity can be enhanced to the few-photon level. Recently, spin-orbit coupling has been demonstrated in this platform \cite{SalaAmo2015}. Microcavities plus optical waveguides is an innovative platform suggested by Lepert et al.~\cite{LepertHinds2011} that uses cavities formed by open microcavities closed by coated waveguide tips. An emitter placed in each resonator acts as a two-level atom, while evanescent coupling between waveguides provide the photon hopping mechanism. Photonic crystals are dielectric  devices with periodically patterned holes, where missing holes called defects can act as cavities. When a semiconductor is used to make a crystal, a quantum dot (artificial atom) grown in each defect resonator can allow the resulting system to be described by the JCH model. Majumdar et al.~have experimentally implemented two cavities coupled to a single quantum dot in this system \cite{MajumdarVuckovic2012b}. The platform that is most promising is superconducting circuits. The JC dimer has already been implemented by Raftery et al.~\cite{RafteryHouck2014}  as we have already seen (the BH one has also been implemented by Eichler et al.~\cite{EichlerWallraff2014}), and building a larger lattice seems certainly within reach in the near future with lattices up to 200 resonators in a Kagome geometry already a reality (without atoms yet) \cite{UnderwoodHouck2012a,SchmidtTureci2010,HouckKoch2012a} . In the latter, an experimental method to probe the lattice using scanning defect microscopy has been developed \cite{UnderwoodHouck2015}. 

\section{Simulating strongly correlated effects in 1D with photons in waveguides with EIT-based nonlinearities}

Simulations of strongly correlated effects with photons are not limited to lattice systems. Continuous one dimensional systems can also be simulated by using nonlinear waveguide systems instead of coupled resonator arrays. This section reviews proposals in this young field. We first explain the mechanism to produce strongly interacting polaritons in 1D waveguides and show how an effective Hamiltonian can be engineered. Equipped with this knowledge, we go on to review various proposals on simulating strongly correlated many-body system in 1D continuum. These range from the condensed matter models such as the Tonks-Girardeau gas of polaritons, quantum sine-Gordon model, Bose-Einstein condensation of polaritons, photonic Luttinger liquids and Cooper pairing to interacting relativistic model named the Thirring model. We end the section with a brief survey of experimental progress in realizing relevant nonlinear optical setups and a comment on possible future directions.

\subsection{Background}

Quantum simulations of one dimensional quantum field theories with strongly interacting photons (or more precisely, polaritons) to be reviewed in this section rely on two main techniques. The first is the technique of stationary light, allowing one to trap pulses of light and describe its time evolution via an effective Hamiltonian. The second is creating strong nonlinearity, achieved by introducing specifically structured atoms to mediate interaction between photons. Both are closely tied to the phenomenon of electromagnetically induced transparency (EIT) \cite{HarrisImamoglu1990,HarrisHarris1997,FleischhauerMarangos2005} as we will explain shortly. 

\subsubsection{Stationary pulses of light}
\label{sect:stationary light}

The concept of stationary light has been proposed and demonstrated by Bajcsy et al.~\cite{BajcsyLukin2003}. The main idea is based on dark state polaritons \cite{FleischhauerLukin2000} formed in EIT, which we will briefly explain here. Consider an electric field propagating through an atomic medium. Normally, if the electric field is resonant with an atomic transition, it will be strongly attenuated as it propagates through the medium. However, a laser field driving another transition can induce destructive quantum interference between the excitation pathways and cancel this attenuation, i.e., the extra laser field may induce transparency. In EIT, this is accompanied with a steep dispersion, resulting in a dramatic reduction of the group velocity and enhanced nonlinearity. 

\begin{figure}[ht]
\begin{center}
\includegraphics[width=0.4\columnwidth]{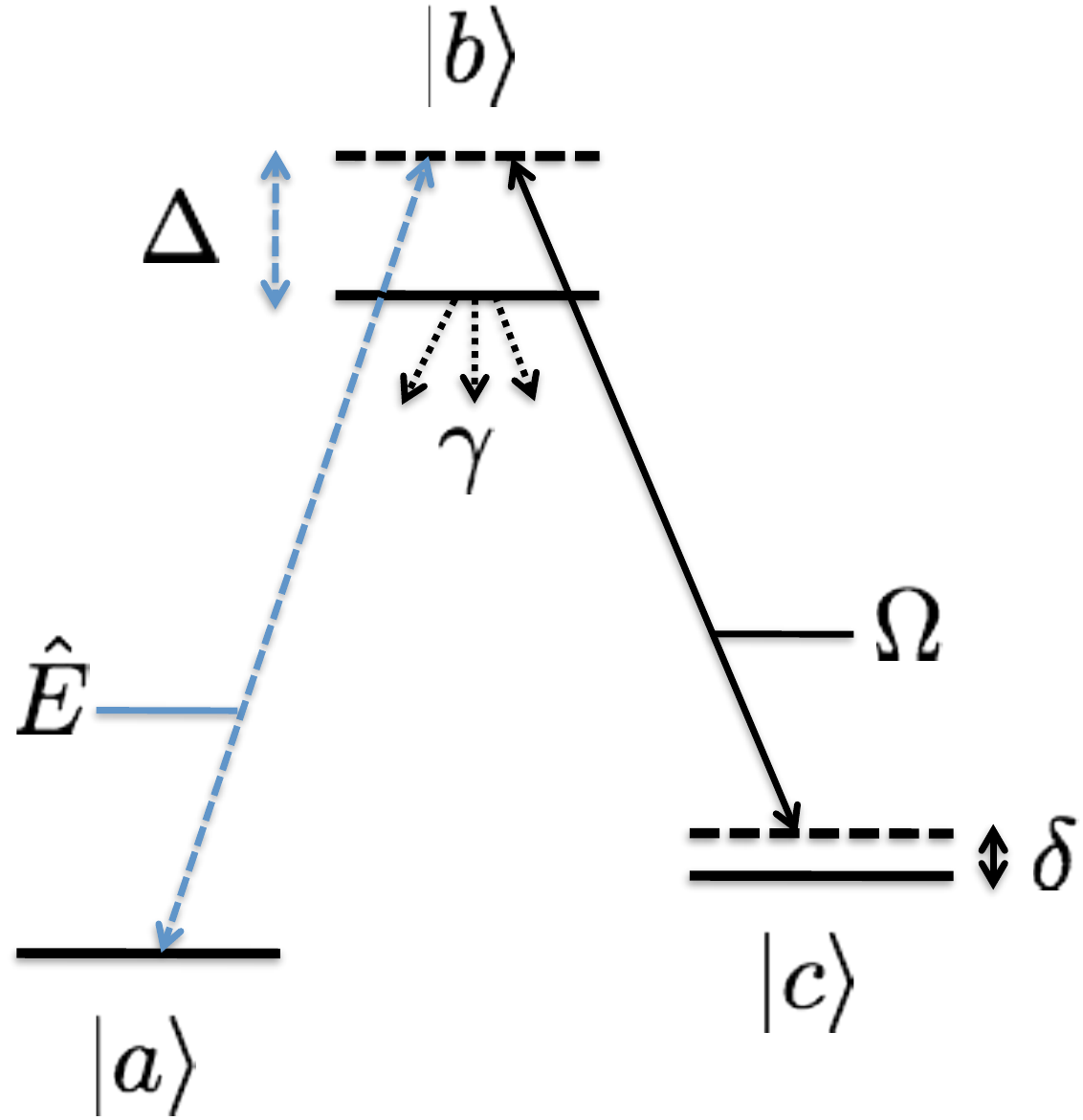}
\caption{Atomic level scheme to create dark state polaritons. $|a\rangle$ and $|c\rangle$ are stable states, $\hat{E}$ is a quantum field, and $\Omega$ is the Rabi frequency of a classical control field.} 
\label{darkpol}
\end{center}
\end{figure}

\textbf{Dark state polaritons}: Dark state polaritons are collective excitations of light and matter formed when a quantum field propagates under specific conditions that allow EIT. A typical atomic level scheme employed in realizing EIT, called a $\Lambda$-scheme, is depicted in figure \ref{darkpol}. It comprises of a dissipative excited state $|b\rangle$ and two steady states $|a\rangle$ and $|c\rangle$. One transition to the excited state is coupled to a quantum field $\hat{E}$ while the other is coupled to a classical control (laser) field of Rabi frequency $\Omega$. 
The Hamiltonian describing the evolution of the atom-field system is
\begin{eqnarray}
 H &=& -n \int dz \bigg [ \Delta \hat{\sigma}_{bb} + \delta\hat{\sigma}_{cc} + g\sqrt{2\pi}\left(\hat{\sigma}_{ba}\hat{E}+ \textrm{h.c.} \right) \nonumber \\
 && + \left ( \Omega\hat{\sigma}_{bc} + \textrm{h.c.}\right)\bigg],
\end{eqnarray}
where $n$ is the linear density of atoms, $g$ is the single photon-atom coupling strength, and $\hat{\sigma}_{ij} \equiv \hat{\sigma}_{ij}(z,t)$ are continuous average spin-flip operators in a small region around the position $z$. The propagation of the electric field is governed by the equation
\begin{equation}
\label{propeqn}
\left(\partial_t + v \partial_z \right)\hat{E}(z,t) = i\sqrt{2\pi}gn\hat{\sigma}_{ab}(z,t),
\end{equation}
where $v$ is the speed of light in the absence of atoms. 

Assuming that atoms are mostly in the ground state $|a\rangle$ and adiabatically eliminating the fast-varying atomic coherence $\hat{\sigma}_{ab}$ in the Heisenberg-Langevin equation, the propagation equation reduces to (for a derivation, see for example \cite{ZimmerFleischhauer2006})
\begin{equation}
\label{dsp}
\left( \partial_t + v_g\partial_z \right)\hat{\Psi} (z,t) = \frac{i2\pi g^2nv_g \delta}{(\Omega^2-i\delta\Gamma)v}\hat{\Psi}(z,t).
\end{equation}
The field operator $\hat{\Psi}$ describes a collective excitation of the quantum field $\hat{E}$ and the atomic excitation $\hat{\sigma}_{ab}$ which is called the dark state polariton.
$\Gamma = \gamma/2-i\Delta$ where $\gamma$ is the decay rate from the excited state $|b\rangle$ and  $v_g \approx v \Omega^2/(\Omega^2+2\pi g^2 n)$ in the limit $\Omega^2 \gg |\delta\Gamma |$, with $\delta$ the two-photon detuning as depicted in figure \ref{darkpol}. The dark state polariton operator reduces to $\hat{\Psi} \approx  g\sqrt{2\pi n}\hat{E}(z,t)/\Omega$ if the energy is mostly stored in the form of atomic excitations, which happens when $2\pi g^2 n \gg \Omega^2$. In such a regime the group velocity $v_g$ is much smaller than the speed of light in the empty waveguide, $v$. 

\textbf{Stationary light}: Stationary light is formed when there are two counter-propagating control fields that form a Bragg grating for the quantum fields propagating in the opposite direction. The dark state polaritons are then trapped in the system while maintaining a finite photonic component, following the intensity profile of the control fields. Being stationary, these polaritons have longer time to interact with each other, which effectively enhances the nonlinearity of the system. To derive the equation of motion for these stationary polaritons, we need to first introduce some notations. Counter propagating fields will be denoted as $\hat{E}_\pm$ and $\Omega_\pm$, where $\hat{E} = \hat{E}_+ e^{ik_c z} + \hat{E}_-e^{-ik_cz}$ and $\Omega = \Omega_+ e^{ik_c z} + \Omega_-e^{-ik_cz}$. $k_c$ is the wavenumber of the control field, which we assume to be similar to the wavenumber of the quantum field. In turn, this induces a slowly varying optical coherence
$\hat{\sigma}_{ab,\pm}$ defined through $\hat{\sigma}_{ab} = \hat{\sigma}_{ab,+} e^{ik_0 z} +\hat{\sigma}_{ab,-} e^{-ik_0 z}$. Because of the wavenumber mismatch, there is an additional term $i\Delta\omega$ in the r.h.s. of the propagation equation (\ref{propeqn}), where $\Delta\omega = v(k_0-k_c)$ (assuming the same speed of light $v$ for both the quantum and classical light) \cite{AndreLukin2005}.

If the atoms are hot enough, we can use the secular approximation, i.e., neglect the fast oscillating contributions proportional to $e^{\pm i2k_0z}$ to obtain  propagation equations similar to (\ref{dsp}) for counter propagating dark state polaritons $\hat{\Psi}_\pm = g\sqrt{2\pi n}\hat{E}_\pm/\Omega_\pm$, but with extra terms that couple the counter propagating fields \cite{BajcsyLukin2003,ZimmerFleischhauer2006}. When $\Omega_+$ = $\Omega_- \equiv \Omega_0$, these equations can be grouped into the equations for the sum, $\hat{\Psi}_S = (\hat{\Psi}_+ + \hat{\Psi}_-)/\sqrt{2}$, and the difference, $\hat{\Psi}_D = (\hat{\Psi}_+ - \hat{\Psi}_-)/\sqrt{2}$ modes:
\begin{eqnarray}
\partial_t \hat{\Psi}_S +2v_g\partial_z \hat{\Psi}_D  &=&  i\left( \delta - 2\Delta\omega \frac{v_g}{v}\right)\hat{\Psi}_S\\
\partial_t \hat{\Psi}_D + v\partial_z\hat{\Psi}_S &=& -\frac{2\pi g^2 n}{\Gamma}\hat{\Psi}_D.
\label{Eq.4.19}
\end{eqnarray}
In the limit of interest, $2\pi g^2 n \gg \Gamma$, the difference mode $\hat{\Psi}_D$ can be adiabatically eliminated to give
\begin{equation}
i\partial_t\hat{\Psi}_S = -\frac{1}{2m_{eff}}\partial_z^2\hat{\Psi}_S + V_{eff}\hat{\Psi}_S,
\label{Schrodingereqn}
\end{equation}
where $m_{eff} = -\pi g^2 n/[vv_g(\Delta+i\gamma/2)]$ and $V_{eff} = 2\Delta\omega v_g/v -\delta$ can be thought of as a (complex) effective mass and an effective potential in analogy to the Schr\"odinger equation. Note that the group velocity of the sum mode is zero (it becomes non-zero for non-identical control fields $\Omega_\pm$), indicating that we have a stationary light.

Assuming $\Delta = -|\Delta| \gg \gamma$, the effective mass becomes real and can be written as $\gamma_{1D}n/4v_g|\Delta|$, where $\gamma_{1D} = 4\pi g^2/v$ can be identified as the spontaneous emission rate of a single atom into the waveguide. $\eta = \gamma_{1D}/\gamma$ then defines a `single-atom cooperativity'--the efficiency of coupling into the waveguide. With these definitions the optical depth of the waveguide is defined as $OD = \eta n l$, where $l$ is the length of the waveguide. $OD$ quantifies the strength of the interaction between photons and the medium and a large value of it ($\approx 1000$) is typically required to get into the strongly interacting regime considered in this section.

\textbf{Preparation and detection of stationary pulses of light}: 
Quantum simulations using stationary light largely consists of three steps: 1) loading a traveling pulse, 2) converting it to a stationary pulse that undergoes a prescribed evolution, and 3) detecting the resulting stationary light. Here let us illustrate this with a typical example. A quantum pulse of light $\hat{E}_+$ is sent in from the left with only the control field $\Omega_+$ initially turned on. In this typical EIT scheme, the quantum pulse travels slowly with the group velocity $v_g \propto \Omega^2$. When the pulse has completely entered the medium $\Omega(t)$ is adiabatically turned to zero, trapping the entire pulse in the system \cite{FleischhauerLukin2000}. Following this loading process, stationary light is created by adiabatically increasing both $\Omega_+(t)$ and $\Omega_-(t)$ simultaneously, creating the long-lived sum mode $\hat{\Psi}_S$ which evolves under an engineered Hamiltonian of choice. After a dynamical evolution or an adiabatic change of the initial state, the pulse is released by turning off $\Omega_-$ which converts the stationary pulse to a propagating one. The out-going pulse can then be detected using standard quantum optical measurements; for example, a photon correlation measurement can detect spatial correlations of the stationary pulse that have been converted into temporal correlations.

\subsubsection{EIT nonlinearities with four-level atoms}

The last essential ingredient in simulating strongly correlated effects with photons is a strong nonlinearity (between photons or more accurately polaritons). Normally, to get photons to interact with eachother strongly, one requires them to be resonant with the atoms, which unfortunately entails strong dissipation. As we have already hinted, an effective way to overcome this problem is to use EIT to suppress the linear absorption while enhancing the Kerr-nonlinearity through a quantum interference effect \cite{FleischhauerMarangos2005}. However, the problem is that the magnitude of the nonlinearity is limited by dissipation in the usual three-lv scheme. 

\begin{figure}[ht]
\begin{center}
\includegraphics[width=0.5\columnwidth]{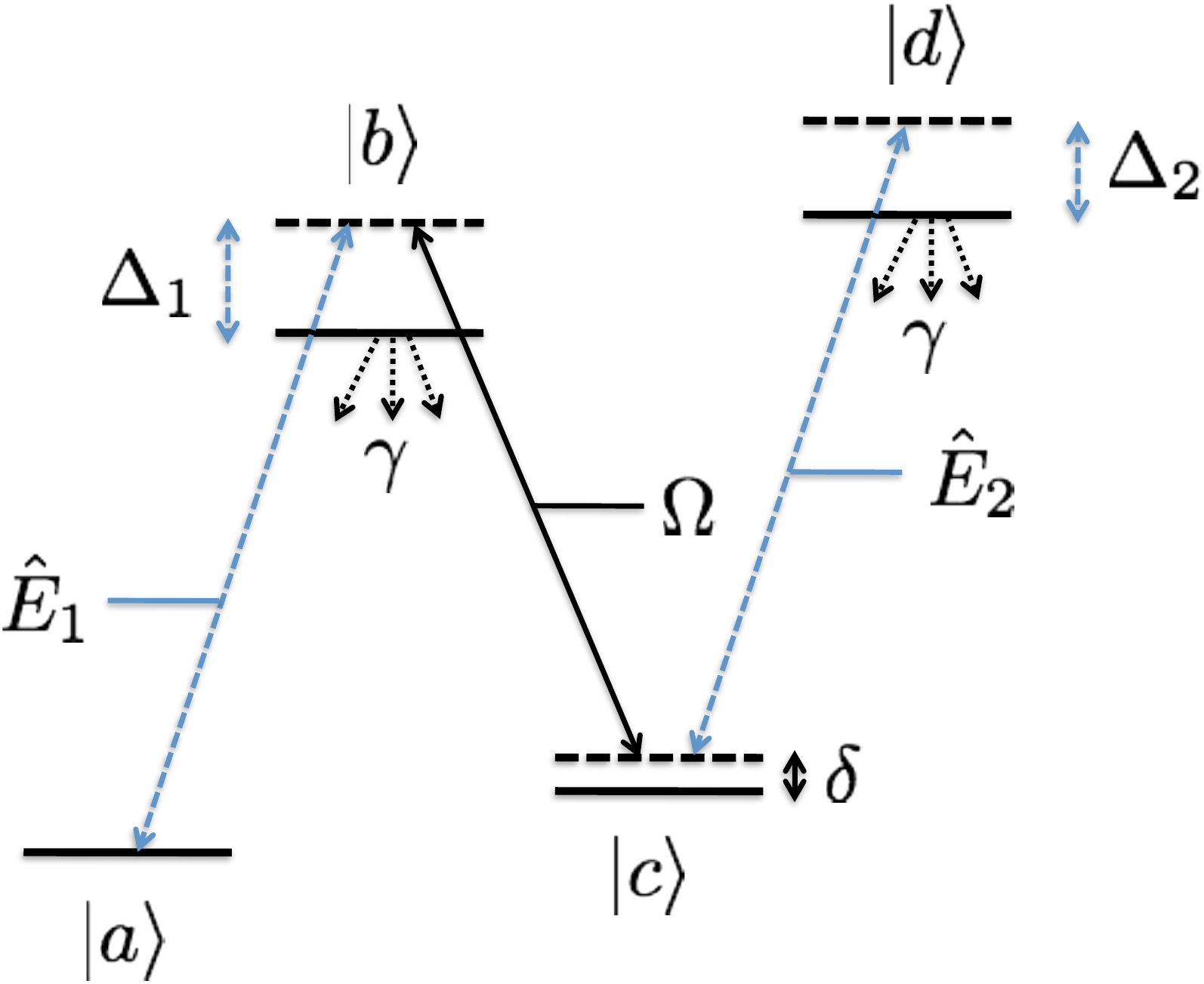}
\caption{Atomic level scheme to generate a giant resonantly-enhanced Kerr nonlinearity.} 
\label{4lvatom}
\end{center}
\end{figure}

A scheme that uses four level atoms to generate a giant resonantly enhanced nonlinearities was first introduced in \cite{SchmidtImamoglu1996}, where the atomic structure depicted in figure \ref{4lvatom} was shown to yield a cross-Kerr nonlinearity between the fields $\hat{E}_1$ and $\hat{E_2}$. The imaginary part of the third-order susceptibility $\chi^{(3)}$ can be made small by increasing the detuning $\Delta_2$, while leaving the real part finite and, in fact, orders of magnitude larger than those obtained from the usual 3-lv scheme. Self-Kerr nonlinearity can also be obtained by replacing $\hat{E}_2$ with $\hat{E}_1$ \cite{ImamogluDeutsch1997}. These types of nonlinearity are exactly what one needs to simulate strongly interacting fields with photons as we will explain in the remainder of this section. We will continue to adopt the Heisenberg-Langevin equation approach to study the dynamics in the low-loss regime in this review, but a master equation description can also be employed as shown by Kiffner and Hartmann \cite{KiffnerHartmann2010a}.

\subsection{Tonks-Girardeau gas of photons}

Combining the techniques of stationary light and giant Kerr nonlinearity, Chang et al.~\cite{ChangDemler2008} showed that photons can crystallize and mimic the behaviour of strongly interacting one-dimensional bosons called the Tonks-Girardeau (TG) gas \cite{Tonks1936, Girardeau1960}. The system consists of an ensemble of four-level atoms interfaced with a tight-confining waveguide such as a tapered optical fiber or a hollow-core optical fiber. There has also been a proposal to observe a TG gas induced by a purely-dissipative interaction \cite{SyassenDurr2008} in the same setting \cite{KiffnerHartmann2010}, but we will not discuss it here. Let us start with a gentle introduction to the physics of 1 dimensional Bose gas before delving into the proposal of Chang and coworkers.

Studies have shown that In 1 (spatial) dimension, effects of quantum fluctuations are enhanced and interacting particles show behaviour dramatically different from the higher dimensional analogues \cite{Giamarchi}. A famous example is the TG gas, where strongly interacting bosons are effectively fermionized. Dynamics of the simplest interacting bosons in 1D are governed by the Lieb-Liniger model \cite{LiebLiniger1963}, described by the Hamiltonian
\begin{eqnarray}
-\int dz \Psi^\dagger(z) \frac{1}{2m}\partial_z^2\Psi(z) + U\Psi^\dagger(z)\Psi^\dagger(z)\Psi(z)\Psi(z), 
\label{LiebLiniger}
\end{eqnarray}
which for a fixed linear density $n$ is characterized by a single dimensionless parameter $\gamma_{LL} = nU/(n^2/2m)$. When $\gamma_{LL} \gg 1$, the bosons are impenetrable, which means that there is an effective exclusion principle and the system behaves in many ways like a non-interacting Fermi gas. 
This fermionization can be observed in the second-order density-density correlation function $g^{(2)}(z,z') = \langle I(z)I(z')\rangle/(\langle I(z)\rangle\langle I(z')\rangle)$, for example, as pointed out by Chang and coworkers. The TG gas has been realized with cold atoms in optical lattices \cite{KinoshitaWeiss2004,ParedesBloch2004}, but a realization with photons brings in a new set of techniques such as the standard photon-correlation measurements.

\begin{figure}[ht]
\begin{center}
\includegraphics[width=0.65\columnwidth]{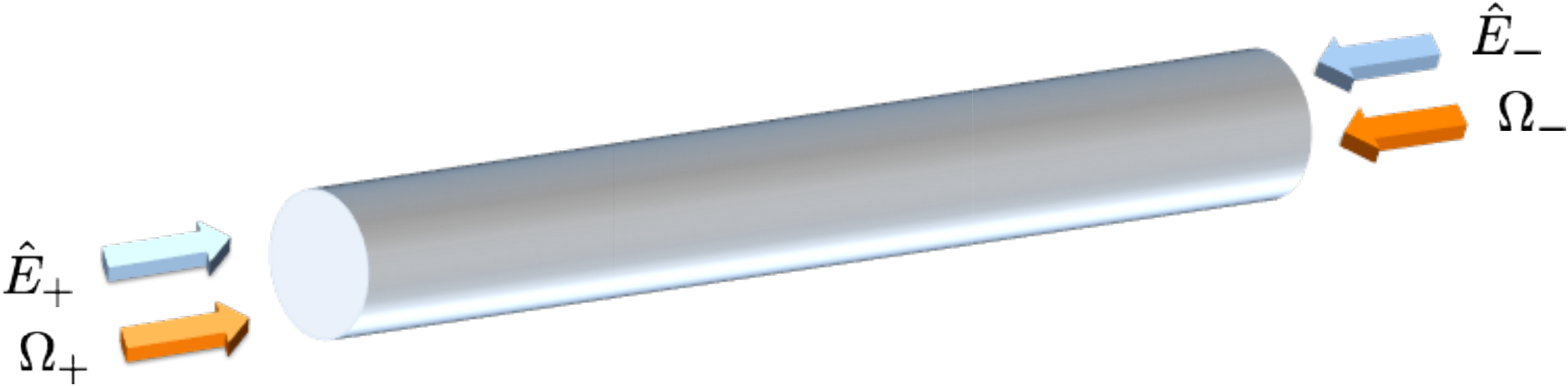}
\hspace{1cm}
\includegraphics[width=0.5\columnwidth]{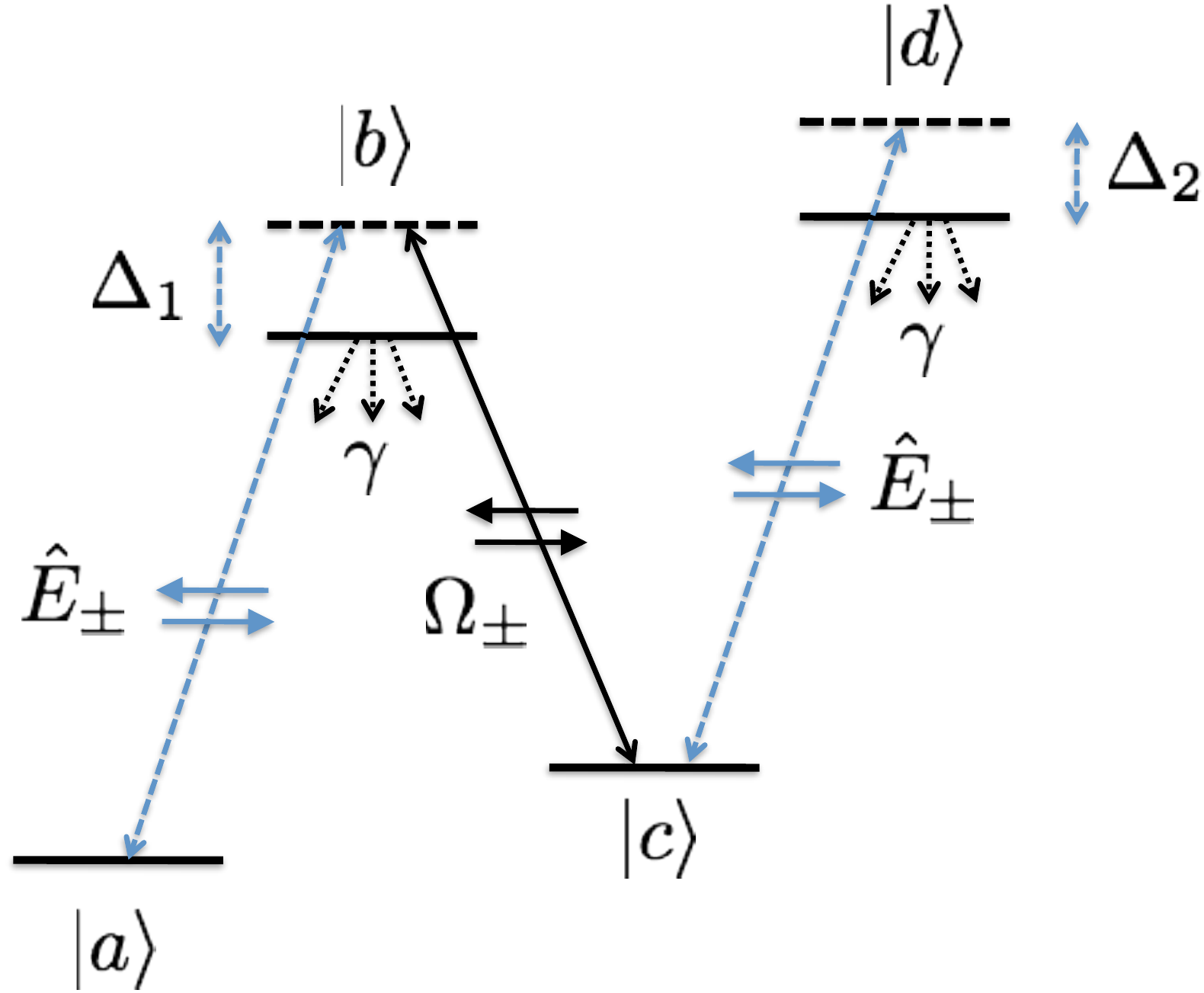}
\caption{Nonlinear waveguide setup simulating the Lieb-Liniger model with stationary pulses of light. The waveguide contains 4-lv atoms whose coupling scheme with the quantum and control fields is indicated.} 
\label{ChangNatPhys}
\end{center}
\end{figure}

A schematic diagram of the proposed setup to realize a TG gas of photons is depicted in figure \ref{ChangNatPhys}. The two-photon detuning $\delta$ is assumed to be zero and each atomic coherence is coupled to two counter-propagating fields as in the stationary light setting. Furthermore, the atomic transition $c \leftrightarrow d$ is coupled to the same quantum fields as the transition $a \leftrightarrow b$ so that the self-Kerr nonlinearity is induced. In the absence of the fourth lv $|d\rangle$, we have already seen that the sum mode obeys the Schr\"odinger equation (\ref{Schrodingereqn}). With the fourth level, this equation becomes a nonlinear Schr\"odinger equation (assuming $V_{eff} = 0$) \cite{ChangDemler2008}
\begin{equation}
i\partial_t\hat{\Psi}(z,t) = -\frac{1}{2m_{eff}}\partial_z^2\hat{\Psi}(z,t) + 2\tilde{g} \hat{\Psi}^\dagger(z,t)\hat{\Psi}^2(z,t),
\end{equation}
where  $m_{eff} = -\pi g^2 n/[vv_g(\Delta_1+i\gamma/2)]$ as before and $2\tilde{g} = 2\pi g^2v_g/[(\Delta_2+i\gamma/2)v]$. We have assumed for convenience that the two excited states have the same decay rate $\gamma$. Because the dark state polariton operators obey the bosonic commutation relations, the above equation describes the evolution under the Lieb-Liniger Hamiltonian (\ref{LiebLiniger}). In particular, the dimensionless parameter $\gamma_{LL}$ is written in terms of bare optical parameters as
\begin{equation}
\gamma_{LL} = \frac{m_{eff}\tilde{g}}{n_{ph}} = -\frac{\pi^2 g^4}{(\Delta_1+i\gamma/2)(\Delta_2+i\gamma/2)}\frac{n}{n_{ph}},
\end{equation}
where $n_{ph}$ is the linear density of photons at the centre of the pulse. Notice that there is a non-zero imaginary part, which will be neglected on the basis that $|\Delta_{1,2}| \gg \gamma$.

Among the characteristic features of a TG gas is an oscillating density-density correlation function, similar to the Friedel oscillations in electronic systems with an impurity \cite{Friedel1958}. In the TG limit, the second-order intensity correlation function $g^{(2)}(z,z') = \langle I(z)I(z')\rangle/(\langle I(z) \rangle\langle I(z') \rangle)$ of the ground state takes the form \cite{Lenard1966}
\begin{equation}
g^{(2)}_{\textrm{TG}}(z,z') = 1-\left( \frac{\sin k_F(z-z')}{k_F(z-z')} \right)^2,
\end{equation}
where $I(z) = \hat{\Psi}^\dagger(z)\hat{\Psi}(z)$ and $k_F = \pi n_{ph}$ is the `Fermi momentum' of the photonic TG gas.

The $2k_F$ oscillations can be observed experimentally by following the preparation and detection procedure outlined in subsection \ref{sect:stationary light}: An initial pulse of coherent state is loaded into the waveguide and turned into a stationary pulse. Then the dimensionless interaction parameter $\gamma_{LL}$ is increased from zero to a large value by changing, for example, the detuning. The initial coherent state, which is an eigenstate of the free Hamiltonian, remains in an eigenstate of the Hamiltonian (with non-zero interaction strength) and therefore exhibits the Friedel oscillations. The stationary pulse can then be converted to a traveling pulse, whose temporal correlation function reveals the density-density correlation function of the simulated quantum state. 

\begin{figure}[ht]
\begin{center}
\includegraphics[width=0.6\columnwidth]{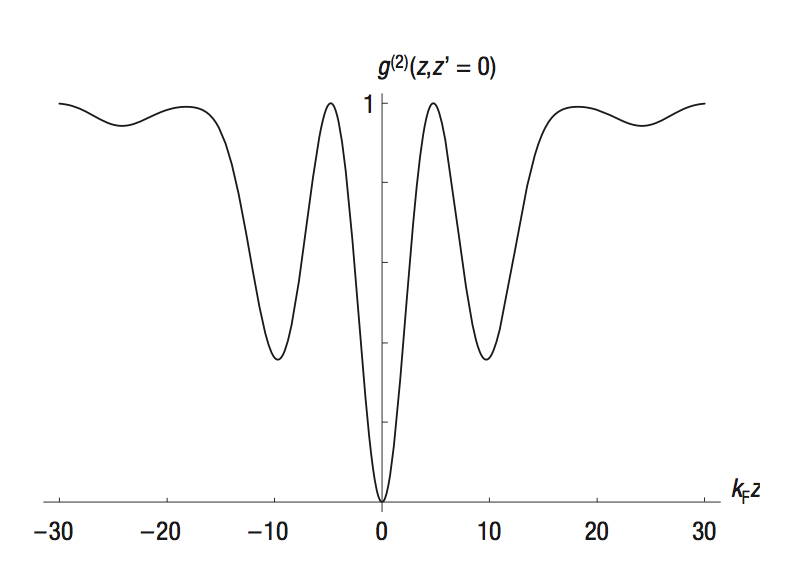}
\caption{Density-density correlation function $g^{(2)}(z,z'=0)$ of the final TG gas of photons. Initial density profile is $n_{ph}(z) = n_0(1-z^2/z_0^2)^{1/2}$, $N_{ph}= 10$, $z_0 = 5k_F^-1$, and $t=10\omega_f^{-1}$. Reprinted with permission from Chang et al.~\cite{ChangDemler2008}.} 
\label{ChangFig2}
\end{center}
\end{figure}

An example is shown in figure \ref{ChangFig2}, taken from \cite{ChangDemler2008}, for an initial density profile $n_{ph}(z) = n_0(1-z^2/z_0^2)^{1/2}$ under an exponentially increasing interaction parameter $\gamma_{LL}$. The distance is in units of the inverse effective Fermi wavevector $k_F$ and the final time is $t=10\omega_f^{-1}$, where $\omega_f$ is the Fermi energy on the order of $n_{ph}^2/m_{eff}$. For an optical depth of $\approx 2000$ and the cooperativity $\eta \approx 0.2$, the achievable maximum $\gamma_{LL}$ is around 10. 

\subsection{Introducing an effective potential--Bose Hubbard and sine-Gordon models, and the `pinning' transition}
Looking back to equation (\ref{Schrodingereqn}), it is not too difficult to see that a spatially dependent two-photon detuning would induce an effective potential for the stationary light. Indeed, this fact has been noticed in a proposal to simulate the Klein tunnelling with stationary light \cite{OtterbachFleischhauer2009}. Upon introducing another transition, one can induce a small spatially-dependent two-photon detuning and as a consequence the resulting nonlinear Schr\"odinger equation would have an effective potential term. Another method is to induce a small change in the spatial profile of the atomic density \cite{HuoAngelakis2012b}. The latter requires cold atomic gas so that the density profile does not change appreciably during the experiment. However, evolution of a stationary light in a cold atomic ensemble is different to that in a hot vapour because of the secular approximation discussed earlier in subsection \ref{sect:stationary light} \cite{HansenMolmer2007,NikoghosyanFleischhauer2009} and one has to resort to a different atomic level scheme of a double-lambda type. Here, we discuss the effects of introducing a lattice potential to strongly interacting photons assuming that the lattice has been created via one of the two methods and defer a discussion on stationary light in cold atomic gases to the next subsection.

\begin{figure}[ht]
\begin{center}
\includegraphics[width=0.3\textwidth]{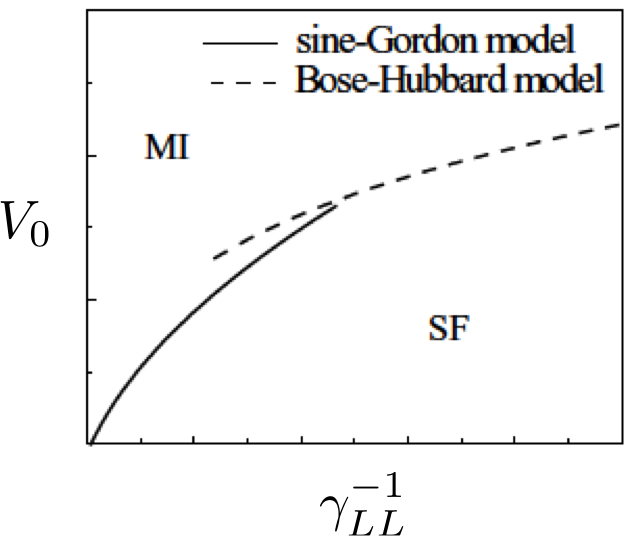}
\caption{A figure showing the phase diagram of the Lieb-Liniger model in the presence of a potential. The interaction strength and the lattice depth is changed by tuning an optical parameter which is not specified here. Reprinted from Huo and Angelakis \cite{HuoAngelakis2012b}.} 
\label{sGBH}
\end{center}
\end{figure}

We follow Huo and Angelakis \cite{HuoAngelakis2012b} and consider the Lieb-Liniger model with an additional potential term of the form $V(z) = V_0 \cos^2(\pi n_{ph}z)$. The periodicity of the lattice potential is set such that the number of sites equals the number of photons in the waveguide. Depending on the lattice depth and the interaction strength, the system shows either the Mott insulating phase or the superfluid phase \cite{BuchlerZwerger2003} as depicted schematically in figure \ref{sGBH}.

There are two interesting limits in the phase diagram: i) weakly interacting particles ($\gamma_{LL} < 1$) in a deep potential ($V_0 \gg E_r$) and ii) strongly interacting particles ($\gamma_{LL} > 1$) in a shallow potential ($V_0 \sim E_r$), where $E_r =  \pi^2 n_{ph}^2/2m_{eff}$ is the recoil energy. For case i), the physics is well described by the well-known Bose-Hubbard model \cite{FisherFisher1989,JakschZoller1998}, where the lowest localised Wannier mode in each lattice site is coupled to the modes in its nearest neighbour sites with depth-dependent on-site interaction strength. As we have seen earlier, the Bose-Hubbard hamiltonian reads
\begin{equation}
H_{BH} = -J \sum_{\langle i,j \rangle} a_i^\dagger a_j + \frac{U}{2} \sum_i a_i^\dagger a_i^\dagger a_i a_i,
\end{equation}
where $a_i$ is now interpreted as the annihilation operator for the Wannier mode at site $i$, $\sqrt{\pi}J/E_r = 4(V_0/E_r)^{3/4}e^{-2\sqrt{V_0/E_r}}$, and $U/E_r = \sqrt{2/\pi^3}(V_0/E_r)^{1/4}\gamma_{LL}$. Note that the effective parameters $J$ and $U$ only depend on $V_0$, $E_r$, and $\gamma_{LL}$, all of which can be tuned by changing quantum optical parameters. By tuning the value of $U/E_r$, through changing the value of the detuning $\Delta_2$ for example, one can observe the Mott-to-superfluid transition.

There is also an interesting quantum phase transition called a `pinning' transition in case ii), when the interaction is strong. In the weak lattice limit, the particle density is weakly modulated and one is able to separate out the density perturbation and phase as canonically conjugate observables. Haldane was the first to consider this case and show that the quantum sine-Gordon model correctly describes the physics in this limit \cite{Haldane1981}. Following the work of B\"uchler, Blatter, and Zwerger \cite{BuchlerZwerger2003}, we briefly reproduce the derivation of the sine-Gordon Hamiltonian here. 

First consider a hydrodynamic approximation where one describes the bosonic field operator in the low energy limit as $\Psi \approx \sqrt{\rho_0+\partial_z\theta/\pi}\exp(i\phi)$, where $\rho_0$ is the mean density of the atoms, $\partial_z\theta/\pi$ is the long-wavelength density fluctuation, and $\phi$ is the phase field. The latter two obeys the boson commutation relation $[\partial_z\theta(z),\phi(z')] = i\pi\delta(z-z')$. Neglecting the lattice potential for now, the low-energy physics is described by the Hamiltonian
\begin{equation}
H_0 =  \frac{1}{2\pi}\int dz \left[ v_J(\partial_z\phi)^2 + v_N(\partial_z\theta)^2 \right].
\end{equation} 
The first part comes from the kinetic energy where $v_J = \pi\rho_0/m$ and the second term derives from the interaction energy with the coefficient $v_N = \partial_n\mu/\pi$ determined by the inverse compressibility. To obtain the Hamiltonian due to the lattice potential, one needs to modify the density operator to account for its discrete character:
\begin{equation}
\rho(z) = \left[ \rho_0 + \frac{1}{\pi}\partial_z\theta\right] \left[ 1+2\sum_{l=1}^{\infty} \cos(l\theta + l\pi\rho_0z)\right].
\end{equation}
In the commensurate limit (one particle per lattice cell), the dominant part of the potential arising from the lowest harmonic in the above equation takes the sine-Gordon form
\begin{equation}
H_V = \frac{n_{ph}V_0}{2}\int dz \cos(2\theta).
\end{equation}

The Hamiltonian $H_0 + H_V$ is the 1 dimensional quantum sine-Gordon equation which is known to exhibit the so-called `pinning' transition \cite{HallerNagerl2010}. This phase transition can best be described by the dimensionless parameter $K=\sqrt{v_J/v_N}$ that is related, in the absence of the periodic potential, to the Lieb-Liniger parameter $\gamma_{LL}$; $K(\gamma \le 10) \approx \pi/\sqrt{\gamma_{LL}-(1/2\pi)\gamma_{LL}^{3/2}}$ and $K(\gamma \gg 10) \approx (1+2/\gamma_{LL})^2$. When $K > K_c=2$, the weak potential is irrelevant and the bosons are in the superfluid phase whereas for $K < K_c$ an arbitrarily weak potential can pin the bosons to lattice sites, achieving the Mott insulating state. 

The procedure to observe these phases in the nonlinear waveguide system with a lattice potential is as follows. First a pulse containing a certain number of photons is loaded into the waveguide. Then one turns up the interaction strength and then the lattice potential adiabatically. This step can be realized experimentally by changing the detunings $\Delta_2$ and $\delta$. By the end of this step, the photonic state would have adiabatically followed the Hamiltonian and consequently would exhibit either the superfluid or the Mott-insulating behaviour which can be observed by releasing the stationary pulse and performing correlation measurement on them as in the observation of a TG gas of photons.

\subsection{Bose-Einstein condensation of stationary-light polaritons}

\begin{figure}[ht]
\begin{center}
\includegraphics[width=0.8\columnwidth]{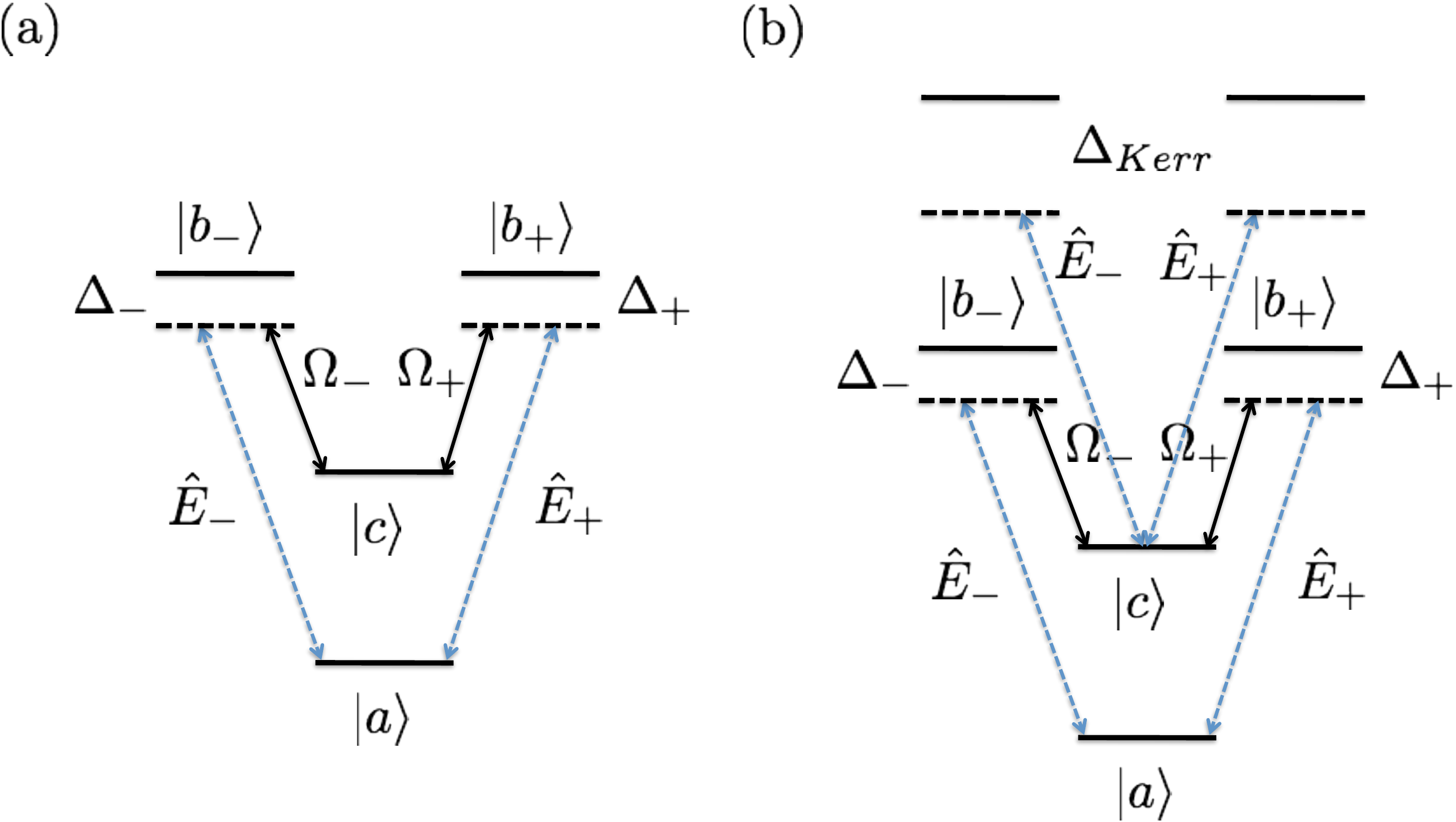}
\caption{(a) Atomic level scheme for creating dark state polaritons using cold atoms. (b) Modified atomic level scheme to induce the Kerr interaction between dark state polaritons, allowing for them to Bose condense.} 
\label{becscheme}
\end{center}
\end{figure}

The realization of strongly interacting photons in the previous subsection relied on having a hot vapour of atoms so that the secular approximation holds. However, it is also possible to create stationary light with cold atoms \cite{MoiseevHam2006,ZimmerFleischhauer2008} as demonstrated in \cite{LinYu2009}.  In particular, Zimmer et al.~\cite{ZimmerFleischhauer2008} showed how the linear Schr\"odinger equation can be obtained with cold atoms using a slightly different atomic level scheme shown in figure \ref{becscheme}(a). In this scheme, $\hat{E}_+$ and $\hat{E}_-$ couple to different atomic transitions unlike in the usual $\Lambda$ scheme where the two couple to the same transition, so the secular approximation no longer needs to be invoked. For equal control field strengths, $\Omega_+ = \Omega_-$, the stationary dark state polariton is defined as 
\begin{equation}
\hat{\Psi}({\bf r},t) = \frac{\cos\theta}{\sqrt{2}}\left( \hat{E}_+({\bf r},t)+ \hat{E}_-({\bf r},t)\right) - \sin\theta\hat{S}({\bf r},t),
\end{equation}
where $\tan^2\theta = 2\pi g^2n/\Omega^2$, $\Omega^2 = \Omega_+^2 + \Omega_-^2$, and $\hat{S} = \sqrt{n}\hat{\sigma}_{ac}({\bf r},t)$. The Schr\"odinger equation obeyed by this field operator reads
\begin{equation}
i\partial_t \hat{\Psi}({\bf r},t) = -\left[ \frac{1}{2m_\parallel}\partial_z^2 + \frac{1}{2m_\perp}\Delta_\perp \right] \hat{\Psi}({\bf r},t),
\end{equation}
where $m_\perp = mv_{rec}/v_g$ and $m_\parallel^{-1} = m_\perp^{-1}k_p(\gamma v/2\pi g^2n)(2\Delta/\gamma+i)$. Here $k_p$ is the wavenumber of the probe field and $v_{rec} = k_p/m$ is the recoil velocity of an atom of mass $m$.

Introduction of additional levels as in figure \ref{becscheme}(b) adds the Kerr-type interaction between the dark state polaritons whose elastic collision rate can be much greater than the loss rates and thereby allow thermalization of polaritons without significant losses \cite{FleischhauerUnanyan2008}. The optical depth required to reach this regime is only modestly high at about $OD > 30$. For realistic parameters, the critical temperature to create a BEC out of dark state polaritons (as compared to the critical temperature for ideal bosonic atoms) is $T_c/T_c^{atoms} \approx 6\times 10^5$ which is in the $mK$ regime. The latter is in the same range as the temperature limit of about 1 $mK$ imposed by the finite frequency window of EIT. Therefore, it is possible to observe the BEC of stationary light polaritons at temperatures several orders of magnitude higher than the critical temperature of atomic BECs. 

\begin{figure}[ht]
\begin{center}
\includegraphics[width=00.9\columnwidth]{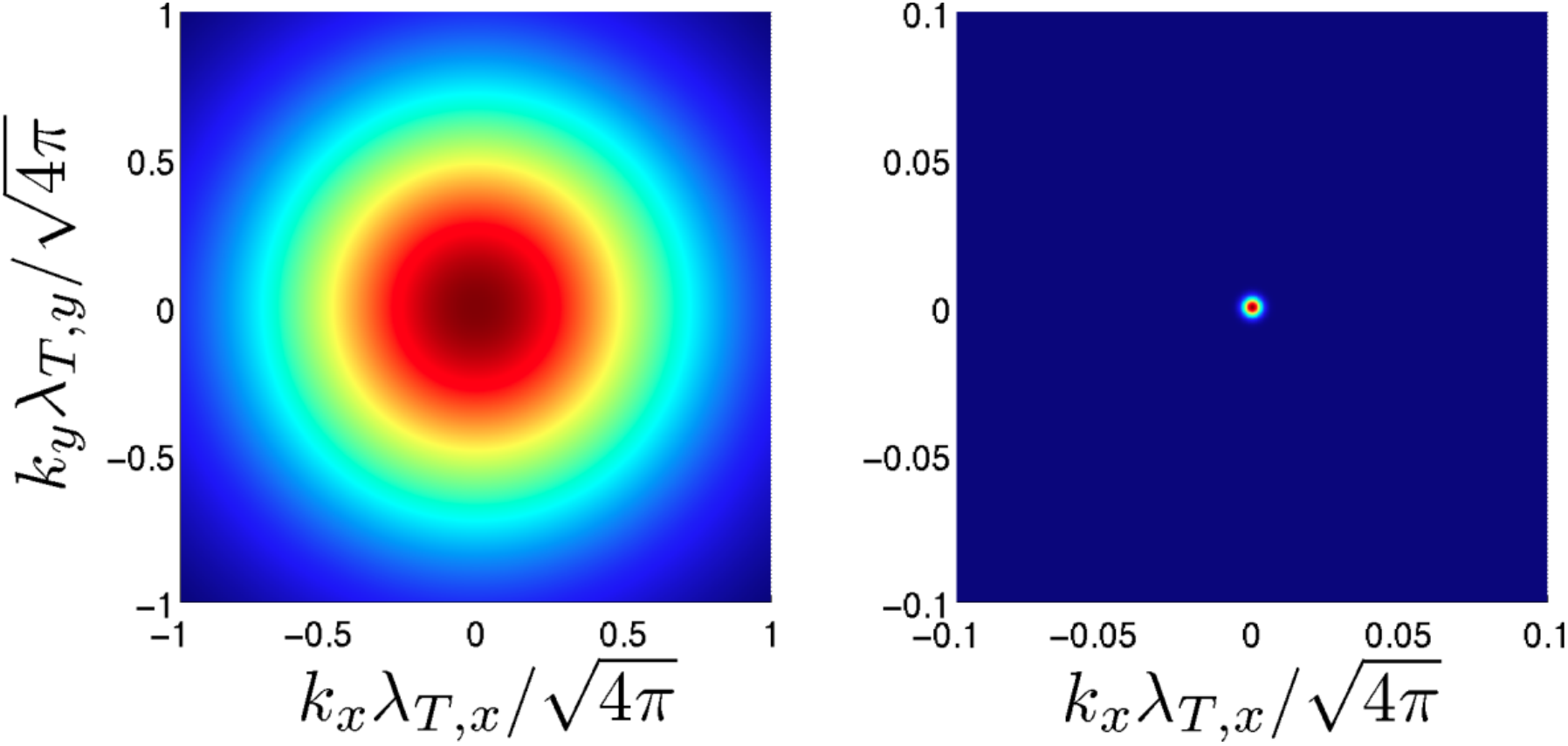}
\caption{Transverse emission profile above (left) and below (right) the condensation temperature in arbitrary units, assuming an ideal quasi-homogeneous Bose-gas of width $178 \lambda_{T,x}$, where $\lambda_{T,x}$ is the transversal de Broglie wavelength at temperature $T$. Reprinted with permission from Fleischhauer et al~\cite{FleischhauerUnanyan2008}.} 
\label{BECpolaritons}
\end{center}
\end{figure}

In the stationary light setup, a BEC will be created not by reducing the temperature of the polariton gas, but by increasing the critical temperature of condensation through changing the effective mass. The steps to observe Bose-Einstein condensation of polaritons is then as follows. i) A pulse of coherent light is trapped as stationary polaritons using the same method described in the previous subsection. Initially, the critical temperature is below the ambient temperature so there is no condensation. ii) As the critical temperature is increased, by tuning the strength of the control laser, the polaritons thermalize through elastic collisions and reach BEC before significant losses occur. This causes the macroscopic occupation of ${\bf k}=0$ momentum state, which can be iii) observed by converting the stationary polariton into a traveling one, whose transverse profile shows the onset of BEC. The latter is illustrated in figure \ref{BECpolaritons}, where the transverse emission profiles are shown at temperatures above (left) and below (right) the critical temperature. A similar proposal with Rydberg atoms--to induce dipolar interaction between dark state polaritons--have been made by Nikoghosyan et al.~\cite{NikoghosyanPlenio2012}. 

\subsection{Two-component bosonic models: Luttinger liquids, spin-charge separation and Cooper pairing with photons}

So far in this section, we have reviewed works on quantum simulations of strongly interacting bosons when there is only a single species of bosons. Soon after the pioneering work on a TG gas of stationary light, Angelakis et al.~realized that the physics of two interacting bosons can also be simulated in a similar waveguide setup \cite{AngelakisKwek2011a,HuoKwek2012a}. Here, we describe how this leads to a realization of the two-component Lieb-Liniger model, which could allow a clear experimental observation of spin-charge separation. Furthermore, by introducing a lattice potential, one can also simulate a two-component Bose-Hubbard model, which exhibits interesting behaviours such as (a 1D analogue of) the famous BEC-BCS crossover \cite{ParedesCirac2003,HuoAngelakis2012a}.

\begin{figure}[ht]
\begin{center}
\includegraphics[width=0.5\columnwidth]{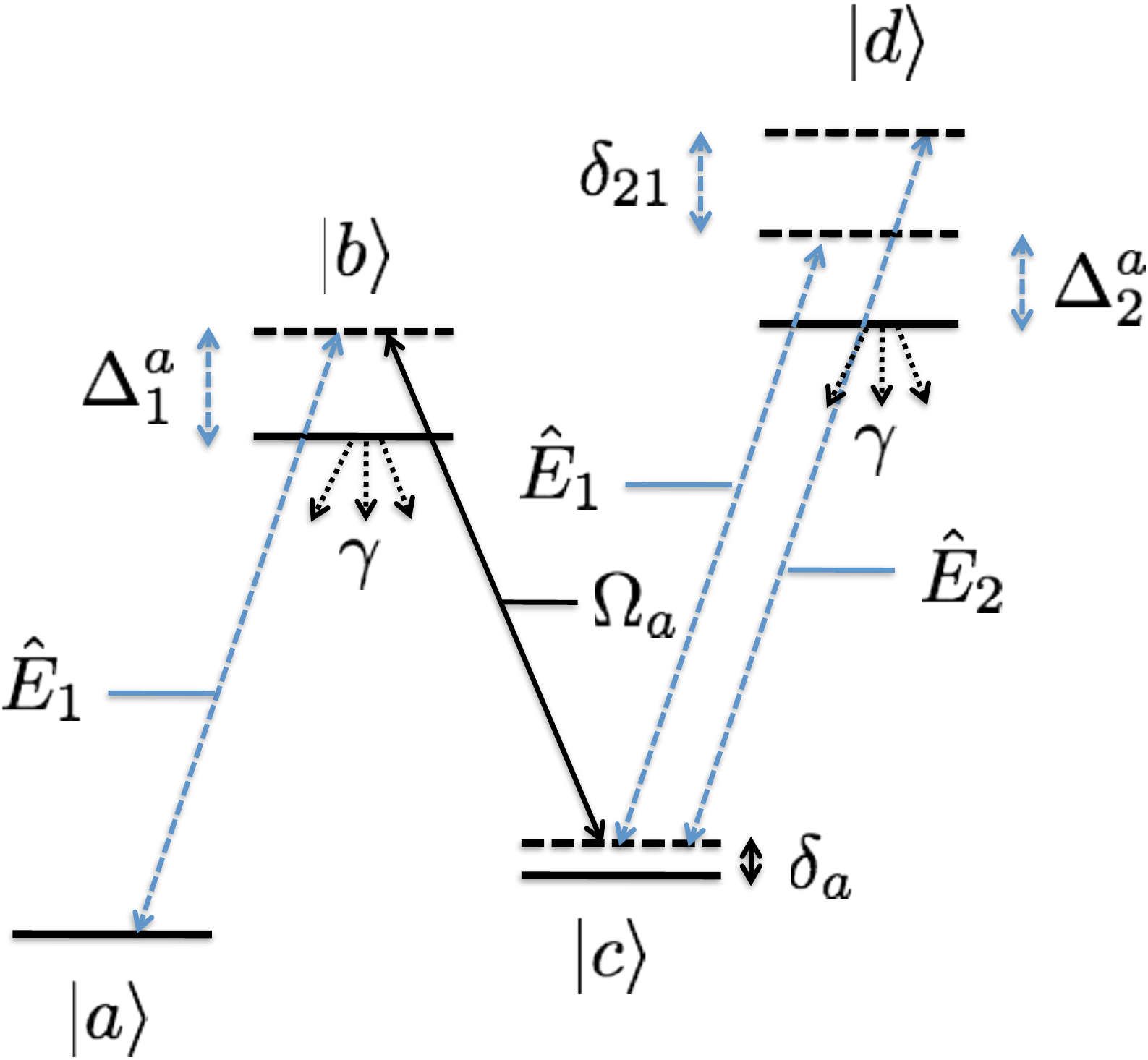}
\caption{Atomic level scheme to realize the two-component Lieb-Liniger model, for the type-$a$ atoms. Type-$b$ atoms have the role of $\hat{E}_1$ and $\hat{E_2}$ reversed and the super- and sub-script $a$ replaced by $b$, as well as $\delta_{21} \rightarrow \delta_{12}$.} 
\label{tclvscheme}
\end{center}
\end{figure}

A schematic diagram of the atomic level scheme that yields the two-component Lieb-Liniger model is shown in figure \ref{tclvscheme}. Two different quantum fields $\hat{E}_1$ and $\hat{E}_2$ with frequencies $\omega_{qu}^{(1)}$ and $\omega_{qu}^{(2)}$ interact with two different species of 4-lv atoms labeled $a$ and $b$ and forms two polariton operators via the usual relation $\hat{\Psi}_{j,\pm} = g\sqrt{2\pi n^{x_j}}\hat{E}_{j,\pm}/\Omega_{x_j}$. The figure depicts the level scheme for type-$a$ atoms; atoms of type $b$ have a similar coupling scheme except that the roles of $\hat{E}_1$ and $\hat{E}_2$ are reversed and the detunings are accordingly labeled by replacing $a$ with $b$. We use $j={1,2}$ to denote the polariton species, $x_1 = a, x_2=b$ to label the atomic species, and we have assumed that the atom-photon coupling strength is independent of the atomic species and transition. $n^a$ ($n^b$) denotes the density of atomic species $a$ ($b$) and $\delta_{ij} = \omega_{qu}^{(i)} - \omega_{qu}^{(j)}$. By setting the detunings such that $\hat{E}_1$ ($\hat{E}_2$) is only coupled to atoms of type-$b$ ($a$) through the transition $|3\rangle \leftrightarrow |4\rangle$, one obtains a cross-species interaction term on top of the self-interaction terms. The dynamics of the polaritons are thus governed by the two-component Lieb-Liniger model
\begin{eqnarray}
H_{\rm tcll} &=& \int dz \sum_j \Big[ -\frac{1}{2m_j}\hat{\Psi}_j^\dagger(z) \partial_z^2 \hat{\Psi}_j(z) \hspace{3cm}
 \nonumber \\  &+& \frac{U_j}{2}\hat{\Psi}_j^\dagger(z)\hat{\Psi}_j^\dagger(z)\hat{\Psi}_j(z)\hat{\Psi}_j(z) \Big ]  + \frac{V}{2}\rho_1(z)\rho_2(z), \hspace{-1.0cm}
\label{tcll}
\end{eqnarray}
where $\rho_j = \hat{\Psi}_j^\dagger(z)\hat{\Psi}_j(z) $. $m_j$ and $U_j$ are defined as $m_{eff}$ and $4\tilde{g}$ introduced earlier, with species-dependent optical parameters, whereas the inter-species interaction parameter $V$ is new. For simplicity, we will assume that all species-dependence are suppressed. Then the inter-species interaction strength is compactly written as  
\begin{equation}
V = \frac{16\pi g^2 n v_g}{\Delta_2}.
\end{equation}
An alternative scheme where the two species correspond to different polarizations has also been proposed in \cite{HuoKwek2012a}. We will come back to this setup in Sect.~\ref{thirring} where a quantum simulation of an interacting relativistic model will be reviewed.

\subsubsection{Luttinger liquid and spin-charge separation}
We have already seen how Lieb-Liniger bosons can be described, in the low-energy sector, in terms of the long-wave density function $\partial_z\theta$ and the phase field $\phi$. It turns out that the same decomposition applied to the two-component Lieb-Liniger model separates the Hamiltonian into two independent components, describing spin and charge components that travel in different speeds. Let us assume that the two species have identical parameters, i.e., $m_j = m$, $U_j = U$, and $\langle \rho_j \rangle =\rho$. Following the notations in  \cite{KleineSchollwock2008} (see also \cite{OrignacGiamarchi1998}), the Hamiltonian (\ref{tcll}) becomes $H = H_c + H_s$, where
\begin{equation}
H_c = \frac{1}{2\pi}\int dz v_cK_c(\partial_z\theta_c)^2 + \frac{v_c}{K_c}(\partial_z\phi_c)^2
\end{equation}
and
\begin{eqnarray}
H_s &=& \frac{1}{2\pi}\int dz v_sK_s(\partial_z\theta_s)^2 + \frac{v_s}{K_s}(\partial_z\phi_s)^2 \nonumber \\
&+& \frac{V}{(\pi \rho^{-1})^2} \cos(\sqrt{8\phi_s}).
\end{eqnarray}
Here the subscripts $c$ and $s$ refer to the charge and spin components and $v_{c,s} = v_0\sqrt{1\pm VK/\pi v_0}$ and $K_{c,s} = K/\sqrt{1\pm VK/(\pi v_0)}$, where analytic expressions $v_0 = \sqrt{U\rho/m}$ and $K = \sqrt{\pi^2\rho/(mU)}$ hold in the weakly interacting regime. 

The physics of this system is fully determined by the velocities $v_{c,s}$ and the so-called Luttinger parameters $K_{c,s}$. Note that the speed of the spin and charge degrees of freedom can be different and could even have different signs. In terms of polaritons, the spin and charge densities correspond to the difference and the sum of the densities of two stationary polariton components. Therefore one could create an excitation in the original polariton field 
and observe as it splits into two components. One would go through the preparation stage discussed before, then let the two components evolve under different speeds, followed by the conversion into traveling slow-light which can be detected via standard optical measurements and post-processed to determine the velocities of the spin and charge parts, for example. In \cite{AngelakisKwek2011a}, it was shown that the optical depth of $\approx 3000$ is sufficient to create the difference $v_c/v_s = 2$ and observe spin-charge separation by measuring the single particle spectral function, which can be determined from the first order correlation function of one of the fields, say $\hat{E}_1$ at a certain quasi-momentum. 

\subsubsection{BCS-BEC like crossover}

\begin{figure}[ht]
\begin{center}
\includegraphics[width=0.4\columnwidth]{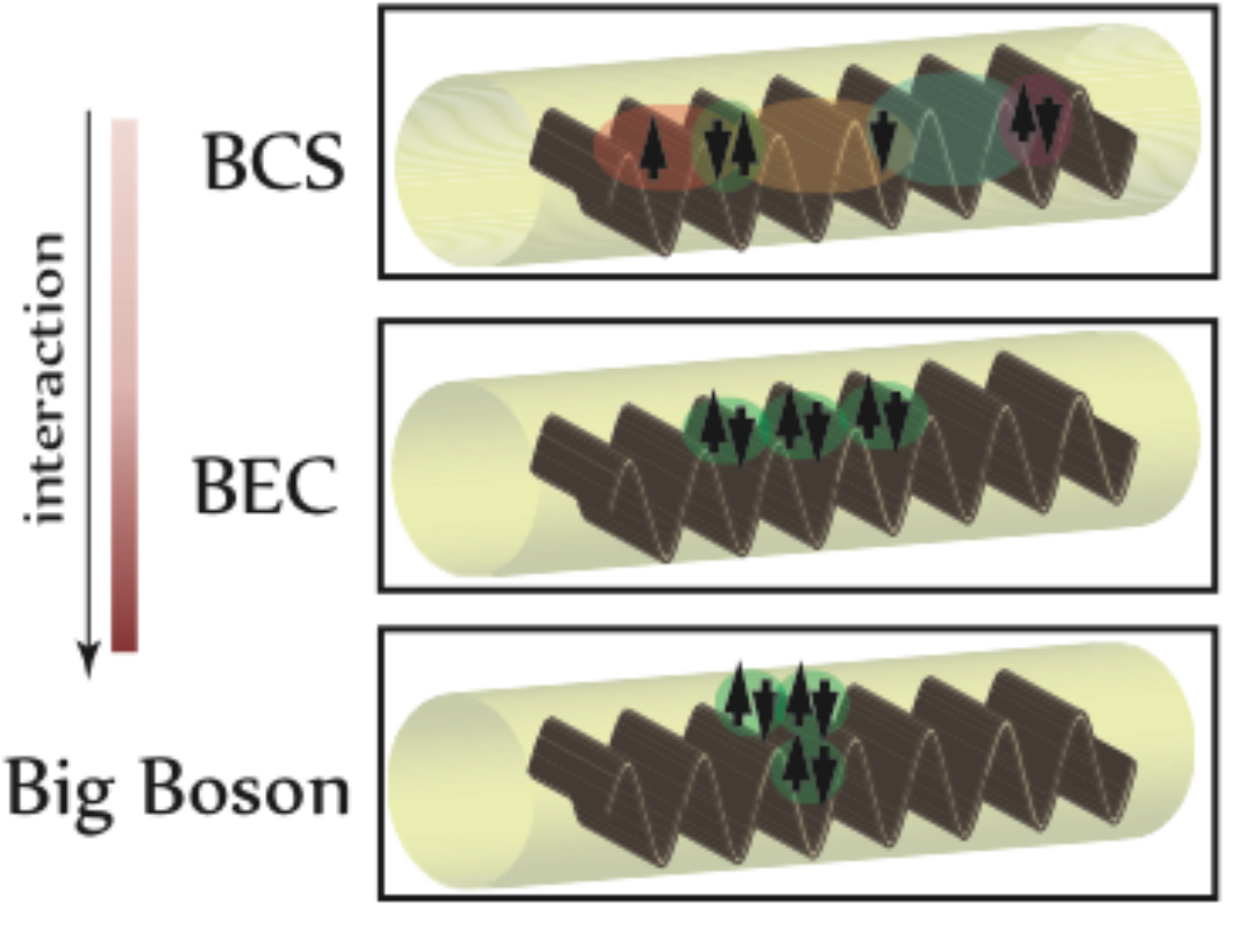}
\includegraphics[width=0.5\columnwidth]{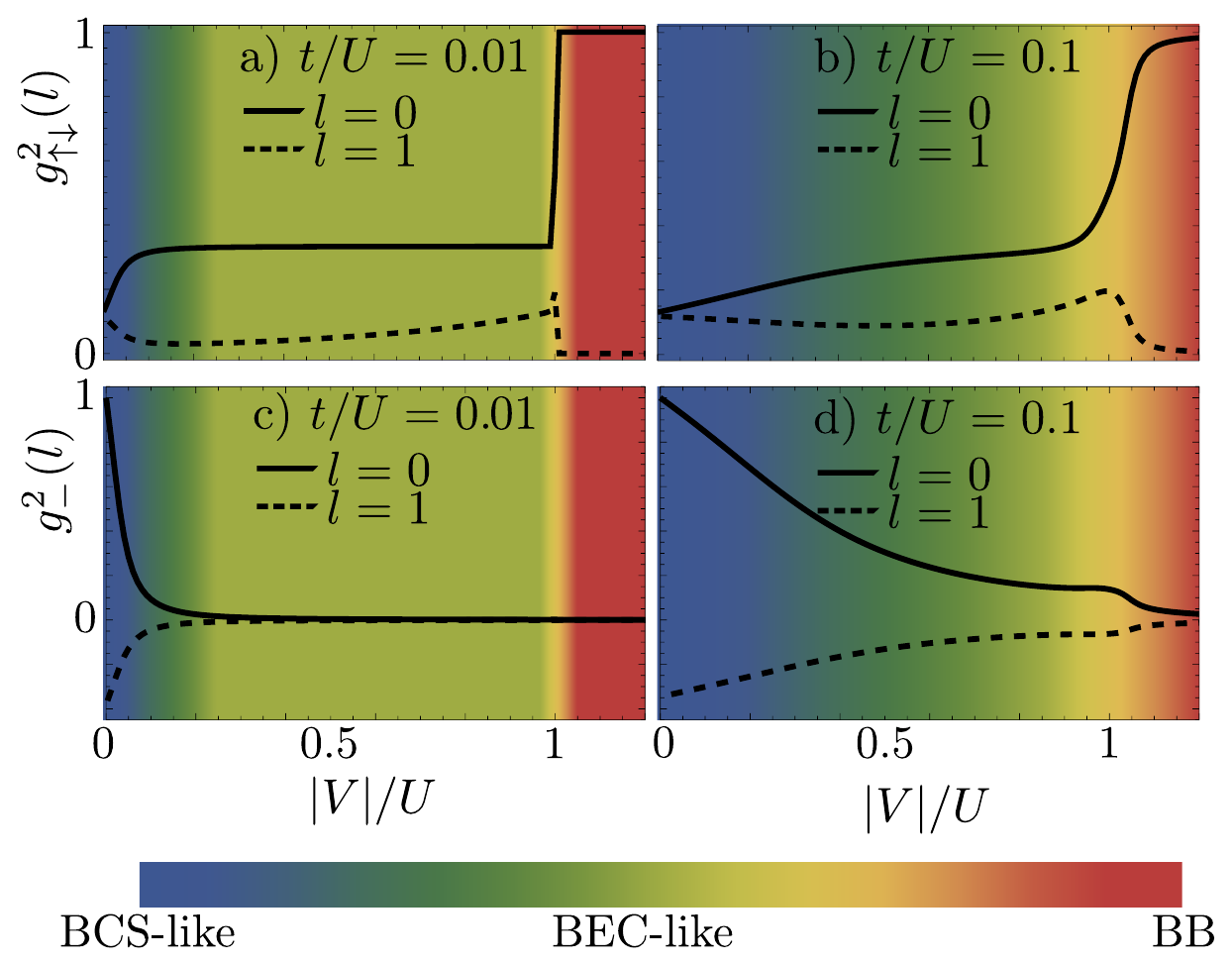}
\caption{Left: A schematic representation of the three regimes achievable in the system for different inter-species interaction strength. A crossover from a BCS-like to a BEC-like phase occurs for small inter-species strengths compared to the intra-species repulsion, whereas in the opposite limit all the stationary polaritons gather at one site to form the so-called Big-Boson state. Right: Cross-species density-density correlation function, $g^2_{\uparrow\downarrow}$, and the density-density correlations between the cross-species population difference, $g^2_-$, as functions of the inter-species interaction strength. The left column is deeper into the effective Fermi-Hubbard regime compared to the right column, displaying sharper crossovers and transitions. Reprinted from Huo et al.~\cite{HuoAngelakis2012a}.} 
\label{tcbhphases}
\end{center}
\end{figure}

Introducing an effective lattice potential to the two-component setup, one can also experimentally realize a 1 dimensional two-component Bose-Hubbard model \cite{HuoAngelakis2012a}, which exhibits some interesting properties. For example, upon inducing strong repulsive interaction between the same species, one can go into an effective Fermi-Hubbard regime where a 1 dimensional analogue of the famous BCS-BEC crossover can be observed. This requires an attractive inter-species interaction, which is easily achieved in the stationary light setup by changing the sign of a single-photon detuning. In the strongly attractive case, the two species tend to pair up to form a molecule although no more than a pair is allowed at a single site because of the effective Pauli exclusion due to the strong intra-species repulsion. This is reminiscent of the short-range-correlated bosonic molecules forming in the BEC regime. In the weakly attractive case, two species tend to stay away from each other and form a long-range correlated pair analogous to the Cooper pair. Thus, by continuously tuning the inter-species interaction strength, it is possible to observe a 1 dimensional analogue of the BEC-BCS cross-over. A schematic illustration of these phases is provided in figure \ref{tcbhphases}.

An optical depth of approximately 1000 was shown to be sufficient for simulating the crossover, where tuning the frequency difference between the quantum fields changes the ratio between the inter- and intra-species interaction strengths. Going through the standard preparation and release processes, one could measure the cross-species intensity correlation function to clearly distinguish all three (BCS-like, BEC-like, and the Big-Boson) phases. In figure \ref{tcbhphases} we plot two types of correlation functions: $g^2_{\uparrow\downarrow}(l) = \sum_i\langle n_{i\uparrow}n_{i+l\downarrow}\rangle$ and $g^2_-(l) = \sum_i\langle (n_{i\uparrow}-n_{i\downarrow})(n_{i+l\uparrow}-n_{i+l\downarrow})\rangle$. The same-site cross-species density-density correlation function $g^2_{\uparrow\downarrow}(0)$ clearly shows the cross-over behaviour upon changing the inter-species interaction strength, whereas a sharp transition is observed as the inter-species attraction becomes stronger than the intra-species repulsion. Correlations in the population difference, $g^2_-$, is sensitive to the BCS-BEC cross over, but not the transition to the Big-Boson phase.

\subsection{Interacting relativistic theories with slow light}
\label{thirring}
\begin{figure}[ht]
\begin{center}
\includegraphics[width=0.7\columnwidth]{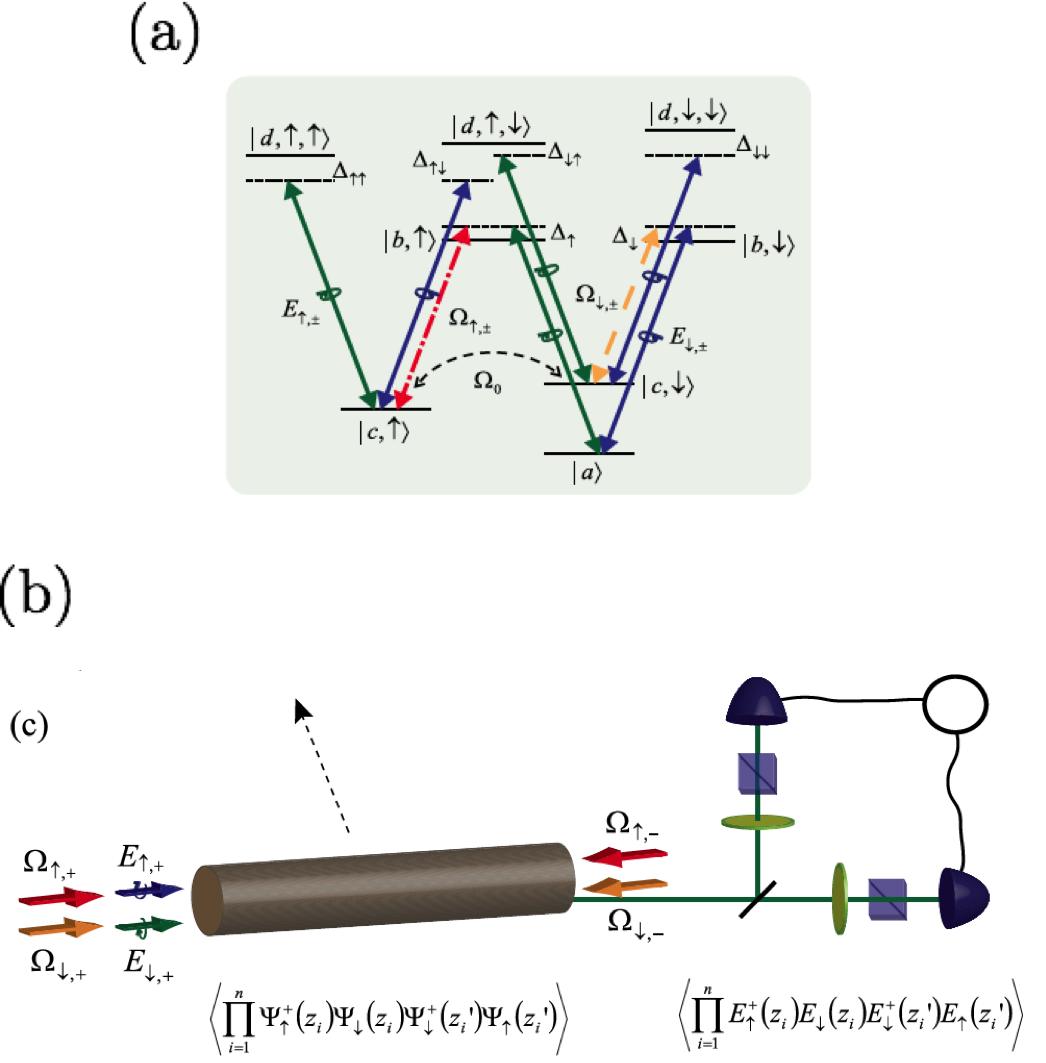}
\caption{(a) Proposed atomic level scheme to realize the Thirring model using two polarized stationary light pulses. (b) Experimental setup illustrating photon correlation measurements at the output that facilitates the direct measurement of the n-point correlation functions. Reprinted from Angelakis et al.~\cite{AngelakisKorepin2013}.} 
\label{thirringsetup}
\end{center}
\end{figure}

Simulations of quantum field theories using nonlinear waveguides can go beyond non-relativistic models. In particular, it was shown that the Thirring model (TM) \cite{Thirring1958}, a relativistic quantum field theory describing the self-interaction of a Dirac field in 2 spacetime dimensions, can be implemented \cite{AngelakisKorepin2013}. The atomic level scheme required is shown in figure \ref{thirringsetup}(a). It consists of two sets of the standard four-level arrangements $\{ |a\rangle, |b,s\rangle, |c,s\rangle, |d,s,s'\rangle \}$, with $s,s'=\{ \uparrow, \downarrow \}$ denoting the two oppositely-polarized quantum pulses $\hat{E}_\uparrow$ and $\hat{E}_\downarrow$. Two self-interacting stationary polaritons $\hat{\Psi}_s$ appear in exactly the same manner as discussed earlier and interact through the extra level $|d,s,\bar{s}\rangle$, $\bar{s}\ne s$, with strength dependent on $\Delta_{s,s'}$. Making the usual assumptions, these polaritons can be shown to obey the two-component Lieb-Liniger model \cite{HuoKwek2012a}. To allow for the linear momentum term required in relativistic models, a small but finite difference between the forward and backward propagating control fields is necessary. Accommodating this change, the (quasi-) stationary lights are now defined as $\hat{\Psi}_s = \alpha_{s,+}\hat{\Psi}_{s,+} + \alpha_{s,-}\hat{\Psi}_{s,-} $ with $\alpha_{s,\pm} = \Omega^2_{s,\pm}/(\Omega^2_{s,+} + \Omega^2_{s,-})$ and $\hat{\Psi}_{s,\pm} = \sqrt{2\pi n}g\hat{E}_{s,\pm}/\Omega_{s,\pm}$. The dynamics of these fields are governed by 
\begin{eqnarray}
i\partial_t\hat{\Psi}_s &=& -\frac{1}{2m_{nr,s}}\partial_z^2\hat{\Psi}_s + i\eta_s\partial_z\hat{\Psi}_s + \Omega_0\hat{\Psi}_{\bar{s}} \nonumber \\ 
&+& \chi_{ss}\hat{\Psi}_s^\dagger\hat{\Psi}_s^2 + \chi_{s\bar{s}}\hat{\Psi}_{\bar{s}}^\dagger\hat{\Psi}_{\bar{s}}\hat{\Psi}_s,
\end{eqnarray}
where $m_{nr,s} = -\bar{\Omega}_s^2/(4v_s^2\Delta_s\sin^22\varphi_s)$, $\eta_s = -2v_s\cos2\varphi_s$, $\chi_{ss}=4\bar{\Omega}_s^2/(\Delta_{ss}n)$, and $\chi_{s\bar{s}} = 2\bar{\Omega}_s^2[2+\cos(\varphi_{\bar{s}}-\varphi_s)]/(\Delta_{s\bar{s}}n)$. $v_s$ is the reduced group velocity for the dark state polaritons, $\varphi_s$ is determined from the relation $\tan^2\varphi_s = \Omega_{s,-}^2/\Omega_{s,+}^2$, and $\bar{\Omega}_s^2 = (\Omega_{s,+}^2+\Omega_{s,-}^2)/2$, and $\Omega_0$ is the coupling rate between $|c,\uparrow\rangle$ and $|c,\downarrow\rangle$ as shown in figure \ref{thirringsetup}(a).

\begin{figure}[ht]
\begin{center}
\includegraphics[width=0.6\columnwidth]{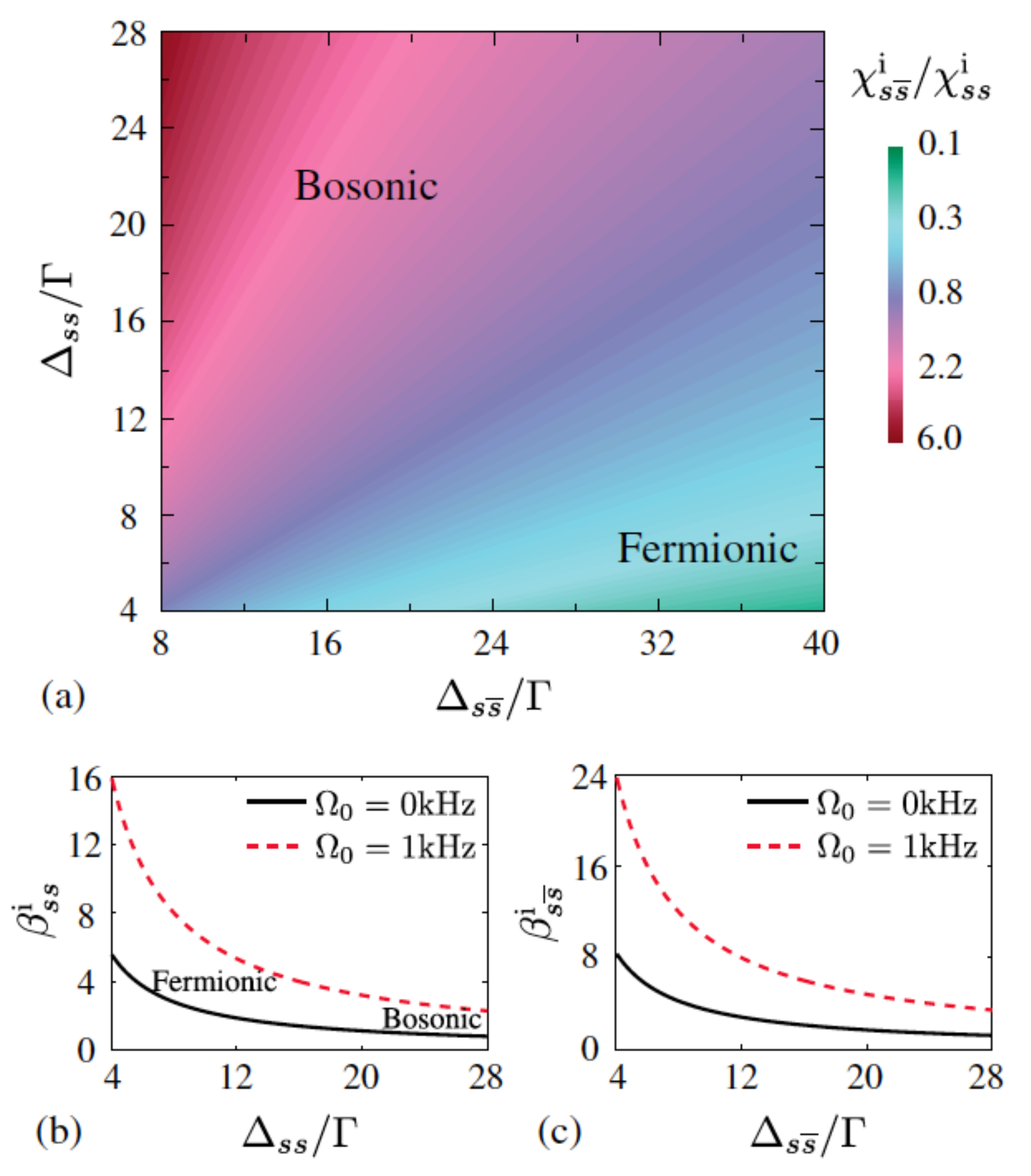}
\caption{(a) The ratio of inter- to intra-species interaction strength as a function of two experimentally controllable detunings for $m_0 = 0$. (b) and (c): The ratio of intra- and inter-species interaction energy to kinetic energy, respectively. Reprinted from Angelakis et al.~\cite{AngelakisKorepin2013}.} 
\label{thirringregime}
\end{center}
\end{figure}

To transform the above equation into the bosonic TM, we need to throw away the quadratic dispersion term and set $\eta_\uparrow = -\eta_\downarrow = \eta$. The latter can be achieved by setting $v_\uparrow = v_\downarrow$ and $\cos2\varphi_\uparrow = -\cos2\varphi_\downarrow$, whereas the former requires the non-relativistic kinetic energy to be much smaller than the relativistic kinetic energy, which can be achieved by an appropriate tuning of the single photon detunings $\Delta_s$. Defining a spinor field ${\bf \Psi} = (\hat{\Psi}_\uparrow,\hat{\Psi}_\downarrow)^T$, the Hamiltonian generating the equation can be written as 
\begin{eqnarray}
H &=& \int dz \bar{\bf{\Psi}}(-i|\eta|\gamma_1\partial_z + m_0\eta^2){\bf \Psi} \nonumber \\ &+& \sum_s\frac{\chi_{ss}}{2} \hat{\Psi}^\dagger_s\hat{\Psi}^\dagger_s\hat{\Psi}_s\hat{\Psi}_s +\frac{\chi}{2}\bar{\bf{\Psi}}\gamma^\mu\bf{\Psi}\bar{\bf{\Psi}}\gamma_\mu\bf{\Psi},
\end{eqnarray}
where $\bar{\bf{\Psi}} = {\bf \Psi}^\dagger\gamma_0$ with the gamma matrices defined as $\gamma_0 = \gamma^0 = \sigma_x$, $\gamma_1 = -\gamma^1 = i\sigma_y$ and $m_0 = -\Omega_0/\eta^2$. Thus, neglecting the small quadratic dispersion term, the dynamics of the stationary polaritons is governed by the bosonic TM. Furthermore, by entering the hardcore regime in which the self-interaction energy dominates over all other terms, one can get into the fermionic regime where, as discussed earlier, the one-dimensional bosons behave in many ways as fermions. Figure \ref{thirringregime} illustrates that this regime (small kinetic energy + hardcore) can be reached in a possible experiment, assuming achievable values for the optical parameters: $|\cos2\varphi_s| = 0.004$, $n = 10^7m^{-1}$, $\bar{\Omega}_2 \approx 1.5\gamma$, $0 < \chi/|\eta| < \pi$, and the single atom cooperativity $\gamma_{1D}/\gamma = 0.2$.

One of the methods to extract interesting physical information in field theories is to calculate correlation functions. In the massless fermionic TM, the n-point correlation function defined as $\langle 0 | \prod_{i=1}^n\hat{\Psi}_\uparrow^\dagger(z_i)\hat{\Psi}_\downarrow(z_i)\hat{\Psi}_\downarrow^\dagger(z'_i)\hat{\Psi}_\uparrow(z'_1) |0\rangle$ has been explicitly calculated \cite{Coleman1975}, and serves as an excellent check for the quantum simulator. After a satisfactory agreement has been obtained in the massless case, one can increase the value of the mass and explore the correlation functions in the relatively unexplored massive regime. To see how the correlation function can be measured, it is helpful to define the fields $\hat{\Psi}_{x,\pm} = (\hat{\Psi}_\uparrow \pm \hat{\Psi}_\downarrow)/\sqrt{2}$ and $\hat{\Psi}_{y,\pm} = (\hat{\Psi}_\uparrow \mp i \hat{\Psi}_\downarrow)/\sqrt{2}$. When written in terms of these fields, the original correlation function becomes a combination of various density-density correlation functions involving the densities of $x$ and $y$, which can be straightforwardly measured. The detection scheme is depicted in figure \ref{thirringsetup}(b). The waveplates mix the $\uparrow$ and $\downarrow$ components into $x$ and $y$ fields, one of which is then detected by using a polarization beam splitter in each arm. In this way, one could probe the scaling behaviour of the correlation functions and moreover infer the renormalization of mass due to the interaction. 

\subsection{Experimental progress}

There are largely two promising experimental techniques to produce the waveguide-atom interface considered in this section. One is loading atoms in a hollow-core photonic band gap fibers (HCPBFs) and the other is coupling atoms to a tapered-fiber or more recently a two-dimensional photonic crystal. The purpose of this subsection is to discuss briefly the experimental progress in these fields.

The HCPBF is a waveguide which has a hollow-core surrounded by a photonic crystal structure \cite{CreganAllan1999,SmithKoch2003} and can therefore accommodate atoms in its core, where the guided mode resides. Strong confinement of light at the core then facilitates the sought-after strong atom-light interaction. The first experimental effort used pressure difference to pump hydrogen atoms into a HCPBF and thereby demonstrate stimulated Raman scattering \cite{BenabidRussell2002}. It was soon followed by the generation of optical solitons in a xenon-filled fiber \cite{OuzounovGaeta2003}. A step towards strong light-matter interaction has been made by demonstrating EIT using acetylene molecules \cite{GhoshGaeta2005} and rubidium molecules \cite{LightLuiten2007}. Optical guiding of thermal \cite{TakekoshiKnize2007} and cold atoms \cite{VorrathSengstock2010} through a HCPBF and trapping of ultracold atoms in a HCPBF \cite{ChristensenPritchard2008} have also been achieved. Most importantly for the subject of this review, guiding and trapping of laser cooled atoms was achieved using a magnetic guide \cite{BajcsyLukin2009}. In the same experiment, authors demonstrated all-optical switching using the four-level giant cross-Kerr effect introduced in Sect.~\ref{sect:stationary light} with a moderate value of optical depth $\approx 30$. More recently, the use of hollow-optical-beam has allowed more efficient guiding of atoms, leading to an improved optical depth of 180 \cite{BajcsyLukin2011}. 

The second promising system consists of atoms coupled to the guided mode of a tapered fiber. When an optical fiber is drawn out such that its diameter is smaller than the wavelength of the guided light \cite{TongMazur2003}, light mostly resides outside the fiber in the form of an evanescent wave, which enables easy coupling between the guided light and surrounding atoms. Coupling trapped cold atoms to the guided mode of a tapered fiber has been demonstrated \cite{NayakHakuta2007,SagueRauschenbeutel2007} and a hot vapor of atoms has been used to demonstrate EIT with a very low pump power \cite{SpillaneShahriar2008}. Trapping and interrogation of cold atoms on the surface of a tapered-fiber by creating an optical lattice above the nanofiber surface has been achieved with an optical depth as high as $\approx 30$ \cite{VetschRauschenbeutel2010,DawkinsRauschenbeutel2011}. Most recently, state-insensitive trapping of atoms has been achieved with an optical depth of 66. 

More recently, people have started looking at two-dimensional nanophotonic structures involving photonic-crystal patterns \cite{HungKimble2013,GobanKimble2014}. These types of systems allow strong light-atom coupling and may lead to an alternative platform to realize the models reviewed in this section. Moreover, other types of models can be readily simulated in this setup, such as the conventional cavity-QED physics \cite{ChangKimble2012} and tunable many-body spin models \cite{DouglasChang2015}. As such, the platform is also a good candidate for a quantum information processor.

\subsection{Outlook: Rydberg polaritons}
Before we close the section, we would like to briefly mention another interesting paradigm to realize strong photon-photon interaction that has attracted much interest recently--Rydberg EIT. 
In Rydberg EIT, the strong long-range Rydberg interaction between atoms are converted to an effective photon-photon interaction through the appropriate use of EIT: one chooses the third state ($|c\rangle$) of the atom to be a Rydberg state (typically requiring the level scheme to be of the ladder type). Due to an energy shift induced by the interaction between Rydberg states, the presence of a single photon in the system then creates a range (called the blockade radius) around the excitation where light propagation is forbidden. This means that two Rydberg excitations cannot travel within the blockade distance, or in other words they repel each other. The nonlinear interaction not only allows one to build photonic quantum information processors \cite{FriedlerKurizki2005,PetrosyanFleischhauer2008,MullerZoller2009,ShahmoonPetrosyan2011,GorschkovLukin2011}, but also to investigate many-body physics in such systems. For example, there has been a proposal to create a BEC of stationary-light dark state Rydberg polaritons \cite{NikoghosyanPlenio2012}, as well as to observe Wigner Crystallization of photons in cold Rydberg ensembles \cite{OtterbachFleischhauer2013}. Experimental progress in the field has been equally amazing; there have been demonstrations of EIT in a Rydberg medium \cite{PritchardAdams2010}, a strong Rydberg blockade \cite{PeyronelVuletic2012,ParigiGrangier2012,DudinKuzmich2012,HofmannWeidemuller2013}, and even a microwave-controlled polariton-polariton interaction \cite{MaxwellAdams2013}. This rapidly growing field would not only provide a means to achieve strong photon-photon interaction but also a chance to simulate different types of phenomena by utilizing the available long-range interaction.   

\section{Summary and outlook}

In this review, we have summarized numerous theoretical proposals and a few experimental demonstrations of strongly-interacting many-body photonic systems. We have started with the early proposals that realized  that atom-induced photon-photon interactions used in tandem with coupled resonator structures can give rise to Hubbard-like models often studied in condensed matter physics and cold atoms in optical lattices. 

Section II reviewed the properties of the many-body ground states of closed coupled-resonator arrays as well as the possible equilibrium phases as calculated via mean-field approaches. We also presented proposals to simulate quantum spin-models and fractional quantum Hall states with photons. 

Section III  focused on works that go beyond equilibrium approaches, exploring experimentally valid situations where the system is naturally out of equilibrium. The differences and similarities between the driven-dissipative photonic Bose-Hubbard and Jaynes-Cummings-Hubbard systems were first highlighted, followed by reviews on early investigations on interaction-driven fermionization and crystallization of photons in different scenarios.  We also discussed two `quenching scenarios', where transient effects of an initially prepared state are probed to reveal the underlying eigenstates of the system in an open but not driven system. Apart from observing the signatures of Mott-superfluid transition, an interesting `self-trapping' was shown to occur, which has been verified experimentally in a superconducting circuit setup. We then summarized a couple of proposals to implement more exotic models. The first is an extended Hubbard model with the cross-Kerr interaction between nearest neighbours which, within the mean-field description, was shown to exhibit a supersolid-like phase. The second is a bosonic Kitaev chain with localized Majorana like modes near the end cavities. Such proposals reveal the versatility of CRA simulators. The section ends with a brief overview of interesting works that we could not cover in detail as well as a short survey on possible experimental platforms.

Section IV presented a very different paradigm which is more suitable for simulating continuum 1D quantum models, but nevertheless shares the main theme in that the central players are interacting photons. The system consists of an ensemble of 3- or 4-level atoms coupled to a 1D waveguide, with EIT or EIT-like coupling schemes. We showed how stationary states of light are formed, whose dynamics can be engineered to a large degree. This way, one can realize a photonic Tonks-Girardeau gas and furthermore add an effective potential whose depth can be varied to allow tuning between the Bose-Hubbard and quantum sine-Gordon limit, which in turn allows the `pinning transition' to occur. We then reviewed a proposal to achieve the Bose-Einstein condensation of dark-state stationary-polaritons if the 2D nature of the cross-section of the waveguide becomes important and interaction is sufficiently strong. Next we discussed a generalization of the 1D system involving two species of atoms or two polarizations for the input pulses, which allows simulations of Luttinger liquids physics, spin-charge separation and BEC-BCS-like crossover. We then showed that one can also implement an interacting relativistic theory, namely the Thirring model, by engineering the dispersion relations of the propagating pulses of light and exploiting the strong EIT-based interactions. Strong interactions between intra-species photons enforce an effective Pauli principle in this 1D system, so that a fermionic model can be implemented in the photonic system. The section finishes with a mini survey on experimental progresses in doing EIT physics with strongly-coupled atom-waveguide interface and a more recent promising platform in this direction using Rydberg-EIT .

Hopefully we have convinced the reader that there have been many exciting--and still ongoing--developments in this young field. Although tremendous progress has been made in the last decade, there is still plenty of room for development. Experimentally, a large-size coupled resonator array with strong atom-resonator coupling in each site has not yet been fabricated, although impressive technological developments in circuit-QED indicate that we are very close to seeing one. Further developments in tools to probe and measure the non-equilibrium many-body states are necessary both experimentally and theoretically. Current theoretical tools are not sufficient to study even a moderate-sized driven-dissipative 2D arrays as we have discussed already, and developing a good numerical or analytical tool is of fundamental importance for further progress. One can ask many questions such as: Will developments in the tensor-networks help? Can we apply techniques from other area such as the Keldysh formalism to analytically find meaningful quantities? What are the optimal observables and ways to characterize the out-of-equilibrium phases in driven-dissipative photonic systems? \\
Regarding simulations of field theories using photonic systems, one of the key questions is whether the EIT-based platform can be generalized to higher dimensions or a different platform is needed. A straightforward option is to utilize the two-dimensions within the waveguide (as we have seen in the BEC of dark state polaritons), but one difficulty is in maintaining a large optical depth while doing so. One can think of increasing the atom-photon coupling strength for this purpose or utilizing the strong interactions between the Rydberg atoms. The latter looks to be a promising avenue with exciting prospects for further discoveries.

\section{Acknowledgements}
We would like to acknowledge the financial support provided by the National Research Foundation and Ministry of Education Singapore (partly through the Tier 3 Grant ``Random numbers from quantum processes''), and travel support by the EU IP-SIQS. We would like to thank the hospitality of the KITP during the program``Many-body physics with light"

\bibliography{review_ref_H.bib}

\end{document}